# EXPLORING EXOPLANETS
## with Interferometry

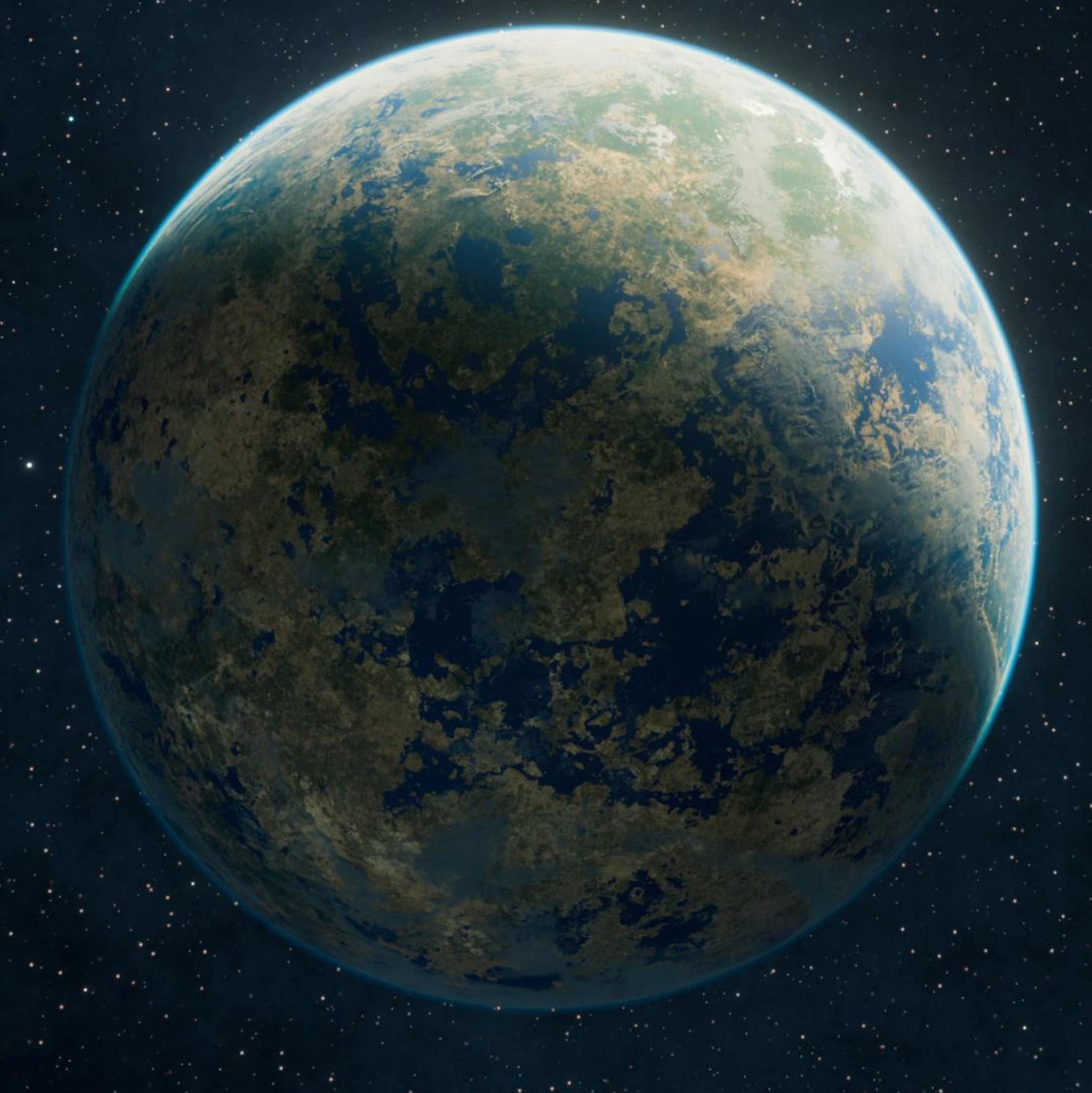

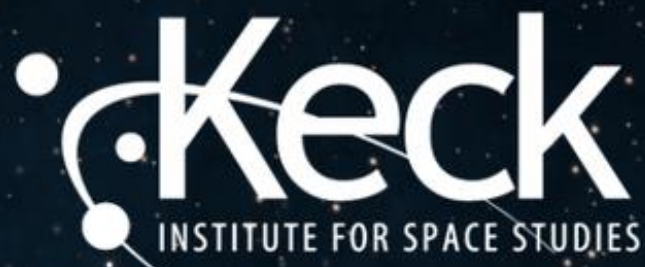

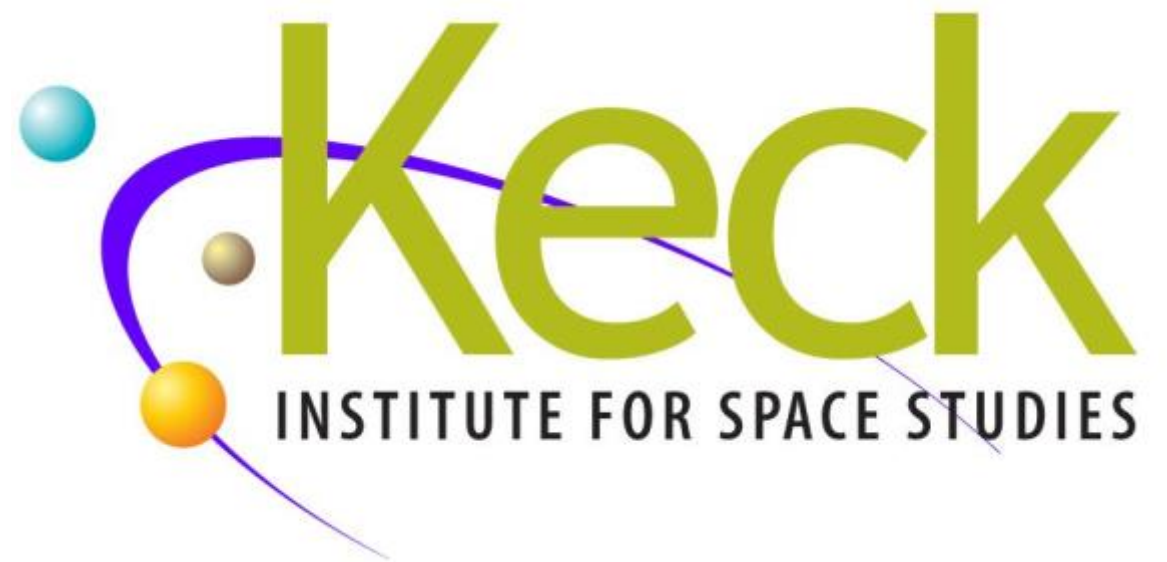

# EXPLORING EXOPLANETS WITH INTERFEROMETRY

Study Workshop: November 28–December 2, 2022

## Study Leads

**Sascha P. Quanz**

ETH Zurich

**Bertrand Mennesson**

Jet Propulsion Laboratory, California Institute of Technology

**Charles Beichman**

Jet Propulsion Laboratory & IPAC, California Institute of Technology

## Pre-Decisional Information—For Planning and Discussion Purposes Only

## Participants

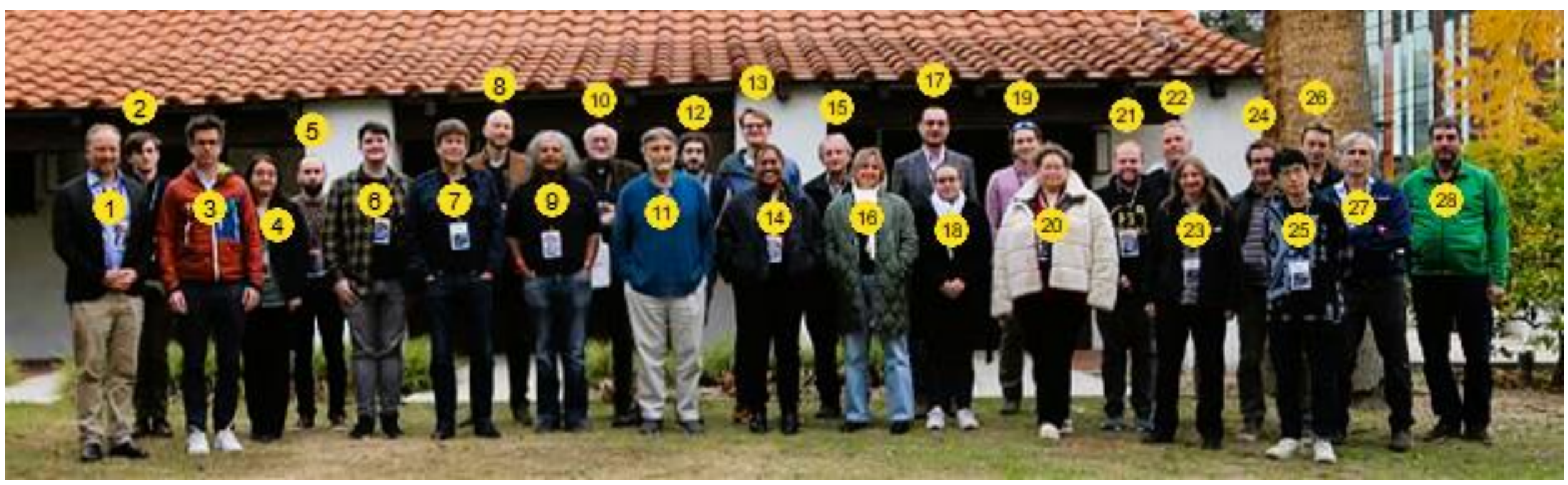

| 1 | Michael Ireland | Australian National University |
|---|---|---|
| 2 | Nicholas Beltsten | Massachusetts Institute of Technology |
| 3 | Sascha Quanz | ETH Zurich |
| 4 | Eleonora Alei | ETH Zurich (now at NASA Goddard Space Flight Center) |
| 5 | Leonid Pogorelyuk | Massachusetts Institute of Technology (now at Rensselaer Polytechnic Institute) |
| 6 | William O. Balmer | Johns Hopkins University |
| 7 | Bertrand Mennesson | Jet Propulsion Laboratory, California Institute of Technology |
| 8 | Denis Defrère | KU Leuven |
| 9 | Gautam Vasisht | Jet Propulsion Laboratory, California Institute of Technology |
| 10 | Malcolm Fridlund | Chalmers University of Technology; Leiden Observatory |
| 11 | Charles Beichman | Jet Propulsion Laboratory & IPAC, California Institute of Technology |
| 12 | Romain Laugier | KU Leuven |
| 13 | Felix A. Dannert | ETH Zurich |
| 14 | Tiffany Kataria | Jet Propulsion Laboratory, California Institute of Technology |
| 15 | Eugene Serabyn | Jet Propulsion Laboratory, California Institute of Technology |
| 16 | Andrea Fortier | University of Bern |
| 17 | Steve Ertel | University of Arizona |
| 18 | Hélène Rousseau | University of Arizona |
| 19 | Kevin Wagner | University of Arizona |
| 20 | Rhonda Morgan | Jet Propulsion Laboratory California Institute of Technology |
| 21 | Jonah T. Hansen | Australian National University (now at ETH Zurich) |
| 22 | Gerard T. van Belle | Lowell Observatory, Northern Arizona University (now at Northern Arizona University) |
| 23 | Gail H. Schaefer | CHARA Array of Georgia State University |
| 24 | Jean-Philippe Berger | IPAG, University Grenoble Alpes |
| 25 | Taro Matsuo | Nagoya University |
| 26 | Ewan Douglas | University of Arizona |
| 27 | John D. Monnier | University of Michigan |
| 28 | Adrian M. Glauser | ETH Zurich |

## Not pictured:

Dimitri Mawet, California Institute of Technology

Michael R. Meyer, University of Michigan

## Recommended citation (long form)

S. P. Quanz, B. Mennesson, C. Beichman, E. Alei, W. O. Balmer, N. Beltsten, J.-P. Berger, F. A. Dannert, D. Defrère, E. Douglas, S. Ertel, A. Fortier, M. Fridlund, A. M. Glauser, J. T. Hansen, M. Ireland, T. Kataria, R. Laugier, D. Mawet, T. Matsuo, M. R. Meyer, J. D. Monnier, R. Morgan, L. Pogorelyuk, H. Rousseau, G. H. Schaefer, E. Serabyn, G. T. van Belle, G. Vasisht, K. Wagner 2026. "Exploring Exoplanets with Interferometry." S. P. Quanz, B. Mennesson, C. Beichman (Eds.) Report prepared for the W. M. Keck Institute for Space Studies (KISS), California Institute of Technology.

## Recommended citation (short form)

S. P. Quanz, B. Mennesson, C. Beichman (Eds.). 2026. "Exploring Exoplanets with Interferometry." Report prepared for the W. M. Keck Institute for Space Studies (KISS), California Institute of Technology.

## Acknowledgments

Study Report prepared for the W. M. Keck Institute for Space Studies (KISS)

Micheal R. Meyer (MRM) gratefully acknowledges support from the International Space Science Institute through a Johannes Geiss Fellowship. Part of this work has been carried out within the framework of the National Centre of Competence in Research (NCCR) PlanetS supported by the Swiss National Science Foundation (SNSF) under grants 51NF40_182901 and 51NF40_205606. Sascha P. Quanz (SPQ), Felix A. Dannert (FAD), Eleonora Alei (EA) and Jonah T. Hansen (JTH) acknowledge the support from the SNSF. Kevin Wagner (KW) acknowledges support from NASA through the NASA Hubble Fellowship grant HST-HF2-51472.001-A awarded by the Space Telescope Science Institute, which is operated by the Association of Universities for Research in Astronomy, Incorporated, under NASA contract NAS5-26555. Denis Defrère (DD) has received funding from the European Research Council (ERC) under the European Union's Horizon 2020 research and innovation program (grant agreement CoG-866070). This work has received funding from the Research Foundation-Flanders (FWO) under the grant number 1234224N. Part of this work was carried out at the Jet Propulsion Laboratory, California Institute of Technology, under a contract with the National Aeronautics and Space Administration (80NM0018D0004). EA's work research was partly supported by an appointment to the NASA Postdoctoral Program at the NASA Goddard Space Flight Center, administered by Oak Ridge Associated Universities under contract with NASA. This research was supported by funding from the Australian Research Council Grant No. DP200102383. JH acknowledges the financial support of the ARC, as well as support from the Australian Government Research Training Program. Michael Ireland (MI) was supported by the Australian Research Council (DP190101477). Jean-Philippe Berger (JPB) acknowledges support by the LabEx FOCUS ANR-11-LABX-0013 and by the Action Spécifique Haute Résolution Angulaire (ASHRA) of CNRS/INSU co-funded by CNES

We wholeheartedly thank the staff at the Keck Institute for Space Studies—Michele A. Judd, Harriet K. Brettle, and Janet W. Seid—for their outstanding support throughout the preparation, execution, and completion of the workshop.

# Executive Summary

Humanity stands at the threshold of answering one of its most profound questions: Does life exist beyond Earth? Ongoing and upcoming space missions, together with powerful ground-based instruments, have prepared the way for a transformational next step—the detailed characterization of Earth analogs orbiting Sun-like and other stars and the search for atmospheric biosignatures that may indicate life.

Within this context, the European Space Agency's Voyage 2050 process has identified the direct detection of thermal emission from temperate terrestrial exoplanets in the mid-infrared (mid-IR) as a top scientific priority. The Large Interferometer For Exoplanets (LIFE)—a space-based, mid-IR nulling interferometer—is designed to meet this goal. LIFE will be capable of detecting climate-relevant gases such as $CO_2$ and $H_2O$, identifying classical biosignatures like $O_3$ and $CH_4$, and probing additional, non-classical biosignatures. It will also provide key data for determining planetary radius, albedo, and temperature, which are essential for assessing habitability.

In parallel, the U.S. National Academy has recommended a complementary mission now called the *Habitable Worlds Observatory* (HWO)—a ~6-meter space telescope equipped with advanced coronagraphs to suppress starlight by a factor of $\sim 10^{10}$ across the visible and possibly into the near-infrared and near-ultraviolet. Together, LIFE and HWO offer synergistic capabilities, enabling a comprehensive and robust assessment of the prevalence of life-bearing exoplanets in our galactic neighborhood—a first in human history.

LIFE builds upon the heritage of NASA's Terrestrial Planet Finder Interferometer (TPF-I) and ESA's Darwin mission studies from the early 2000s. Although no fundamental technological barriers were identified at the time, major advances in exoplanet science, instrumentation, and relevant technologies now make such a mission timely and feasible. We are ready to develop and implement a LIFE-like mission.

The LIFE science team has defined clear objectives and Level 1 requirements for wavelength coverage, spectral resolution, sensitivity, angular separation, and target characterization. Technical studies are underway to translate these into engineering specifications for baseline length, aperture size, and nulling configuration.

Ongoing developments—including work with the Very Large Telescope Interferometer (VLTI) and Center for High Angular Resolution Astronomy (CHARA) interferometers, progress in photonic components, formation flying, and affordable launch systems—will further enable LIFE's realization. Over the coming years, the international LIFE community aims to establish a preliminary mission design and a Design Reference Mission (DRM) to demonstrate its scientific potential, including the detection of habitable-zone planets around late K and M stars, which remain beyond the reach of HWO.

By uniting an international and interdisciplinary community of scientists and engineers, LIFE offers a credible pathway toward the direct detection and characterization of potentially habitable—and even inhabited—worlds.

## Main Recommendations

Detailed recommendations on the various topics discussed during the workshop are presented in the following sections. At the highest level, we advocate for the following actions:

1. **Advance mission development**: The development and implementation of a mid-infrared (mid-IR) nulling interferometry mission—such as LIFE—for the detection and characterization of temperate terrestrial exoplanets should be vigorously pursued.

2. **Quantify scientific potential and synergies**: The unique science opportunities enabled by such a mission should be further quantified, and scientific as well as operational synergies with reflected-light missions should be systematically explored.

3. **Support and coordinate technology development**: Programs for the development and maturation of required technologies (including components, subsystems, and precursor missions) should receive sustained financial support and be globally coordinated. Synergies with related technology development efforts for other missions should also be leveraged.

4. **Leverage ground-based experience**: Lessons learned from past, ongoing, and planned ground-based interferometric instruments should be consolidated, shared, and—where applicable—integrated into the space mission development program.

5. **Build and sustain a global community**: A strong, vibrant international interferometry community should be fostered—one that shares the scientific potential of interferometric observations with the next generation of scientists and engineers and strengthens global collaboration.

# Contents

# 1 Motivation for a mid-infrared LIFE-like mission

## 1.1 General context

Are we alone in the universe? Humankind has been pondering over this question for centuries. We are the first generation that has not only the scientific vision and aspiration, but also the technological means to implement space missions that will start providing long-sought answers based on empirical data. In the context of exoplanet science, these missions are not simply "the next exoplanet missions" on an already impressive list of ground-breaking space missions; these missions have the potential to truly revolutionize our understanding of our place in the cosmos and trigger a "Copernican revolution 2.0."

Today, more than 6000 exoplanets—planets orbiting stars other than our Sun—are known[1]. Statistical analyses suggest the occurrence rate of planets with radii between 0.5 and 1.5 Earth radii and orbital separations that could, in principle, allow for the existence of liquid water on their surface, is of order 10%. This is true for planets orbiting stars with similar effective temperature as the Sun (i.e., G and K dwarfs), but also for planets around much fainter, low-mass M dwarfs (e.g., Bryson et al., 2020, Dressing & Charbonneau, 2015). In the above-mentioned radius range, the planets are expected to be "rocky;" i.e., their bulk density is dominated by silicate rocks and metals (like the terrestrial planets in the solar system) and not by a hydrogen/helium-rich atmosphere (e.g., Rogers, 2015). In parallel to quantifying how many stars may host exoplanets that are similar in size and/or energy influx as Venus, Earth, and Mars, the first steps towards investigating the *existence* of atmospheres around such exoplanets are underway. Because these studies apply to transit spectroscopy and/or secondary eclipse observations (Seager and Deming, 2010), they are limited to exoplanets *transiting* small, (very) low-mass M dwarfs, as only for these objects the achievable signal-to-noise ratio provides some first quantitative constraints (Wordsworth and Kreidberg, 2022). Key examples include results from the TRAPPIST-1 system (Gillon et al., 2017) and K2-18 b (Montet et al., 2015) using data from NASA's *Hubble Space Telescope*, *Spitzer Space Telescope* (e.g., De Wit et al., 2016; Moran et al., 2018; Benneke et al., 2019; Ducrot et al., 2020) and, most recently, the *James Webb Space Telescope* (JWST; Greene et al., 2023; Zieba et al., 2023; Lincowski et al., 2023; Lustig-Yaeger et al., 2023).

There is a common understanding that, even for these most favorable systems, JWST will not allow for a detailed analysis of their atmospheric composition, and in particular, the search for biosignatures, i.e., atmospheric imprints of gases indicating the possible presence of a biosphere on the planet (e.g., Schwieterman et al., 2018), is beyond reach (e.g., Krissansen-Totton et al., 2018; Lustig-Yaeger et al., 2019; Meadows et al., 2023). However, providing empirical evidence that low-mass planets orbiting cool, low-luminosity M-dwarf stars *can retain* an atmosphere despite the prolonged pre-main sequence time of these stars and their generally high level of activity (e.g., high UV fluxes with frequent coronal mass ejections) would be an important finding as it would significantly increase the sample of planets to be investigated in the future. Furthermore, JWST has the potential to get a first glimpse at non-traditional biosignatures in gas-rich exoplanets, such as dimethyl sulfide (DMS), but potential first detections and their interpretation (Madhusudhan et al., 2023) are debated in the community (e.g., Wogan et al., 2024). In addition to JWST, the Ariel mission from the European Space Agency (ESA; Tinetti et al., 2018), a dedicated exoplanet characterization mission for transiting exoplanets with a current launch date in 2029, will significantly expand our understanding of exoplanet atmospheres. Given its relatively small aperture size[2], Ariel will focus on hot and warm, gas dominated planets but has the potential to investigate the spectra of hundreds of transiting exoplanets at wavelengths between 0.5 and 7.8 $\mu$m.

[1] https://exoplanetarchive.ipac.caltech.edu/index.html

To unlock the scientific potential of investigating the atmosphere of *non-transiting* exoplanets, which represent the vast majority of the exoplanet population, new observational approaches relying on direct detection techniques are needed. In other words, all things being equal, transits are so rare that non-transiting examples of potentially habitable worlds will be six times closer than their transiting counterparts. A systematic atmospheric study of tens of Earth analogs and a quantitative assessment concerning the possible existence of biosignature gases requires an optimized direct detection space mission. Only in space can the three key ingredients needed to directly detect and scrutinize the orders of magnitude fainter planets at small angular separations from their host stars can be realized: sufficient spatial resolution, contrast, and sensitivity.

As an ambitious future exoplanet space mission, the Large Interferometer For Exoplanets (LIFE) is driven by the scientific objective to search for life beyond the solar system and investigate the diversity of other worlds (Quanz et al., 2022a,b). LIFE will allow us to take a major step forward in understanding our place in the cosmos—and we are ready, scientifically and technologically, to take this step.

## 1.2 LIFE science objectives and capabilities

### *1.2.1 LIFE: A short description*

Exoplanets can be directly detected either in reflected light, where light from the host star bounces off the object and is subsequently detected and analyzed, or in thermal emission, where the intrinsic emission due to the object's temperature is used as a probe. The *Habitable Words Observatory* (HWO) will probe exoplanets and their atmospheres in reflected light and we refer to Section 1.3 below and to study reports from earlier but similar mission concepts (HabEx and LUVOIR; Gaudi et al., 2020; The LUVOIR Team, 2019) for a detailed description of the science drivers and the characterization potential of HWO. LIFE will follow an alternative approach and measure the thermal emission of exoplanets in the mid-infrared (mid-IR) wavelength regime (cf. Figure 1.1). The exact wavelength range and required spectral resolution is still to be determined, but preliminary studies suggest a range of at least ~6-16 $\mu$m (better ~4-18.5 $\mu$m) and a spectral resolution of R~100 (Alei, Konrad, Angerhausen, et al., 2022; Konrad et al., 2022, 2023, 2024). As a space-based nulling interferometer located in L2 and consisting of several collector spacecraft separated by tens to hundreds of meters and a beam combiner spacecraft (Figure 1.2), LIFE will provide sufficient

[2] Ariel will feature an off-axis Cassegrain telescope having a 1.1 x 0.7-meter primary mirror providing a collecting area of 0.61 m$^2$.

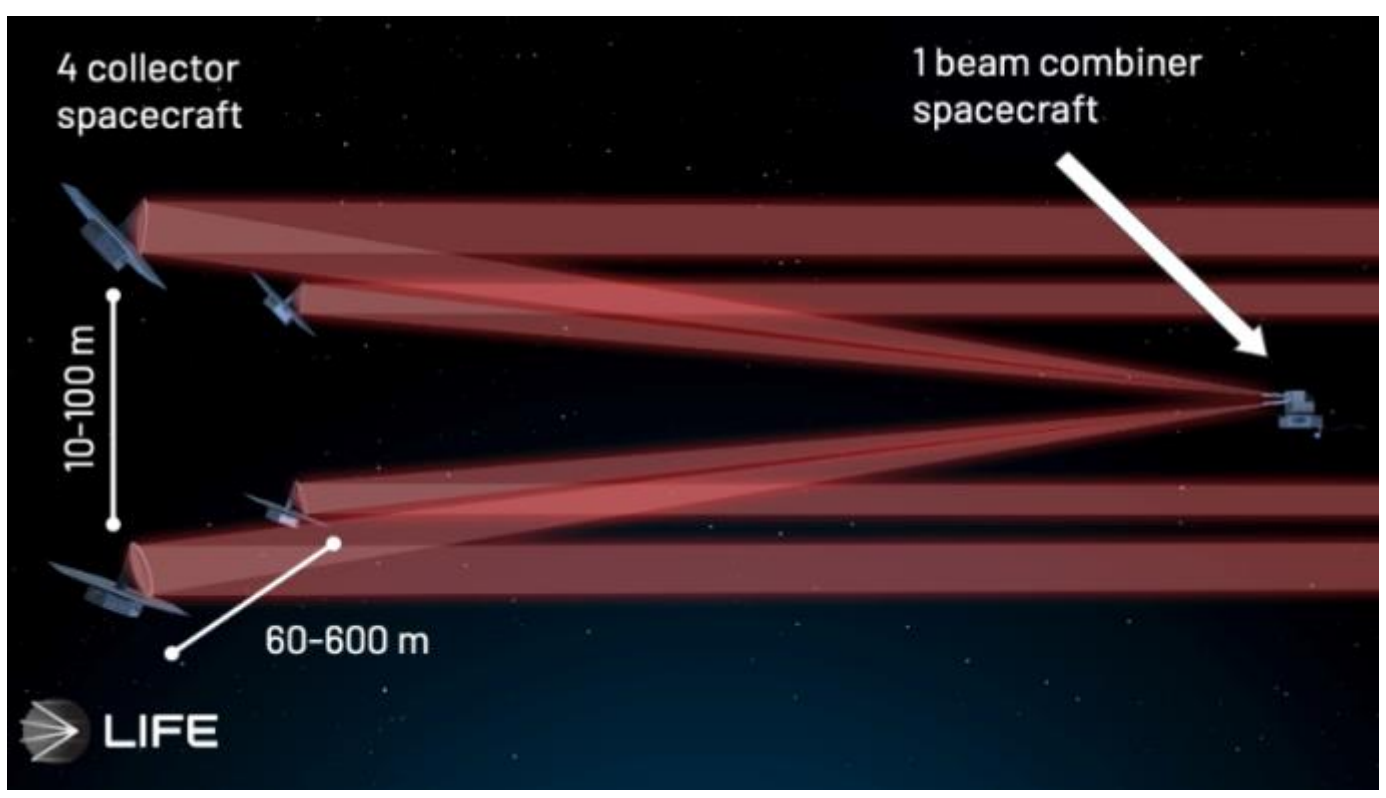

Figure 1.2: Artist impression of the LIFE mission. In this case, a mission concept consisting of four collector spacecraft and one beam combiner spacecraft located outside the plane of the collector spacecraft is shown. Credit: The LIFE Project/ETH Zurich

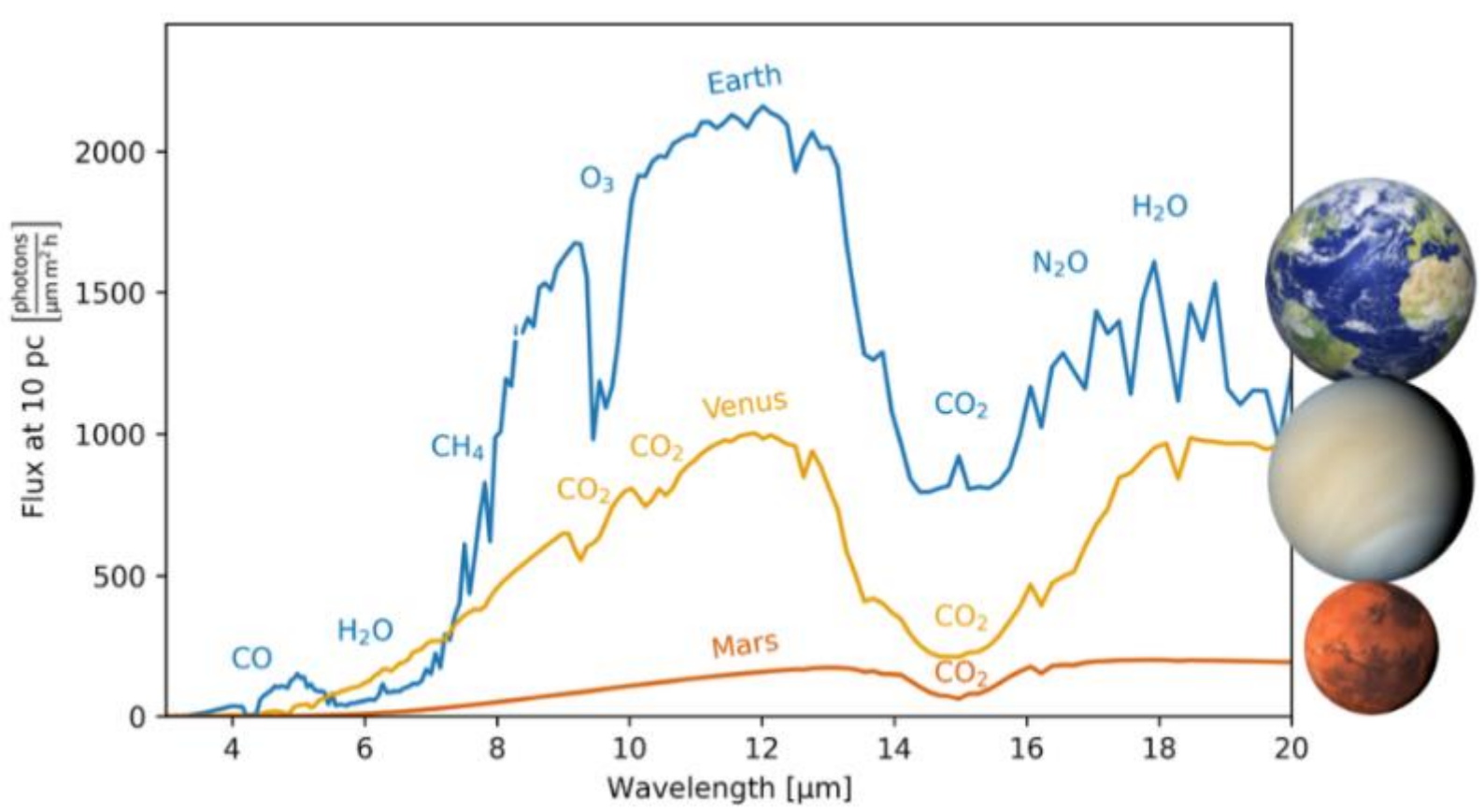

Figure 1.1: Thermal emission spectra of Earth, Venus and Mars illustrating that the planets' thermal emission peaks in the mid-IR regime that LIFE will cover. The Earth is the brightest of all solar system planets at 10 µm. Several absorption bands of important atmospheric constituents are also accessible in the wavelength range including the biosignature gases $CH_4$, $O_3$ and $N_2O$ in the Earth spectrum. Credit: B. Konrad/The LIFE Project

spatial resolution, contrast performance, and sensitivity to detect and characterize the atmospheres of hundreds of nearby extrasolar planets, including dozens that are similar to Earth, assuming a total mission lifetime of 5–6 years (Kammerer et al., 2022; Quanz et al., 2022a,b). A preliminary mission baseline concept is described in Glauser et al., 2024. At this point in time, it is assumed that, prior to launch, a significant fraction of the best-possible targets within 20–25 pc from the Sun will *not* be known. Additional efforts are needed to push ground-based radial velocity surveys to reach the required precision to reliably detect Earth-like planets in Earth-like orbits around nearby solar-type stars. Similarly, there is no current plan for a space astrometry mission capable of reaching that level of precision. This is not a showstopper, however, as the LIFE mission will feature a "detection campaign" and search nearby stars for yet undiscovered planets. As the mission progresses, the fraction of time spent on in-depth spectroscopic characterization will increase and we refer to this as the "characterization campaign." Importantly, both campaigns do not necessarily follow each other sequentially as there is a large number of different types of known nearby exoplanets that LIFE can start characterizing from the first day of the mission (Carrión-González et al., 2023).

### *1.2.2 Science drivers for LIFE*

LIFE is proposed to be an optimized platform for carrying out a broad range of specific experiments to investigate the existence of life beyond the solar system. These experiments can be grouped into at least three themes:

(1) **Earth-like biospheres**, i.e., temperate, terrestrial exoplanets similar to Earth orbiting a star similar to the Sun and showing atmospheric biosignatures similar to those that have been present in large amounts throughout the evolution of our home planet (e.g., $CH_4$, $O_3$).

(2) **Non-Earth-like biospheres**, i.e., exoplanets that show atmospheric biosignatures other than those found in significant and detectable abundances in Earth's atmosphere. Depending on the molecules of interest, these targets may also include other types of exoplanets such as super-Earths or sub-Neptunes (e.g., for DMS). Furthermore, they may orbit also other types of host stars, e.g., low-luminosity M dwarfs.

(3) **Technosignatures**, i.e., exoplanets that show atmospheric constituents (e.g., $CF_4$, $SF_6$) hinting at the existence of an extraterrestrial intelligence that has been actively manipulating the atmospheric composition.

These mission-driving science themes must be translated into specific science objectives from which high-level science requirements have to be derived. These include, for instance, the number and types of exoplanets that need to be investigated, any spectral-type dependency that should be taken into account, and also the observed atmospheric signals and the metric that would be used to quantify whether life has been detected (cf. Catling et al., 2018; Schwieterman et al., 2018). It is worth pointing out that preliminary studies in the context of Earth-like atmospheric biosignature detection with LIFE have already been carried out (e.g., Alei et al., 2022; Konrad et al., 2022, 2024) and also that the search for non-standard biosignatures and technosignatures with LIFE was already addressed in recent publications (e.g., Angerhausen et al., 2023, 2024; Schwieterman et al., 2024). These studies serve as an excellent basis for deriving quantitative requirements.

We strongly emphasize that the identification and interpretation of atmospheric biosignature or technosignature gases will require additional contextual information about the candidate planet itself, the stellar system it is part of, but also about the broader exoplanet population that provides reference cases for comparison. Hence, it will be crucial to infer as much information as possible about the exoplanets in question and to detect and investigate as large a sample as possible. Since the detection of warmer or larger exoplanets is easier than the detection of temperate, terrestrial exoplanets, and since, according to the exoplanet statistics from the Kepler mission (e.g., Bryson et al., 2020), warmer/larger planets are expected to be found frequently, the search campaign of LIFE will reveal a large number of planets covering a wide parameter space (Quanz, Ottiger, et al., 2022; Figure 1.3). This unique sample will offer unprecedented insights into physical and chemical properties of exoplanet atmospheres either directly from first epoch data during the search campaign, provided the signal-to-noise of the data is sufficient, or from dedicated second-epoch observations. Classifying diverse exoplanet atmospheres into groups that exhibit important atmospheric chemical cycles (e.g., Mills et al., 2021) could be a vital step in understanding the outcomes of planet formation and evolution in terms of composition and physical conditions, which in turn will be crucial for the successful identification of biosignatures.

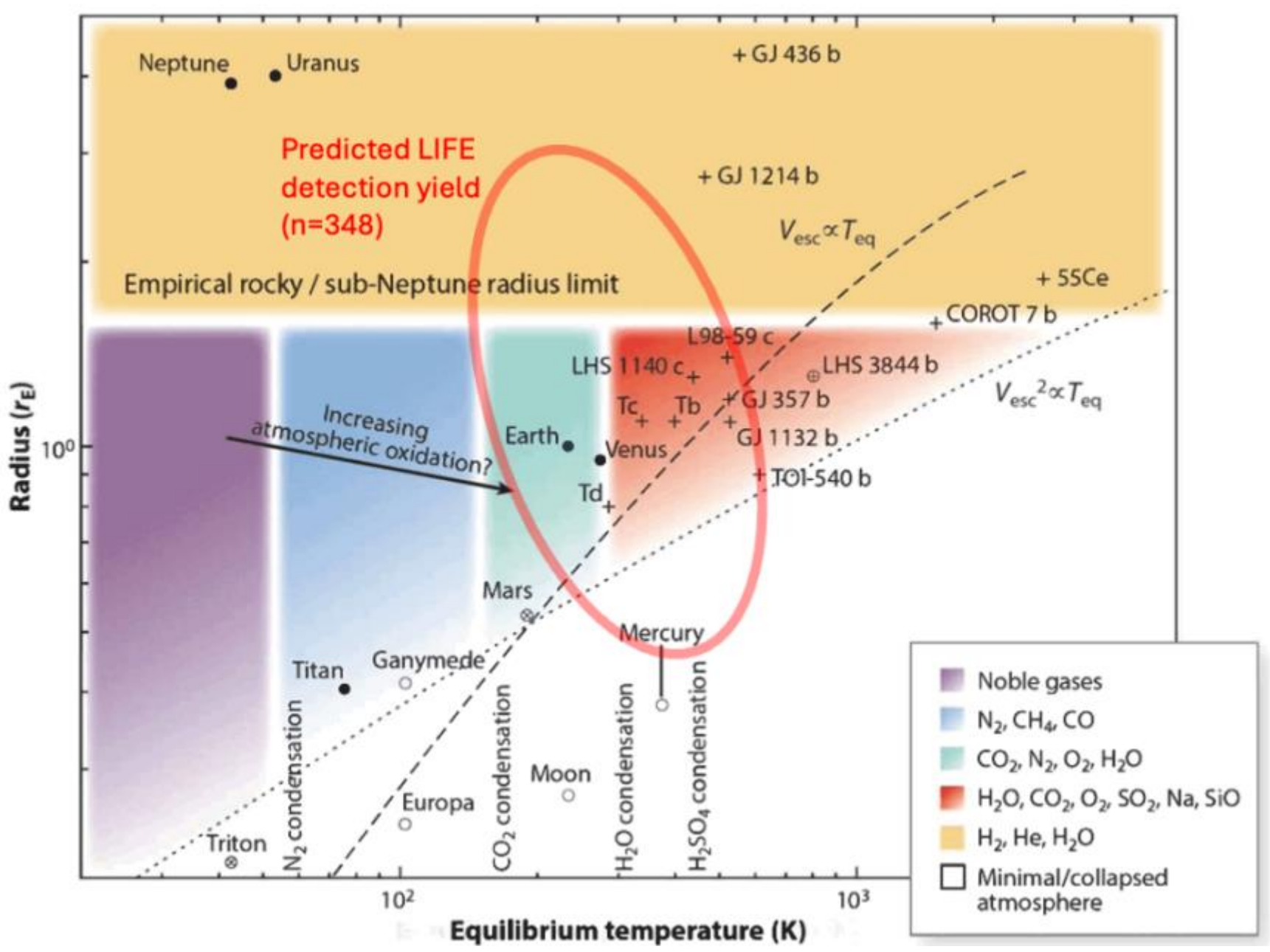

Figure 1.3: Plausible rocky planet atmospheric compositions as a function of equilibrium temperature (assuming an albedo of 0.5) and radius. Filled circles represent solar system planets and moons with stable atmospheres, crossed circles represent those with collapsed atmospheres, and open circles represent those without atmospheres. Plus symbols (+) represent the somewhat limited number of exoplanets amenable to near-future characterization with JWST. They occupy a region of parameter space that is not observed in the solar system. The red ellipse indicates the primary discovery space of LIFE resulting from a 2.5-year search phase and revealing ~350 planets. Abbreviations: Tb, Tc, and Td = TRAPPIST-1 b, c, and d, respectively; 55Ce = 55 Cancri e. (Plot adapted from Wordsworth & Kreidberg, 2022 with permission; red ellipse based on search phase detection yield computations presented in Quanz, Ottiger, et al., 2022.)

Finally, we note that, in principle, a mid-IR interferometer in space may also benefit several non-exoplanet science cases. As the search for life beyond the solar system will drive the overall mission requirements and design, a careful trade needs to be conducted between the expected science return and the potential increase in mission complexity and risk, if additional observing modes and science cases are considered. We discuss this point in Section 8.

### *1.2.3 Unique measurement capabilities in the mid-IR with a LIFE-like mission*

A dedicated mid-IR mission that is optimized for characterizing the thermal emission spectra of dozens of nearby temperate terrestrial exoplanets and searching for atmospheric imprints of biological activity will provide a unique set of measurement capabilities that will clearly set it apart from any other mission or instrument. LIFE will:

- directly constrain the **pressure-temperature structure** of exoplanet atmospheres.
- access (multiple) atmospheric absorption bands of **major molecules** such as $H_2O$ and $CO_2$, as well as collision-induced absorption from $N_2$ and $O_2$.
- search for numerous **atmospheric biosignatures** in the context of terrestrial exoplanets and gas-dominated super-Earths and sub-Neptunes (e.g., $O_3$ and $CH_4$, but also $N_2O$, $PH_3$, $NH_3$, and $CH_3Cl$; see, e.g., Schwieterman et al. (2018); Schwieterman & Leung, 2024 for an overview).
- search for **atmospheric technosignatures** such as $NF_3$ and $SF_6$ (e.g., Seager et al., 2023; Schwieterman et al., 2024).

Many of the molecules mentioned above are more readily detectable at mid-IR wavelengths than at shorter wavelengths. In particular for the searches for biosignatures, the primary objective of LIFE, one should be as agnostic as possible and have access to as many features as possible acknowledging that biospheres on exoplanets may arise and evolve very much differently than on Earth. Figure 1.4 summarizes the strength of mid-IR observation for the detection of biosignatures.

In addition, it is important to highlight the following opportunities of the mid-IR wavelength range as they directly relate to the detection and characterization efficiency of the mission in particular in comparison to a reflected light mission. LIFE will:

- constrain directly the **effective temperature** of exoplanets as well as their bolometric luminosities, and thanks to the known distance to all target stars, provide estimates of their **radii;**
- be **less affected by clouds** and, depending on the atmospheric composition, provide **direct constraints on surface conditions** such as pressure and ground temperature;
- secure a higher detection completeness during search phase as it is **less affected by the orbital phase function** of the exoplanets' emission compared to reflected light missions;
- immediately start characterizing already known small, temperate exoplanets around nearby M stars previously detected from ground-based radial velocity surveys.

In the absence of an empirical mass estimate for the exoplanets, constraining their radius with sufficient precision is critical—and potentially more discriminant than mass when it comes from separating between super-Earths and sub-Neptunes—as this will be a key parameter for creating a prioritized list for second epoch or characterization observations. According to preliminary results, LIFE will provide data that will allow the community to constrain the radius of temperate rocky exoplanets to within 20% uncertainties from a single epoch observation, i.e., during the search campaign (Dannert et al., 2022). Reflected light data suffer from a radius-albedo degeneracy, which makes the determination of the planetary radii more complicated and uncertain (e.g., Carrión-González et al., 2020).

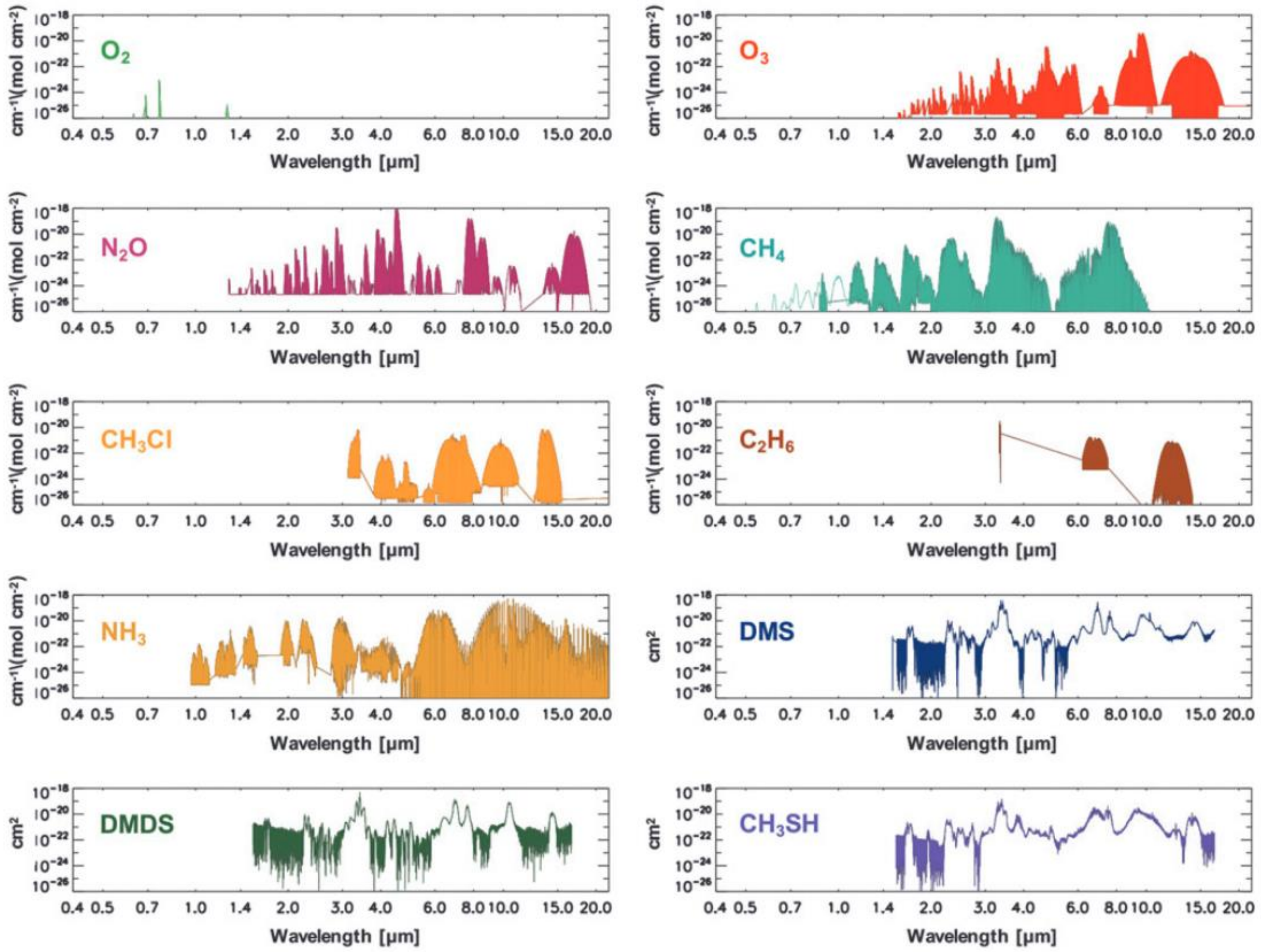

Figure 1.4: Biosignature gas absorption features. Line intensities (cm$^{-1}$/[molecule cm$^{-2}$]) for $O_2$, $O_3$, $N_2O$, $CH_4$, $CH_3Cl$, $C_2H_6$, and $NH_3$ are sourced from HITRAN 2012 (Rothman et al., 2013), while cross-sections (cm$^2$) for DMS, DMDS, and $CH_3SH$ are sourced from the Pacific Northwest National Laboratory (PNNL) spectral database (Sharpe et al., 2004). $C_2H_6$, ethane; $CH_3Cl$, methyl chloride; $CH_3SH$, methanethiol; DMDS, dimethyl disulfide; DMS, dimethyl sulfide. Figure from Schwieterman et al. (2018). © Mary Ann Liebert, Inc.

The last bullet point listed above needs to be emphasized as it means that from the first day of the LIFE mission, there will be a list of targets for direct detection and atmospheric characterization with empirical mass and orbit parameter constraints (Carrión-González et al., 2023).

## 1.3 Synergies with Habitable Worlds Observatory (HWO)

### *1.3.1 Description of HWO and its exoplanet science capabilities*

The Decadal Survey on Astronomy and Astrophysics 2020 (Astro2020; hereafter the Decadal Survey) recommended the creation of a Great Observatories Maturation Program (GOMAP) to advance technology and science in preparation for a suite of great astronomical observatories, starting with the HWO, a ~6-meter class space observatory mission that would be launched in the early 2040s. In response to this recommendation, NASA formed the HWO Technology Maturation Project Office (TMPO) in August 2024, charged with coordinating HWO technical and scientific pre-formulation efforts. HWO shall be optimized for exoplanet direct imaging and spectroscopy at UV, optical and near-IR wavelengths, and designed to provide revolutionary capabilities in many other fields of astrophysics. One of the main scientific goals of HWO is to directly detect habitable zone (HZ) Earth-sized planets ("exo-Earths") in reflected starlight around Sun-like stars, characterize their atmosphere, explore their habitability and search for possible biosignatures using UV/optical/NIR spectroscopy.

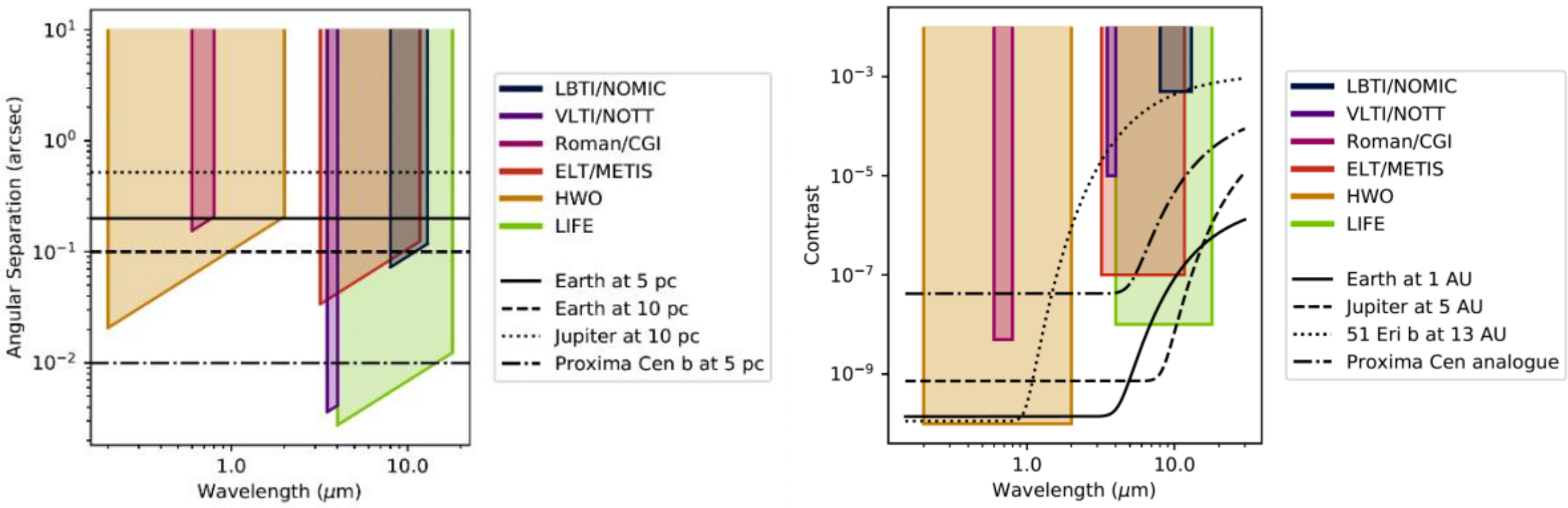

Figure 1.5: Angular resolution (left) and contrast performance (right) of various missions compared to planetary systems. Figure adapted from Hansen (2023).

While the exact science capabilities of HWO remain to be fully defined in the future, the Decadal Survey recommendation, together with previous large mission concept studies such as the Large Ultraviolet Optical Infrared Surveyor (LUVOIR) and the Habitable Exoplanet Observatory (HabEx) (The LUVOIR Team, 2019; Gaudi et al., 2020) provide a useful basis for understanding the promises and limitations of such a reflected light exoplanet direct imaging and spectroscopy mission. HWO will search for exo-Earths and probe their habitability by surveying over a hundred nearby stars, mostly FGK dwarfs.

For each of these planets, it is anticipated that at visible wavelengths alone, HWO would be able to detect the presence of an atmosphere (through Rayleigh scattering short of 0.6 $\mu$m), as well as water vapor (habitability tracer, e.g., around 0.82 and 0.94 $\mu$m), oxygen (biosignature gas, e.g. around 0.76 $\mu$m) if present at present Earth abundance level (PAL), as well as methane if present at Archean Earth levels (e.g., around 0.73 and 0.89 $\mu$m). Detection of oxygen or methane at significantly lower levels will be challenging for HWO. For oxygen, it would require high-contrast observations of exo-Earths in the UV, where planetary photons are scarce, looking for ozone, a by-product of oxygen that still produces a broad UV absorption feature at significantly lower oxygen levels than PAL. For methane, it would require high-contrast observations at near-IR wavelengths, where resolving faint planets from their host star is more difficult than at visible wavelengths, but where methane absorption features are more pronounced (e.g., around 1.0 and 1.69 $\mu$m). Carbon dioxide may also be diagnostic (e.g., around 1.59 $\mu$m) to constrain the origin of any methane detected.

### *1.3.2 Joint detection space*

A first indication about the joint discovery space of LIFE, HWO and potentially other missions and ground-based instruments is presented in Figure 1.5, in which we show a first-order comparison between the spatial resolution and (expected) contrast performance. Only the combination of LIFE and HWO will give access to Earth analogs over a large wavelength regime.

As first proof that the joint detection space between HWO and LIFE is populated, a preliminary stellar target list independently generated for the LIFE mission restricted to only FGK-type stars is compared to a preliminary star list put together in the context of HWO (Figure 1.6). The target stars shared between both missions demonstrate the joint discovery space, which is further supported by the overlap in detection parameter space for known planets (Carrión-González et al., 2023). However, the small number of shared targets (~15% of the total number of targets), when both missions are scheduled independently, highlights the necessity to develop a joint strategy to maximize the overall science return.

### *1.3.3 Joint characterization*

A reflected light mission like HWO and a LIFE-like mission, detecting the thermal emission of exoplanets, have both their unique strengths, and highlighting the unique characterization space for each individual approach will be important. However, their combined diagnostic power for investigating other worlds cannot be overstated and warrants a thorough and comprehensive understanding of how and where synergies can be fully leveraged.

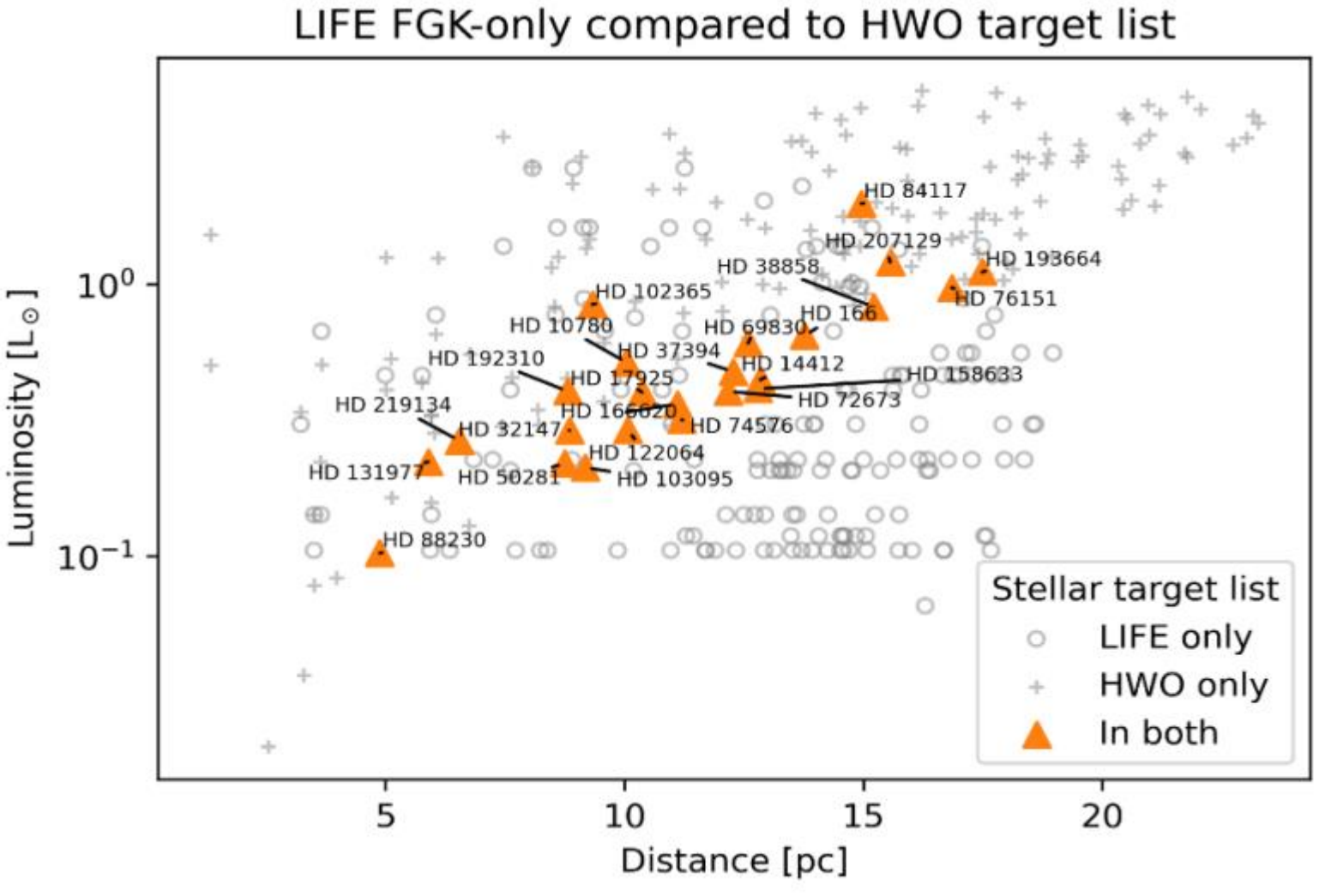

Figure 1.6: Stellar targets common to the LIFE and HWO (NASA ExEP Mission Star List for the HWO, 2023) target list. The LIFE target list is retrieved for a 3-meter aperture setup (equal collecting area to the ~6-meter HWO aperture) and for observing only FGK-type stars. We note that this overlooks the significant discovery potential that LIFE has around M-type stars. Reproduced with permission by F. Dannert & R. Morgan.

Figure 1.7 shows the various atmospheric features an Earth-twin exoplanet displays between 0.2 and 22 $\mu$m, hence capturing the reflected light portion of the spectrum as well as the intrinsic thermal emission. Having access to both parts of the spectrum provides the most holistic way of searching for Earth-like exoplanets. Furthermore, resolving the radius-albedo ambiguity inherent to observations in reflected light enables assessment of the full energy budget of the planet. If the emergent flux suggests an equilibrium temperature in excess of that implied by the albedo, models that incorporate the greenhouse effect can be used quantitatively, incorporating abundances from molecules detected in the spectrum. Discovering an active greenhouse effect on a world outside the solar system would be a major milestone in the field and guide our ability to accurately identify atmospheric biosignatures.

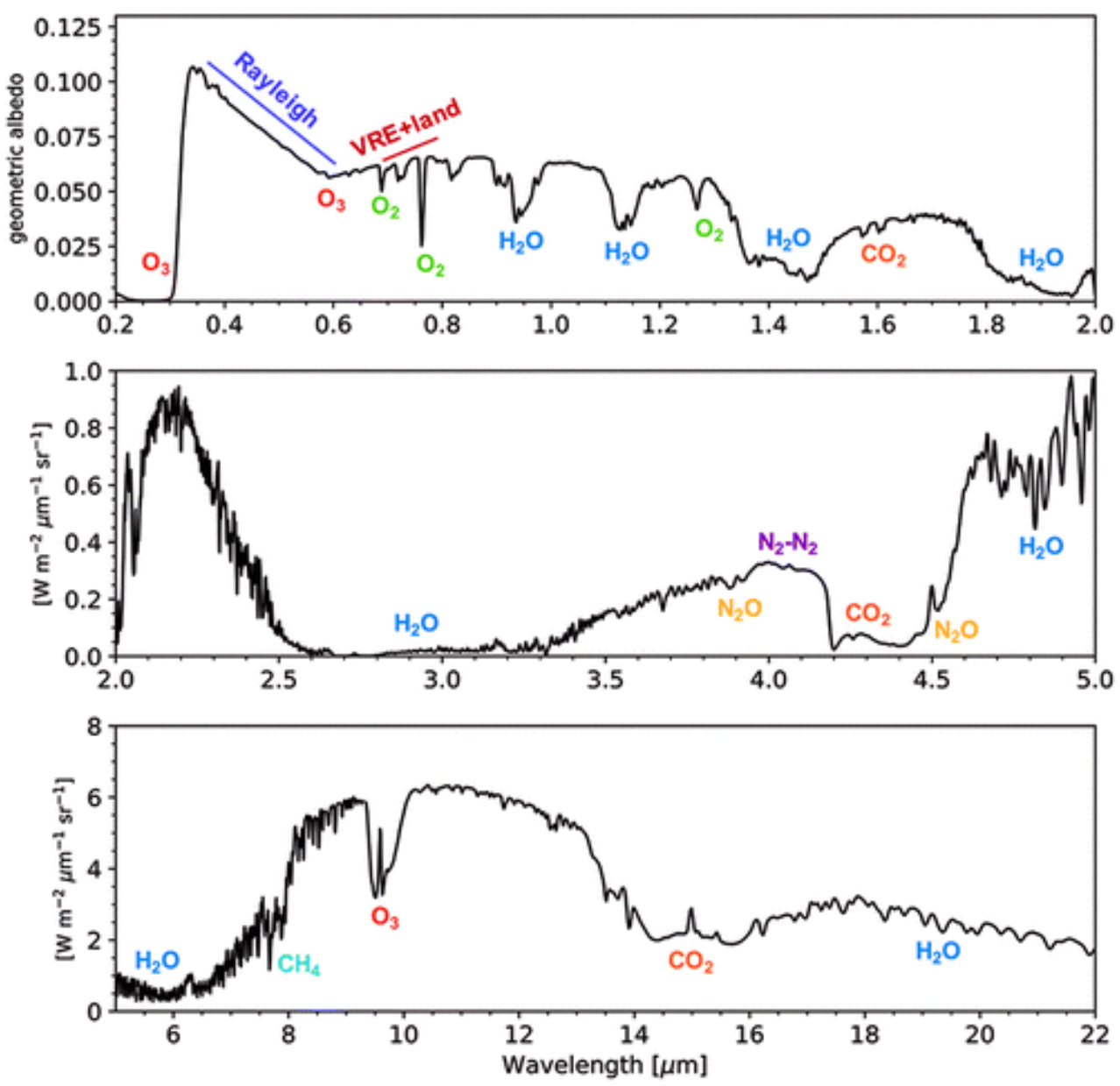

Figure 1.7: Simulated average spectrum of the Earth spanning the visible to the mid-IR wavelength range. Main features and molecular absorption bands are indicated. Rayleigh = spectral slope due to Rayleigh scattering; VRE+land = vegetation red-edge and land reflection. (Figure from Schwieterman et al 2018; reproduced with permission by Mary Ann Liebert, Inc.)

In the absence of a direct measurement, an approximation for the planetary mass can be obtained by combining any observed available upper limits on the mass with mass estimates provided by mass-radius relationships (e.g., Chen & Kipping, 2016; Otegi et al., 2020) where the radius is derived from the mid-IR data. Further improvements towards achieving the required dynamical precisions (< 10 cm/s for RV or < 1 $\mu$as for astrometry) for an empirical mass measurement are beyond the scope of this report.

Going forward, in addition to continuing a more detailed analysis of target sample overlap, it will be important to carry out atmospheric retrieval studies to quantify the increase in science return when both reflected light data as well as thermal emission data are available for (temperate, terrestrial) exoplanets. A first step in demonstrating the scientific potential of joint reflected light and thermal emission retrievals was presented in Alei et al. (2024) for a (cloud-free) modern Earth-twin exoplanet. The presented analyses demonstrate how joint observations constrain the full set of parameters accessible in both wavelength ranges. Additional work will be required to quantify the gain in knowledge of joint retrievals including clouds, assuming different types of exoplanets, and applying more realistic noise models for both mission as the designs mature.

#### *1.3.4 Joint operations*

Beyond scientific synergies, one could also imagine several operational synergies between a reflected light mission like HWO and a LIFE-like mid-IR mission when flown at the same time. For instance, during the search phase, identifying the best targets for in-depth follow-up might be more efficient when access to thermal emission data is available, as this would reduce the number of re-visits needed. This is particularly true for objects that are candidates for being rock-dominated and appear to be located close to the habitable zone (HZ). For instance, projection effects and degeneracies make a robust assessment based on single epoch reflected light data alone very challenging. However, in combination with data constraining the thermal emission of the planet, and hence its radius and temperature, stronger statements about the true nature of the object can be derived. Overall, this could lower the time searching for planets and increase the total number of planets that can be investigated in greater detail. More quantitative work in this direction is needed.

## 1.4 Synergies with ground-based extremely large telescopes (ELTs)

In the coming years and decades, exoplanet science will also be driven forward by several new ground-based observatories and instruments. In combination with new space missions, they will allow us to understand the diversity of planetary bodies and system architectures, providing unprecedented data for planet formation and evolution models.

For the specific goal of directly detecting and characterizing temperate, terrestrial exoplanets and searching for life outside the solar system, flagship-style space missions like LIFE and HWO are, however, indispensable. A remaining gap in discovery space will be filled with the advent of ground-based 30–40-meter extremely large telescopes (ELTs). This gap concerns reflected light observations of small planets around nearby M-type dwarf stars, which have much tighter HZs than Sun-like stars. HWO, with its ~6-meter primary aperture, will not provide sufficient spatial resolution to directly image many HZ planets around these M stars. Given that the contrast required for their detection in reflected light is, however, two orders of magnitude lower than that for the detection of Earth analogs around solar-type stars, they are within reach of high-contrast ELT instruments (e.g., Kasper et al., 2021; Figure 1.8). In a complementary approach, adaptive optics assisted mid-IR instruments on ELTs might be able to image small planets around the very nearest AFGK stars (e.g., Bowens et al., 2021; Quanz et al., 2015). However, their numbers will remain low, and a detailed atmospheric characterization will not be possible except perhaps in the Alpha Centauri system.

## 1.5 Why now?

The time is right to engage in a systematic and coordinated search for life beyond the solar system, develop a mission like LIFE in parallel to HWO, and trigger the “Copernican Revolution 2.0.”

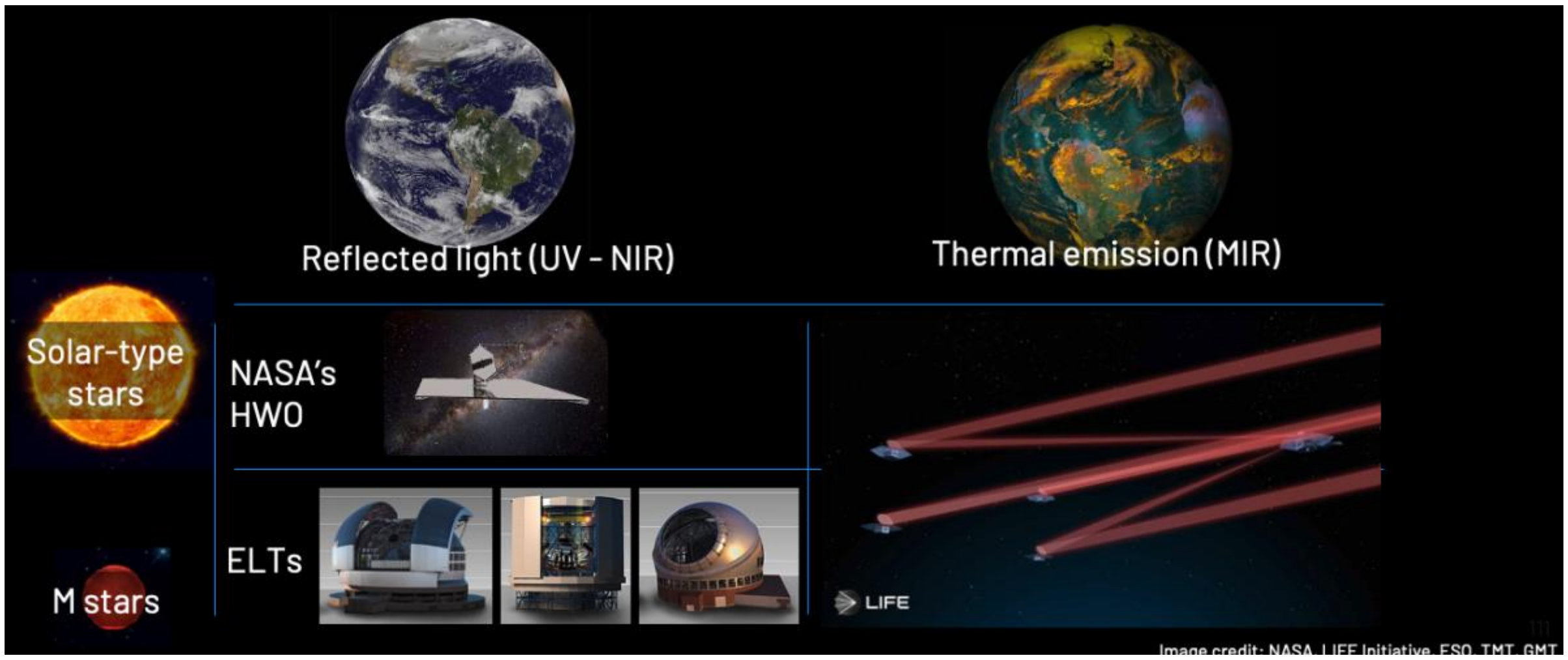

Figure 1.8: Primary discovery space for the direct detection and characterization of temperate, terrestrial exoplanets for different observatories. While a LIFE-like mission will provide thermal emission data irrespective of spectral type of the host star, obtaining reflected light data will require an HWO-like mission for planets orbiting Sun-like stars and ground-based ELT instruments for planets around nearby M stars. Credit: NASA, The LIFE Project, ESO, TMT, GMT

### *1.5.1 Scientific context*

- Our understanding of the exoplanet population has advanced significantly since the study of similar concepts like NASA's TPF-I and ESA's Darwin in the early 2000s. With initial estimates for the occurrence rate of Earth-like planets now available, we can robustly quantify the expected detection yield and science return of a LIFE-like mission.

- The Hunt for Observable Signatures of Terrestrial Systems (HOSTS) survey carried out at the Large Binocular Telescope Interferometer (LBTI) demonstrated that exozodiacal dust disks have, on average, low surface brightness in LIFE's mid-infrared wavelength range, and that the noise they add to data is no longer considered a limiting factor for future exoplanet imaging missions.

- Dozens of exoplanets have been discovered to orbit stars within 10 pc of the Sun. As this population continues to grow with the continuation of ground-based radial velocity searches, it provides an immediate catalog of compelling targets for atmospheric characterization with a LIFE-like mission.

- Significant effort has been put into understanding the concept of atmospheric biosignatures and how to quantify their potential detection given signals may arise from false positives. In addition, statistical frameworks have been put forward to quantify the possible detection of extraterrestrial life. While more work in this direction is needed, the exoplanet and astrobiology communities have identified the need to develop more robust quantitative approaches to prepare the interpretation of data from future missions.

- Understanding the origin of life on Earth and engaging in the quest for life outside our home planet has become a key topic for many academic institutions. The recent establishment of several dedicated research centers worldwide demonstrates a clear global commitment to advancing this field[3].

### 1.5.2 *Advances in technology*

As further detailed in the following sections, significant advances in relevant technologies were made over the last 10–15 years, setting the stage for a more targeted and dedicated technology maturation program for a LIFE-like mission. Examples include:

- On-sky demonstration of interferometric nulling at mid-IR wavelengths with the LBTI and the Keck interferometer
- The ongoing development of a mid-IR nulling instrument for the VLTI
- The development of new beam-combination schemes potentially reducing risk and increasing the scientific performance of future nulling interferometers
- Advances in astro-photonics that allow for a significant reduction in mass and complexity compared to bulk-optics approaches in astronomical instrumentation
- The development of cryogenic, mid-IR instrumentation for space missions such as JWST Mid-Infrared Instrument (MIRI) or Ariel
- Advances in low-noise, highly stable mid-IR detector technologies and the advent of novel photon-counting energy-resolving detectors with possible applications in the mid-IR regime
- Ongoing and upcoming formation-flying missions that will contribute to the maturation of key technologies for autonomous, high-precision formation flying as needed for a LIFE-like mission

### 1.5.3 *Community support*

- The report of the Senior Committee[4] to the Director of Science of ESA in the context of the Voyage2050 process to define the topics for future L-class missions in the ESA Science Program underlines the scientific impact a LIFE-like mission will have by stating that
- “[...] launching a Large mission enabling the characterization of the atmosphere of temperate exoplanets in the mid-IR should be a top priority for ESA within the Voyage 2050 timeframe.”

---

[3] Examples include the Leverhulme Centre for Life in the Universe (University of Cambridge, UK; https://www.lclu.cam.ac.uk/), the Origins of Life Initiative (Harvard University, USA; https://origins.harvard.edu/), the Center for the Origins of Life (University of Chicago, USA; https://originsoflife.psd.uchicago.edu/), the Centre for Origin and Prevalence of Life (ETH Zurich, CH; https://copl.ethz.ch/), or the Earth–Life Science Institute (ELSI, Tokyo; https://www.elsi.jp/en/).

[4] https://www.cosmos.esa.int/web/voyage-2050

- “Being the first to measure a spectrum of the direct thermal emission of a temperate exoplanet in the mid-IR would be an outstanding breakthrough that could lead to yet again another paradigm-shifting discovery.”

- The search for life outside Earth is not only strongly supported by the space science and astrophysics communities of NASA and ESA, but by diverse communities around the world including Australia, Japan, and China. Also, the European ground-based astronomy community considers this topic to be of highest relevance in the coming ~15–20 years based on the outcome of the Scientific Prioritization Community Poll 2020 of the European Southern Observatory (ESO) (see Figure 1.9).

- In contrast to earlier attempts to implement flagship-class missions to search for life outside the solar system, the international communities stand united in the vision that chances of success are only maximized through collaborative international partnerships and coordinated efforts involving more than one space agency and more than one space mission. Only this will allow for a truly holistic and agnostic approach covering a large wavelength range with unparalleled diagnostic potential and involving a large enough number of characterized temperate, terrestrial exoplanets.

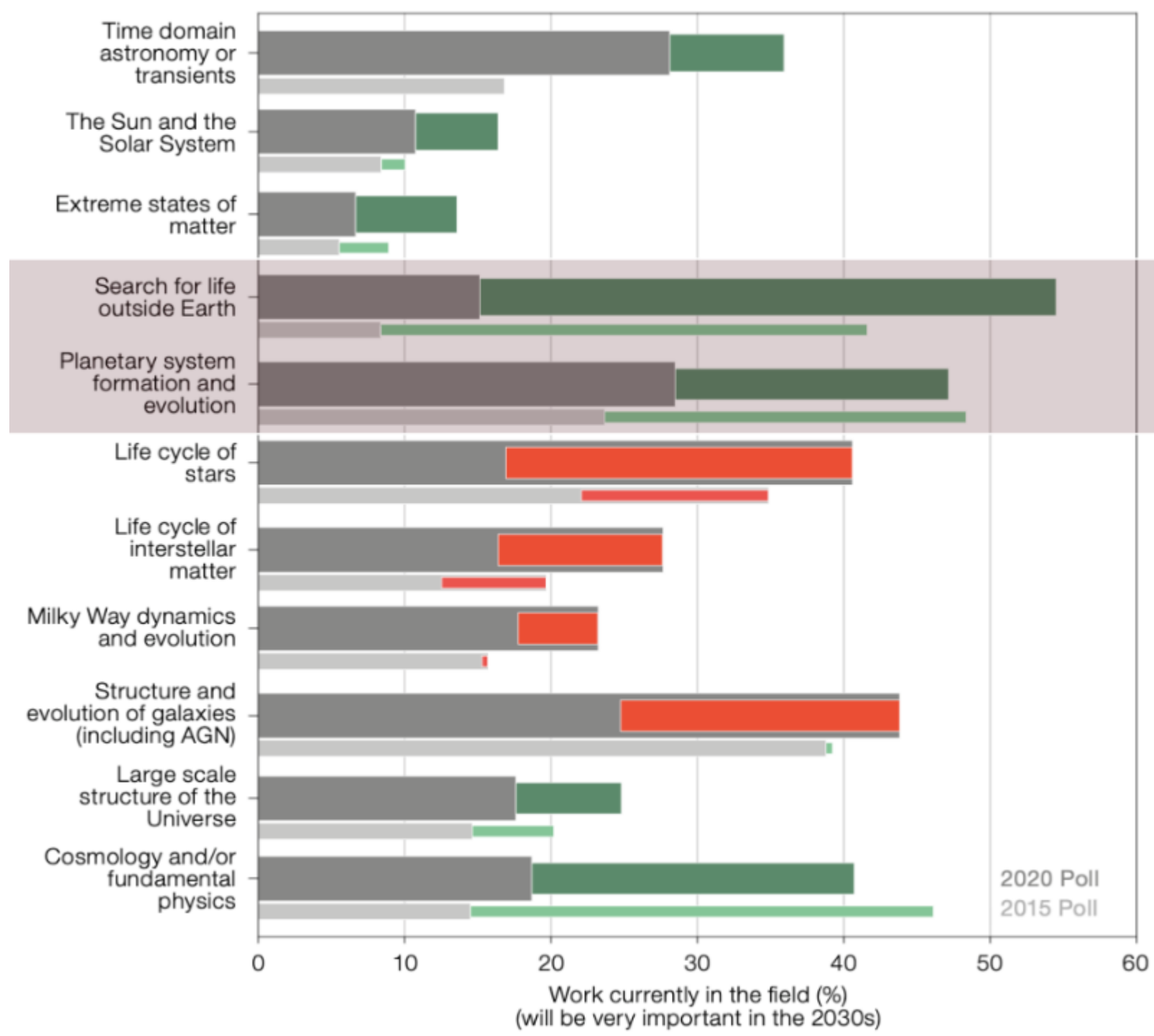

Figure 1.9: Current field of research (gray bars), and fraction of respondents thinking it will be very important in the future (green bar for increase and red bar for decrease). The thinner and lighter color bars show the results of the 2015 poll. Two key topics of the LIFE mission—the search for life outside Earth and the formation and evolution of planetary systems—are of the highest importance in the 2030s. Note that in 2015, it was not asked whether “Time domain...” will be of importance in the coming decade. (From Mérand et al., 2021, © ESO)

# Section 1 Recommendations

| Objectives | Recommendations | Near-term Actions (< 3 years) | Long-term Actions (> 3 years) |
| --- | --- | --- | --- |
| 1. Understand and quantify the unique scientific discovery space of a LIFE-like mission. | • **Perform scientific parameter space studies.**<br>• **Develop and understand which targets are priorities for LIFE.** | • Identify which classes of targets are most uniquely suited for a LIFE-like mission.<br>• Determine the intersection between this unique parameter space and the scientific need.<br>• Develop a target list for a LIFE-like mission. | • Finalize the science requirements for a LIFE-like mission.<br>• Leverage the unique detection space in collaborating with other missions, such as the ELTs, JWST and HWO. |
| 2. Understand the operational synergies and scientific value of combining observations from LIFE and HWO. | • **Perform joint simulations of LIFE and HWO observations**<br>• **Evaluate various observing strategies to maximize the output of LIFE and HWO observations** | • Retrieval on Earth-like simulated and real data assuming simultaneous observations with LIFE and HWO.<br>• Retrievals of Earth-like simulated data with LIFE and HWO, assuming various observing strategies and dynamic sharing of information. | • Retrieval of various terrestrial atmospheres assuming simultaneous observations with LIFE and HWO<br>• Retrievals of varying terrestrial simulated data with LIFE and HWO, assuming various observing strategies and dynamic sharing of information. |

# 2 LIFE precursor science investigations: Theory and observations

## 2.1 Understanding viable biosignatures

When it comes to detecting life beyond the solar system, we are particularly interested in "biosignature" gases, atmospheric species (such as $O_2$ and its photochemical product $O_3$, $CH_4$, and $N_2O$ for an Earth-like atmosphere) whose presence in the atmosphere could be explained by biological activity. Most of these gases, such as $O_2$, could also be produced by "abiotic" (non-biological) sources (e.g., ocean loss and photochemistry with subsequent atmospheric escape). These abiotic mimics are known as "false positives." Therefore, the detection of a single species would not assure the presence of life on the observed planet since other contextual information is needed to discriminate true biosignatures from false positives. For example, the simultaneous detection of two or more biosignature gases would provide a more convincing indication of biological activity. The most notable example is the simultaneous detection of $CH_4$ and $O_2$ (or $O_3$), atmospheric indicators of disequilibrium chemistry, which is difficult to explain without the presence of life on a planet (Lederberg, 1965; Lovelock, 1965). Recently, the simultaneous detection of $CH_4$ and $CO_2$ in the absence of CO has also been claimed as a biosignature for the Archean eon on Earth (4-2.5 billion years ago; Krissansen-Totton, Olson, et al., 2018).

Correctly detecting and interpreting potential biosignatures is a very challenging task for a variety of reasons. As detailed in the NASA Network for Life Detection/Nexus for Exoplanet System Science (NfoLD/NExSS) Standards of Evidence for Life Detection report[5], contextual knowledge is required to correctly interpret a potential biosignature and to rule out abiotic scenarios. The evolution of the environment is, however, strictly coupled to the origin and prevalence of life itself. Earth and its life have evolved together, as a coupled system. Life has caused a gradual oxygenation of the primordial atmosphere of the Earth to the modern levels. For this reason, searching for life elsewhere requires combined perspectives from planetary science, geology, biology, chemistry, and astronomy.

Furthermore, biosignatures can vary depending on the planetary and stellar environment (Meadows et al., 2018; Figure 2.1) due to geochemical effects as well as photochemistry. In this context, studies are being performed on biosignature detection with JWST (Leung et al., 2022; Schwieterman et al., 2022) and the LIFE mission (e.g., Alei et al., 2022, 2024; Angerhausen et al., 2023; Konrad et al., 2024). False positives also change depending on the scenario. Identifying possible false-positive indicators and ensuring their detectability with future missions is as relevant as detecting biosignatures and should be considered when defining the scientific requirements of any future mission. The modeling of abiotic processes is also of prime importance when it comes to reducing the probability of false positives in biosignature detection.

At this point, there is still a lot of uncertainty concerning the origin and prevalence of life on a planet. The origin of life on Earth likely took place between 4.4 and 4.1 Gya, with the first evidence of life on Earth being around 4.1–3.7 Gya. In between these two time periods, multiple origins of life could have happened. Only one of them generated our current tree of life, which led to the development of a community of genetically related cells that represented the Last Universal Common Ancestor (LUCA) of all life on Earth, the most recent common ancestor of all life on Earth.

[5] https://arxiv.org/abs/2210.14293

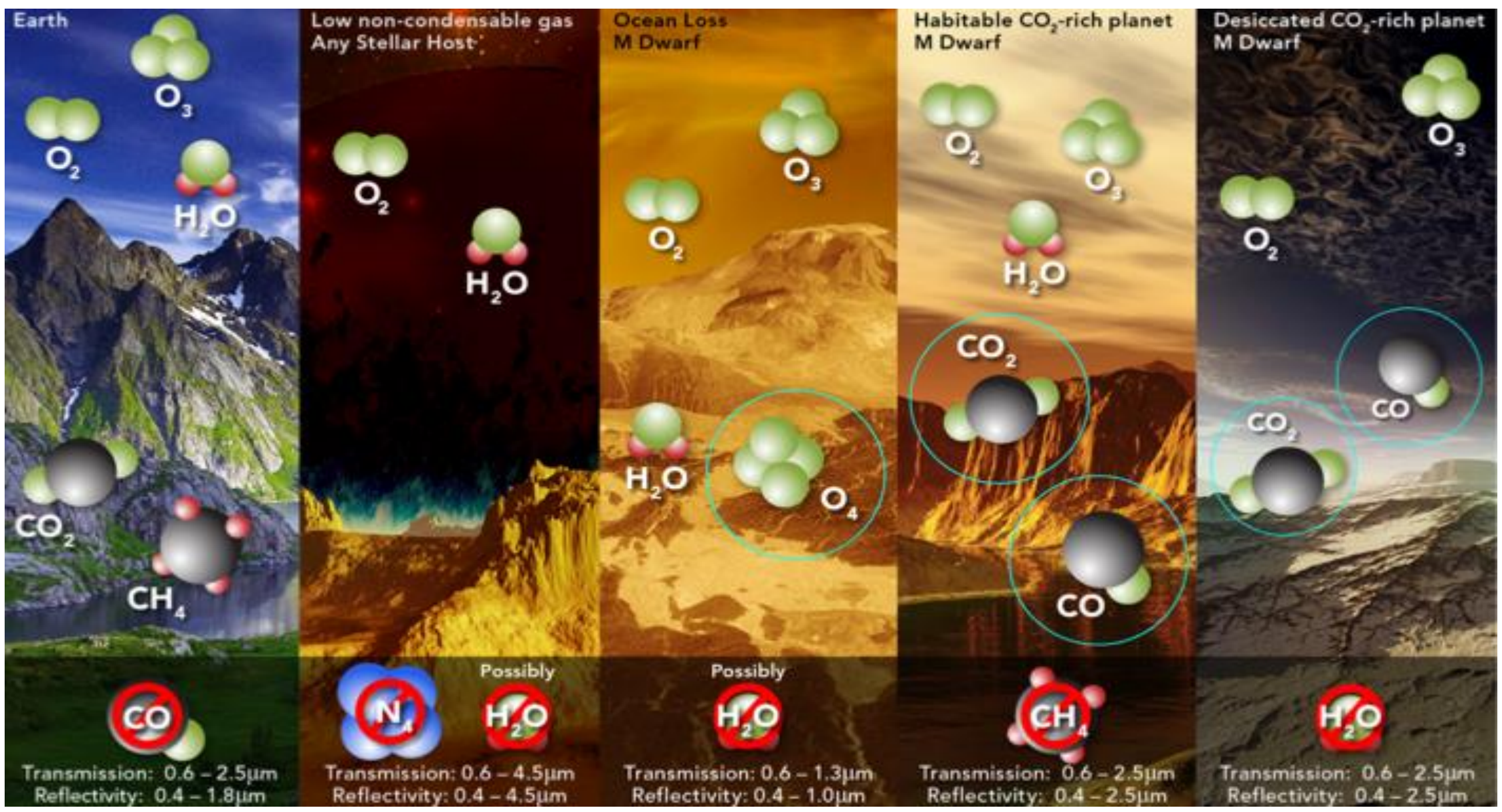

Figure 2.1: Potential false-positive mechanisms for $O_2$ for various scenarios. A detection of any of circled molecules would motivate a false-positive $O_2$ detection. A non-detection of the forbidden molecules (in the bottom, shaded bar) would also reveal the false positive. (Meadows et al., 2018; Reproduced with permission from Ron Hasler and Mary Ann Liebert, Inc.)

Precursor laboratory work is being carried out to understand the origin of life, mainly through bottom-up and top-down approaches. To the first category belong experiments to replicate prebiotic chemistry in early Earth conditions; to the second category belong studies of modern life to infer traits of LUCA and pre-LUCA biology. Ideally, these two approaches should lead to consistent results.

We need to properly define "life" as a process and differentiate life from non-life, thus defining a "biosignature threshold" (Barge et al., 2022; Figure 2.2), which is dependent on the environmental context and relies on a deeper understanding of the abiotic paths to produce the building blocks of life (e.g., ATP, lipids, peptides, RNA, enzymes, etc.). A deep knowledge of the abiotic mechanisms is required to understand if abiotically produced molecules might indeed be considered the main building blocks for life or prebiotic chemistry on early Earth, as described in the 2015 NASA Astrobiology Strategy[6].

[6] https://astrobiology.nasa.gov/about/astrobiology-strategy/

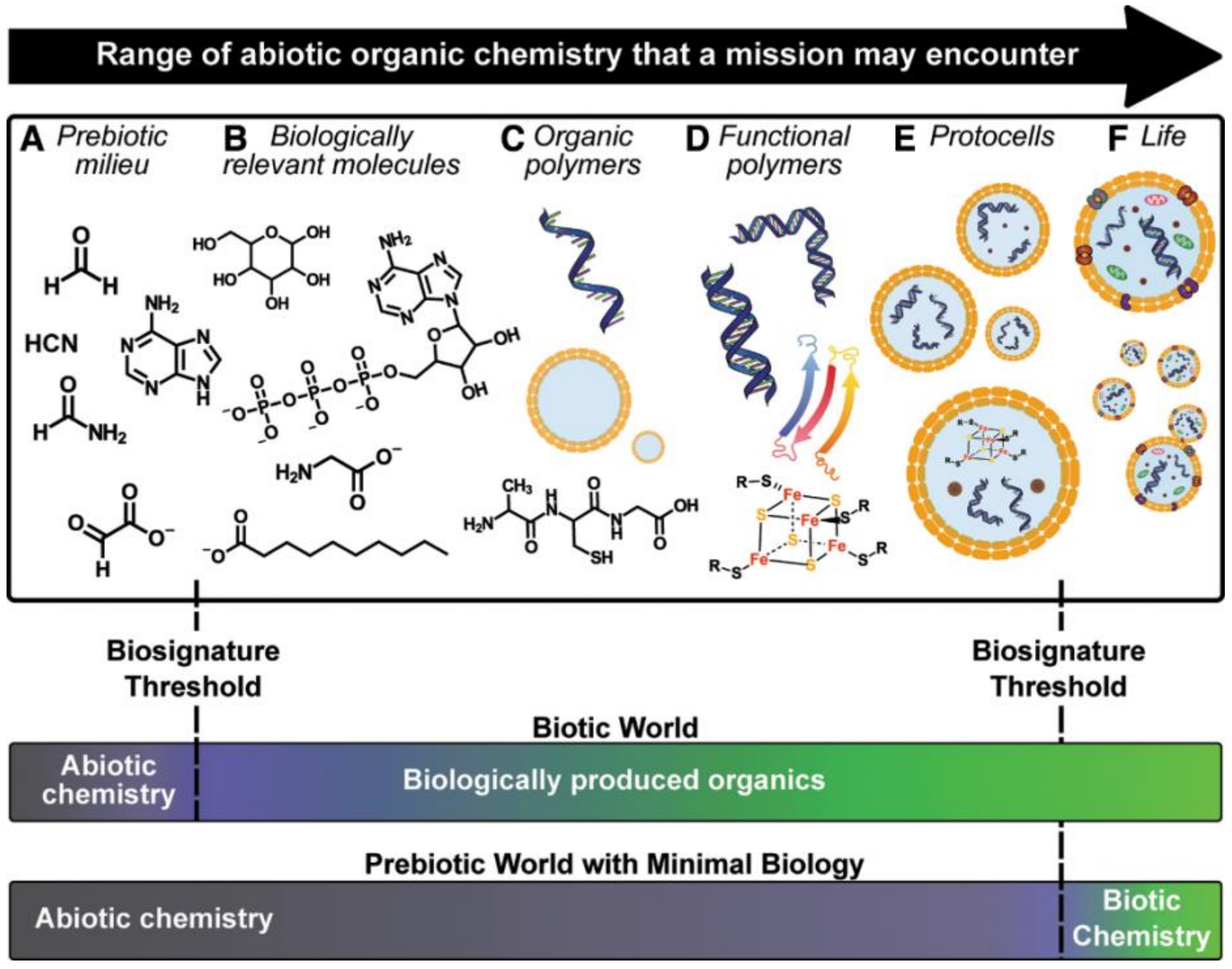

Figure 2.2: The biosignature threshold (i.e., the boundary between "abiotic" and "biotic" organic chemistry) can shift depending on the geobiological state of the planet. In a biotic world, such as modern-day Earth, the threshold where an organic detection is likely to be from life is low since abiotic chemistry is suppressed. In a prebiotic world with advanced abiotic organic processes and minimal (if any) biology, the threshold to identify "life" would be higher. To determine whether an organic detection from a mission is a biosignature, we need a more complete understanding of where the abiotic–biotic boundary lies for the world or environment of interest. Credit: Figure from Barge et al., 2022); reproduced with permission from *Astrobiology*.

Planet Earth is the only data point we currently have for an inhabited planet in the universe. For this reason, we often refer to terrestrial life when looking for signs of life in the universe. However, we should explore alternative paths for life, to broaden the search for life beyond the terrestrial scenario. This could be achieved through the study of agnostic biosignatures, which do not assume any characteristic specific to terrestrial life, but rather focus on the complexity of chemicals and patterns, evidence of energy transfer, or elemental accumulation (cf. Catling et al., 2018; cf. Schwieterman et al., 2018; Seager et al., 2016).

Many molecules that could be tracers of life in the spectrum of an exoplanet show many features in the mid-IR (see Figure 1.4 in Section 1), where also lies the peak of the thermal emission for a temperate exoplanet. This wavelength range allows access to the temperature structure of the atmospheres, which is an essential element to determine if a planet can be considered habitable (cf. Section 1). Studies that branch out from the Earth-like scenario (e.g., Seager et al., 2013; Sousa-Silva et al., 2020) that also take into account novel biosignatures, as well as various stellar classes, are essential to properly conduct trade studies for future missions. In this context, studies are being performed on biosignature detection with JWST/MIRI (e.g., Leung et al., 2022; Schwieterman et al., 2022) and the LIFE mission (Angerhausen et al., 2023, 2024), for instance, for nitrogen dioxide, phosphine and methyl bromide/chloride.

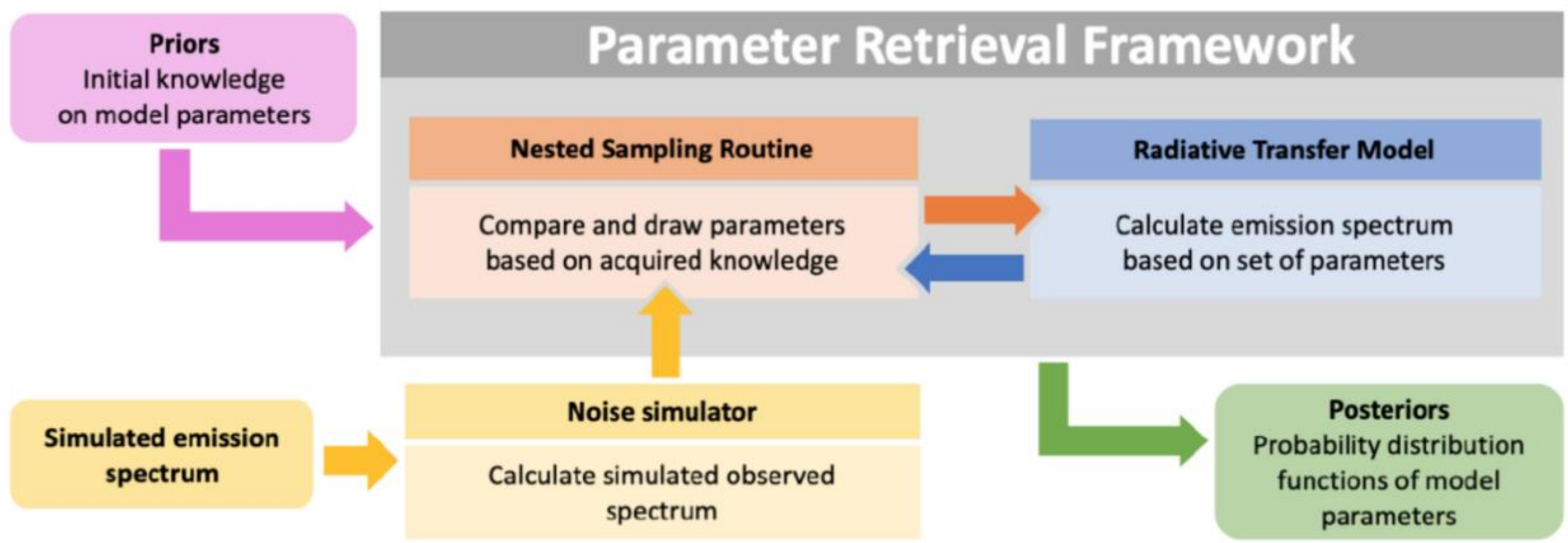

Figure 2.3: Schematic showing the workflow of a generalized Bayesian atmospheric retrieval framework. Credit: E. Alei, B. Konrad

Ultimately, "multiple measurements and approaches, spanning disciplines and missions, will be needed to make a convincing claim" for finding life in the universe (Standards of Evidence Report), with complex trades and a holistic treatment of exoplanets and their environments. A theoretical knowledge of the pathways that can lead to the appearance and survival of life on an exoplanet, how life impacts and is impacted by the environment, and the processes that could mimic the presence of life is vital to defining the science goals of a future mission and technologies that would be needed to achieve those.

## 2.2 Further investigation of spectral retrieval to inform science requirements

Bayesian retrieval frameworks apply Bayes' theorem to infer model parameters based on experimental data, thus providing a statistically sound way to select the best theoretical model that justifies the data. In the context of exoplanet science, atmospheric retrieval frameworks (see Figure 2.3) are widely used as the gold standard to analyze and interpret observations, but also as valuable tools when it comes to designing future missions. Assuming various noise models and architectures of a concept instrument, it is possible to reproduce a simulated observation of known planetary scenarios. By feeding this mock observation to a Bayesian retrieval framework, it is possible to quantify the amount and quality of information that we could infer from the data. This brings two main advantages: first, we have access to the ground input truth, which can be used to benchmark the result; second, performing retrievals on simulated observations allows us to freely change the setup of the instrument. By varying the wavelength range, the noise model, the spectral resolution, and the observing strategy, we can ensure that, if we were to observe the target using a specific instrumental configuration, it would in principle be possible to acquire the desired information. This way, it is possible to optimize the scientific requirements over a specific set of plausible scenarios.

Initial studies have been carried out for generic/idealized versions of HabEx/LUVOIR-type instruments (see, e.g., Damiano & Hu, 2022; Feng et al., 2018; Robinson & Salvador, 2023). For the LIFE mission, retrieval studies allowed us to define first requirements for a space-based mid-IR interferometer to characterize an Earth-like planet (Alei, Konrad, Angerhausen, et al., 2022; Konrad et al., 2022, 2023, 2024).

Typically, such studies focus on specific planetary scenarios (e.g., an Earth analog, the Earth through time, Venus). This is a sensible validation of the minimum requirements for such missions, with the search for habitable (and inhabited) planets, in the context of Earth-like life, being the premium goal. However, the atmospheres of terrestrial exoplanets are believed to be quite diverse depending on the exoplanetary context. In view of the expected discovery space of next-generation missions such as LIFE and the opportunity to finally perform population studies on the sample of terrestrial exoplanets, it is critical to expand the spectral retrieval efforts to include other potentially plausible atmospheres, computed using atmospheric models that consider radiative-convective equilibrium, chemistry and photochemistry. Performing retrieval studies of simulated atmospheres will not only provide further knowledge about atmospheric evolution and characterization, but also on the detectability of various atmospheric signatures, and on the instrument requirements and architecture that enable such characterization.

To date, many retrievals performed on real observations of gas giants and small planets, in transit, secondary eclipse, and/or direct imaging, are not good fits in a reduced chi-squared sense, since the retrieval algorithms rely on a variety of approximations to allow for reasonable computing time. This also means that the results for individual model parameters are likely not robust. For this reason, the flexibility and reliability of retrievals should be improved. A recent example is the introduction of "more physics" (specifically, a variable water abundance profile and a simple cloud model) in a retrieval study that analyzed modern Earth as an exoplanet using data from Earth-observing satellites and yielded very promising results (Konrad et al., 2024). Generally, as shown in recent inter-model comparison studies (e.g., Barstow et al., 2020; Villanueva et al., 2024), results from spectral radiative transfer codes can be extremely model-dependent. Some of the issues have been identified with differences in the opacity treatment and lack of updated opacity sources for some atmospheric species (e.g., Alei, Konrad, Mollière, et al., 2022). Other differences might arise through different parameterizations and treatments of physical processes, such as scattering, which might bias the results. An ongoing inter-model comparison among the available codes is necessary to identify and smooth differences, when possible, to enhance the consistency of the results. Laboratory and theoretical work on the atomic and molecular transitions from which opacity tables can be calculated should also be carried out, preferably in a standardized way.

## 2.3 Future studies of exozodiacal dust disks

Exozodiacal dust refers to hot and warm dust around main-sequence stars distributed inside and near the habitable zone (HZ), respectively (Kral et al., 2017; Wyatt et al., 2025). Both types of dust are detected around ~20% of stars surveyed. It is important to note that the sensitivities of the observations are not homogeneous and depend on stellar type, distance and other factors, and that the estimates of dust masses and location from the data are challenging (e.g., Defrère et al., 2021; Kirchschlager et al., 2018).

To date, sensitive observations of warm exozodi are typically carried out using ground-based nulling interferometry in the N band (Mennesson et al., 2014; Ertel et al., 2020), where the thermal emission from the ~300 K dust peaks and suppressing the bright, otherwise dominant star light. The most sensitive instrument for detecting this dust is currently the Large Binocular Telescope Interferometer (LBTI; Ertel et al., 2018, 2020).

Hot exozodi are also commonly detected in the near-IR using optical long-baseline interferometry (e.g., Absil et al., 2013, 2021; Ertel et al., 2014, 2016; Kirchschlager et al., 2020). The detected emission is most likely thermal and originates from dust at temperatures of 1000 K–2000 K, near the dust sublimation radius of a few to a few tens stellar radii (depending on stellar luminosity). A review of the field of hot exozodiacal dust in the context of exo-Earth imaging has recently been presented by Ertel et al. (2025).

The potential impact of HZ dust on LIFE is particularly important because there is a clear connection between cold debris disk and warm HZ dust (Mennesson et al., 2014; Ertel et al., 2020), and suggests evidence of a connection between cold debris disks and rocky, close-in planets (Marshall et al., 2014) even though it is not conclusive (Wittenmyer & Marshall, 2015). Thus, it is possible the planets LIFE is trying to detect are located specifically in systems with large amounts of HZ dust. Only a tentative observational connection between the presence of a cold disk and of hot exozodiacal dust has been found so far (Absil et al., 2021), but the origin of the material observed as hot dust is likely to be at larger separations from the star similar to the Kuiper or asteroid belts in the solar system.

The strongest constraints on HZ dust have been provided for early-type stars by the Hunt for Observable Signatures of Terrestrial planetary Systems (HOSTS; Ertel et al., 2020). For Sun-like stars, the HOSTS constraints are weaker, and so more numerous and sensitive observations would be beneficial. Quanz, Ottiger, et al. (2022) have shown that HZ dust at the levels constrained by HOSTS is not a dominant source of uncertainty for LIFE if it is smooth and symmetric, although the impact of clumpy dust distributions is still to be investigated. The tentative discovery of variable dust emission from exozodiacal dust with LBTI (Garreau et al., 2025) adds further risk, as past constraints on the dust levels of individual systems may not be valid for future exo-Earth imaging observations.

Stronger constraints on HZ dust by a factor two to three can be derived from existing HOSTS data by more sophisticated background subtraction (Rousseau et al., 2024). Characterization of clumpiness through more observations, including more sensitive LBTI nulling, as well as a quantitative analysis of the impact of clumpy HZ dust on sensitivity, are crucial for LIFE. A proposed detector upgrade of LBTI (Ertel et al., 2022) from the existing AQUARIUS detector (e.g., to a new Teledyne GeoSnap detector) would improve the sensitivity of future observations to provide the strongest constraints possible (Ertel et al., 2020).

While current and future observations of the dust's thermal emission can directly inform the impact of exozodis on the sensitivity of LIFE, which operates in the same wavelength regime, predicting the scattered-light brightness of the dust at visible wavelengths is still uncertain. Direct scattered-light observations at visible wavelengths with the Coronagraphic Instrument (CGI) on *NASA's Nancy Grace Roman Space Telescope* (Roman; Douglas et al., 2022) are thus important for future direct imaging campaigns operating in that wavelength regime, such as HWO. At visible wavelengths, the dust may also produce a significant polarization signal close to the star that may limit the effectiveness of polarimetric starlight suppression methods for coronagraphic exo-Earth imaging.

Hot exozodiacal dust is an issue for LIFE more than for visible-light coronagraphy, because LIFE will resolve the star and innermost regions of a system better than a coronagraphy mission. Furthermore, structures in the hot dust distribution can have a stronger impact on null measurements, while coronagraphy leakage will be more affected by overall flux and size of the emission. Also, hot exozodis will be very bright at LIFE wavelengths (> 1% of the stellar flux), while in the visible they will only contribute scattered light, which is likely << 1% of the stellar flux.

A new nulling instrument for the VLTI, called Nulling Observations for exoplaneTs and dusT (NOTT, Defrère et al., 2024) is under development and predicted to be 50 times more sensitive to hot dust than available observations. This can be used to constrain the hot exozodi luminosity function and search for structures in the dust distribution. It can also be used to better constrain how bright hot exozodis are at LIFE wavelengths (L-N bands). NOTT is, however, mostly designed for the detection of giant exoplanets around young stars. NASA has just funded a project to develop NOTT's capabilities for exozodi observations (PI: S. Ertel), but a funded effort to fully exploit the instrument for exozodi research will be needed.

Finally, the correlation between hot and warm dust with more accessible properties of planetary systems is still an open question. Both observations (e.g., JWST) and theoretical studies (e.g., N-body simulations, collisional models) of the origin of the dust are needed so we can better predict the amount of dust in systems that cannot be observed by LBTI or NOTT, or where the dust is too faint to be detected by current methods.

## 2.4 Planet occurrence rates

Exoplanet demographics (e.g., Gaudi et al., 2021) are an important factor in designing a mission like LIFE. A key factor in assessing the architecture of the LIFE mission is an understanding of the occurrence of Earth-sized planets in the habitable zones of their host stars. This value, called $\eta_{Earth}$, corresponds roughly to the range of orbital separations (stellar insolation) that would permit the presence of liquid surface water in a dense atmosphere (Kasting & Catling, 2003). The inner and outer extent of the HZ is a function of the age and spectral type of the host star, the composition of the planet's atmosphere, and many other factors. In our own solar system, the HZ extends roughly from the orbit of Venus (too hot) to the orbit of Mars (too cold) with the Earth (just right) in the middle. Other factors are also critical in determining whether a stellar system is a suitable host for a habitable planet, such as the architecture of the planetary system: the presence of inner large planets might expel a smaller Earth-sized planet from the HZ; a high level of stellar flaring in UV or X-ray, which might preclude development of a habitable environment; or a gas giant beyond the ice line might be crucial for regulating the volatile content of inner rocky planets.

Extensive observing programs from ground- and space-based telescopes will need to improve our knowledge of the habitability of nearby stars. Continuing analysis of the Kepler dataset augmented by results from *NASA's Transiting Exoplanet Survey Satellite* (TESS) and *ESA's PLAnetary Transits and Oscillations of stars* (PLATO) missions, and further advances in precision radial velocity measurements, will help to refine the value of $\eta_{Earth}$ for FGK stars currently thought to be around 0.1–0.25 with significant

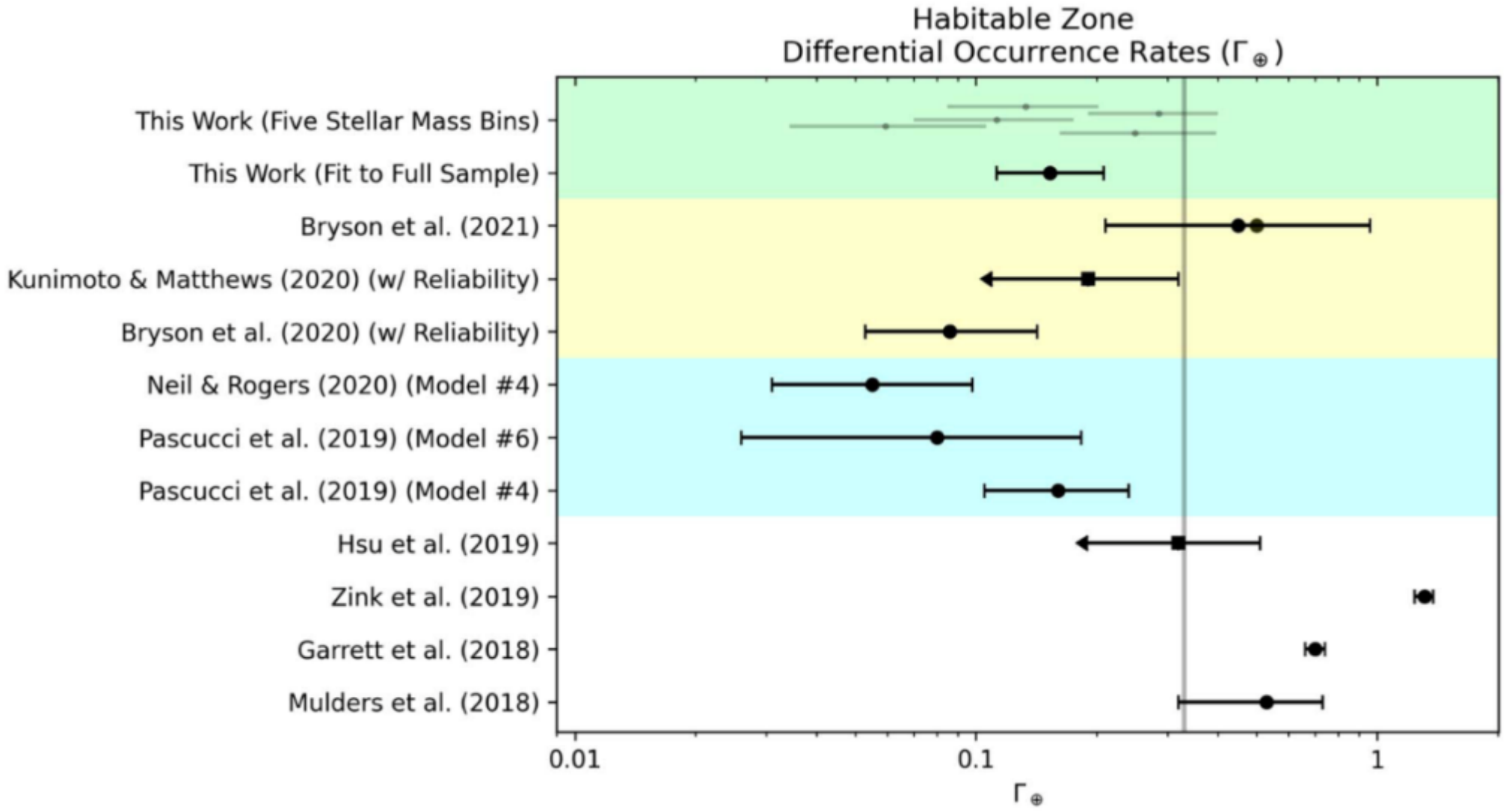

Figure 2.4: Comparison of habitable zone occurrence rates for Earth-sized planets from Bergsten et al. (2022). Note that the parameter shown here ($G_{Earth}$) is akin to calculating $\eta_{Earth}$ in some orbital period and radius bin, then dividing the occurrence by the dimensions of the bin in units of natural log Earth years and radii. For reference: the mean value of the fit to the full sample used in Bergsten et al. (2022) as shown at the top of the figure ($G_{Earth}$~0.15) corresponds to an $\eta_{Earth}$~ 0.09. © AAS

uncertainties (e.g., Bergsten et al., 2022; Bryson et al., 2020; Hsu et al., 2019; Figure 2.4). Whether or not M-dwarf stars host more exoplanets in their HZ is also debated (Dressing & Charbonneau, 2015; Bergsten et al., 2023) and seems to depend also on the spectral type range considered. Microlensing results from Roman will bring important new data on the demographics of Earth-sized planets in orbital separations extending from the HZ to the snowline (Clanton and Gaudi, 2016; Bachelet et al., 2022), particularly for the K and M stars that will be prime targets for LIFE. More generally, advances in our understanding of the formation and evolution of planetary systems as functions of stellar mass, metallicity, stellar and planetary multiplicity, etc., will allow us to assign a probability that any given star is a good LIFE target based upon how likely the star is to host a rocky, habitable zone planet.

Finally, it is worth recalling that not only temperate terrestrial exoplanets are prime targets for LIFE, but other exoplanet types that may feature non-Earth-like biospheres will be investigated. Also, for these objects it will be crucial to better understand their general occurrence rate and any correlations with other properties of their host stellar system.

## 2.5 Mapping out the solar neighborhood

Naturally, both HWO and LIFE will start investigating exoplanets that orbit stars within the immediate solar neighborhood out to distances of about 30 pc (Mamajek & Stapelfeldt, 2024; Menti et al., 2024). To prioritize the best possible targets and maximize the chance for mission success, all potential target stars must be scrutinized in detail. Future high-precision RV surveys must systematically search for planets around these stars and fill the gaps and push the detection limits coming from past or ongoing RV surveys. While the identification of additional temperate, terrestrial exoplanets in (or close to) the HZ of nearby stars must be a main goal, even if no (additional) planets are found, any sensitive upper limits provide important information about the general architecture of the systems.

Furthermore, multiplicity, elemental composition, variability, and UV flux of the target stars are key parameters that must be known as they influence the potential habitability of exoplanets in their systems. Also here, systematic and coordinated high-quality observations and homogeneous data analyses are needed. While many parameters can likely be obtained from the ground via high-precision optical spectra, empirical measurements of UV (and possibly even X-ray) fluxes require space-based observations. Given the large impact these high-energy photons have on planetary atmospheres via photochemistry and atmospheric loss mechanisms, a robust understanding of their flux levels is particularly required for the interpretation of possible biosignature gases.

Finally, in addition to deriving the statistical occurrence rates, flux levels and correlations of hot and warm exozodiacal disks and cold debris disks, a systematic search around the nearest, top-priority target stars will provide additional data that—in combination with other parameters—can inform the final target selection.

## 2.6 Direct and transit mid-IR observations of small planets

With present sensitivity, mid-IR observations with 8–10-meter telescopes can detect warm giant planets in the habitable zones of a few very nearby AFGK-type stars (Mawet et al., 2019; Wagner et al., 2021; Bowens et al., 2022). Current sensitivity estimates within the habitable zone of the closest Sun-like star, Alpha Centauri A, are only a factor of four away from imaging warm rocky planets (Wagner et al., 2021). With ongoing improvements in sensitivity, this may be overcome before the beginning of the LIFE mission. Two possible avenues exist for further progress to ground-based mid-IR exoplanet imaging that may occur independently or concurrently:

- upgrading to newer detector technologies with lower intrinsic noise (e.g., Leisenring et al., 2023)
- implementing these technologies on larger telescopes

The LBT offers a prototype ELT design with twice the collecting area and three times the spatial resolution compared to a single 8-meter telescope. A key program is planned for the LBT to image habitable zone planets around nearby stars in the coming years (Ertel et al., 2022).

From space, JWST will likewise be limited to habitable zone planets around only the nearest stars, and to planets with radii that are larger than those likely to be rocky (Beichman et al., 2019). However, mid-IR transits with JWST are already probing rocky planets around nearby M dwarfs (Greene et al., 2023; Zieba et al., 2023).

The next-generation ELTs will have the sensitivity and angular resolution to detect small rocky planets around the nearest several Sun-like stars. Bowens et al. (2021) describe a 3–13 $\mu$m imaging survey with the Mid-infrared ELT Imager and Spectrograph (METIS) on the ESO 39-meter ELT that yields an expectation value of 1.7 planets detected with radii < 4 $R_{Earth}$ in an optimized 11-night survey. The handful of detections expected will become optimal demonstration targets for the commissioning phase of LIFE. Furthermore, any demographics that can be gleaned from these initial first detections will supplement extrapolations of Kepler occurrence rates (that are based mostly on planets interior to an astronomical unit). This could inform the design and expected yield of habitable zone planets observed during the LIFE mission.

# Section 2 Recommendations

| Objectives | Recommendations | Near-term Actions (< 3 years) | Long-term Actions (> 3 years) |
|---|---|---|---|
| 1. Understand the impact of exozodiacal dust on observations with LIFE. | • **Obtain further constraints on exozodiacal dust brightness and clumpiness.**<br>• **Perform quantitative evaluation of the impact on LIFE performance.** | • Improve the background subtraction in existing HOSTS data. Obtain new LBTI observations constraining the clumpiness of exozodiacal dust.<br>• Assess the possibility of unresolved mid-IR excess detection using JWST spectra of Sunlike stars<br>• Adapt existing simulators to quantitatively evaluate the impact of clumpy dust on LIFE performance. | • Obtain funding for program to exploit VLTI-NOTT for the study of hot exozodiacal dust.<br>• Upgrade LBTI detector for improved sensitivity. |
| 2. Perform a search for initial targets for the LIFE mission | **Perform mid-IR direct imaging and transit observations of nearby small exoplanets.** | Perform direct imaging of closest stars with large AO-assisted ground-based telescopes (VLT-ERIS, LBT, VLT-VISIR) in the mid-IR. | Perform ELT observations at high angular and spectral resolution in the mid-IR. A particular focus should be a survey of the HZ of the nearest stars to detect rocky exoplanets. |
| 3. Improve our knowledge on the LIFE target populations. | • **Conduct large-scale EPRV discovery survey at necessary precision to find potential targets for LIFE.**<br>• **Reduce uncertainties in demographic extrapolations.** | • Support the development of radial velocity to reach 10 cm/sec routinely.<br>• Investigate and implement novel data methods to estimate $\eta_{\oplus}$ from Kepler and other data. | • Run legacy surveys with extreme precision radial velocity on the LIFE target stars. Perform differential astrometry of binaries at micro-arcsecond level.<br>• Assess impact of demographics on habitability (e.g., import of ice-line giants in habitability). |

| Objectives | Recommendations | Near-term Actions (< 3 years) | Long-term Actions (> 3 years) |
|---|---|---|---|
| 4. Understand diversity of LIFE targets. | • **Constraints on planetary architectures and targets.**<br>• **Diversity of exoplanet atmospheres.** | • Specifically investigate whether existence of planets outside the HZ impacts HZ occurrence rates.<br>• Develop model for how to classify atmospheres based on dominant chemical cycles.<br>• Perform JWST mid-IR observations:<br>○ of transiting exoplanets around nearby M-stars.<br>○ for direct imaging around the closest stars. | • Continued characterization of hosts, including constraints on gas giants beyond the ice line.<br>• Use the model outputs to identify observables and refine requirements for a LIFE-like mission.<br>• Leverage survey results from JWST and Ariel to understand atmospheric diversity. |
| 5. Expand our understanding of habitability and biomarkers. | • **Revise the definition of habitability.**<br>• **Laboratory studies on origins of life.**<br>• **Quantify origin of life pathways, life interaction with the environment, and false positives.** | • Consider non-Earth biased metrics for selection of LIFE targets.<br>• Curate a complete list of potential (anti-) biosignatures. Advocate for "origins of life" research. | • Update the LIFE science objectives to reflect the non-Earth-like environments and understand their impact on the mission requirements.<br>• Identify spectral signatures that would unambiguously signify origin of life conditions.<br>• Understand how detections of biosignatures in diverse environments can be interpreted towards the science objectives of LIFE. |
| 6. Ensure LIFE spectra can be leveraged to their full potential using terrestrial direct imaging retrievals. | **Create robust algorithms for (simulated) data analysis** | Run retrievals on generalized simulated atmospheres to determine requirements (ongoing as requirements are refined). | • Improve retrieval routines (e.g., updates on opacities, parameterizations of clouds/pressure-temperature profiles, likelihood treatment) |

| Objectives | Recommendations | Near-term Actions (< 3 years) | Long-term Actions (> 3 years) |
|---|---|---|---|
| | | | • Perform more inter-model comparisons to identify the strengths and weaknesses of the available algorithms. |

# 3 LIFE Technology: space demo, free fliers and metrology systems

## 3.1 Introduction

The technical risk and ultimate mission costs for a mid-IR, formation-flying, nulling space interferometer (like LIFE) could be substantially reduced through precursor missions to demonstrate key technologies. In this section we summarize the progress in formation flying, outline key technology gaps, and propose three mission architectures that could best advance the state of the art.

Early studies in the 2000s of the connected-element astrometric-focused Space Interferometry Mission (SIM; Unwin et al., 2008) and the mid-IR nulling formation flyers Darwin (Cockell et al., 2009) and the Terrestrial Planet Finder Interferometer (TPF-I; Beichman et al., 2004) explored system architectures we still consider today (Glauser et al., 2024). Smaller missions flown in the past decade have demonstrated centimeter-level relative positioning in 3D (PRISMA; D'Amico et al., 2013) while also monitoring inter-spacecraft distances to the nanometer level using laser metrology (GRACE-FO; Abich et al., 2019). The advent of CubeSats and SmallSats have allowed even free-flier science-focused missions, mostly studying the Sun. The formation-flying ESA PROBA-3 mission (Llorente et al., 2013) launched end of 2024 is establishing the required station-keeping requirements necessary for LIFE. Virtual Super-resolution Optics with Reconfigurable Swarms (VISORS; Guffanti et al., 2023) will carry out centimeter-level positioning with CubeSats, and the NASA/JPL SunRISE mission (Lazio, J. Kasper, et al., 2022) will likely be the first true separate-element space interferometer (in radio). Both missions are expected to launch in 2026 or 2027.

Encouraged by recent mission successes, the plummeting cost of small spacecrafts and launches, and the potential for international partnerships, here we propose three potential precursor mission concepts that could advance the technology readiness levels for all relevant subsystems in preparation for the more challenging LIFE mission.

1) A free-flying CubeSat mission with mainly technical goals to demonstrate the required millimeter-level station keeping, transfer of light from collector spacecrafts separated by 10–100 meters, interferometric beam combination at a simple central combiner, and real-time fringe tracking under realistic formation drifts (see Figure 3.1).

2) A short-baseline, connected-element interferometer with a cryogenic nulling payload to demonstrate the necessary mid-IR contrast limits and space-qualify a variety of hardware components.

3) A science-focused, free-flying SmallSat mission supporting the *NASA's Nancy Grace Roman Space Telescope*, with potential goals to

    a. detect exo-Jupiters using a visible nulling combiner and/or

    b. demonstrate phase referencing for faint science, such as measuring the Einstein rings of detected bright microlensing events.

c. Observe stars orbiting Sgr A* and monitor flares, extending VLTI-GRAVITY

We urge NASA, ESA, the Japan Aerospace Exploration Agency (JAXA), and other international space agencies to invest in the disruptive and transformative potential of space interferometry as a means to detect biosignatures (see Section 1) around terrestrial exoplanets, one of the highest-priority goals of recent Decadal Surveys and other long-term planning processes such as Voyage2050 in Europe. To do so, the "Technology Gap" lists curated by NASA should be expanded to include formation-flying and nulling interferometry milestones. Funding for CubeSat and SmallSat missions should be increased with the explicit objective of spurring development of risky, new technologies with strategic long-term importance.

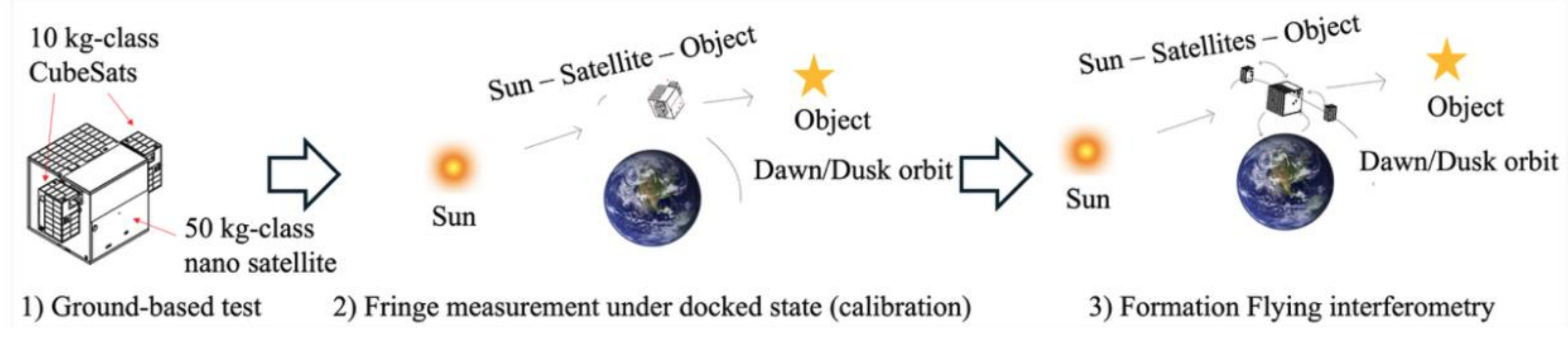

Figure 3.1: Schematic of the Space Experiment of IR Interferometric Observation Satellites (SEIRIOS) mission (Matsuo et al., 2022), a SmallSat mission selected for funding by JAXA in 2024 that aims to be the first fringe-tracking interferometer in space. It is one example of a space mission that would address technology gaps on the pathway to a mid-IR nulling interferometer (such as LIFE). © SPIE

## 3.2 Relevant past and ongoing formation-flying missions

Some technology demonstrations for LIFE have been carried out in other contexts. The TPF-I was a NASA-led project that matured formation-flying (FF) algorithms and sensors for initialization, reconfiguration, collision avoidance, and science operation of precision (Scharf, Keim and Hadaegh, 2010a, 2010b). These FF radio frequency (RF) sensors and algorithms were demonstrated in a lab up to Technology Readiness Level (TRL) 4 to achieve relative centimeter-level and < 10 degrees precision. The TPF-I team also identified Miniature Xenon Ion (MiXI) thrusters as suitable for controlling the 6DOF of the collectors to point them with sub-arcsecond precision (Martin et al., 2008).

Since the TPF-I lab experiments, several of the relevant technologies have been demonstrated in space. PRISMA achieved centimeter-level position stability between two spacecraft using FFRF and (separately) using visual-based relative navigation, even in the highly dynamic environment of low-Earth orbit (LOE) (Bodin, Nylund and Battelino, 2012). Measuring (rather than controlling) the distance between spacecraft has been achieved with a noise level of 1 nm/√Hz above 100 mHz on the Laser Ranging Interferometer (LRI) on the *Gravity Recovery and Climate Experiment* (GRACE) follow-on satellites (Abich et al., 2019). Laser Interferometer Space Antenna (LISA) Pathfinder (Armano et al., 2019) characterized high-precision thrusters suitable for interferometry with an average noise level of 0.17 µN/√Hz, while also demonstrating spacecraft station-keeping at the sub-micron level relative to an inertial frame using an on-board reference mass. Finally, *Starling* (Kruger, Hwang and D'Amico, 2024), launched in 2023, is a NASA-funded tech demonstration of autonomous swarm control consisting of a 4x6U CubeSat in Sun-synchronous orbit (SSO) exploring separations of 60200 km.

## 3.3 Upcoming formation-flying missions

Recent advances in small satellites could open pathways for relatively inexpensive interferometers in space in the near future (Monnier et al., 2019, 2024a). In particular, the ASTERIA CubeSat mission demonstrated 0.5 arcsec absolute pointing for astronomy (Knapp et al., 2020) and CubeSat Laser Infrared CrosslinK (CLICK) will demonstrate similar pointing precision for free-space optical communication (Cahoy et al., 2019). NASA selected the STarlight Acquisition and Reflection toward Interferometry mission (STARI; Monnier et al., 2024a) in August 2024 to demonstrate millimeter-level formation flying using CubeSats, reflecting starlight from one CubeSat to another, and coupling that starlight into a single-mode fiber (Monnier et al., 2024a). Even more recently, the SEIRIOS mission was selected for funding by JAXA in December 2024 and will aim to detect first interferometric fringes on astronomical sources in 2030, using 50 kg-class satellites (Matsuo et al., 2022).

Furthermore, NASA funded two Phase A studies of SmallSats that utilized formation flying. *Miniaturized Distributed Occulter Telescope* (mDot; Macintosh et al., 2019) would consist of a 250 kg mini-sat separated from a CubeSat by 500 meters and would demonstrate starshade technology for stellar coronagraphy, a critical alternative technology for exoplanet detection. This would require +/-10 cm relative navigation. The *Virtual Telescope for X-ray Observations* (VTXO; Krizmanic et al., 2020) was studied to use a lightweight Fresnel lens at a distance of 1 km from the detector telescope, and was also designed around a hybrid of a 100 kg ESPA-class spacecraft with a 6U CubeSat. The science goal is to image compact x-ray objects with 50 milliarcsecond angular resolution and would require +/-5 mm control and +/-1 mm knowledge of relative orientation with a $50 million estimated cost.

The most active science areas currently impacted by formation-flying spacecraft are solar and plasma physics, with many funded projects in active development by NSF, NASA and ESA. The NSF Geosciences Division, for example, is currently funding two different CubeSat formation flyers. *Space Weather Atmospheric Reconfigurable Multiscale Experiment* (SWARM-EX; Palo et al., 2022) will fly 3x3U CubeSats in low-Earth orbit (LEO) to study upper-atmosphere plasma with varying separations of 200 m–1300 km with a nominal launch date in late 2026. NSF is also supporting VISORS that will image the Sun in the extreme ultraviolet using Fresnel optics and also mature key technologies in a CubeSat form, which could be applied to a future SmallSat interferometer, including centimeter-level relative navigation using differential GPS and operation over 40-meter separation in LEO. While these NSF CubeSats missions are relatively inexpensive (~$5 million range), NASA recently green-lit the $250 million HelioSwarm mission (Pecora et al., 2023) consisting of nine spacecraft with 50–3000 km spacings to study the solar wind (expected launch 2029). This will be among the most expensive and ambitious of all the formation-flying projects currently in development.

One of the most highly anticipated formation-flying missions is the approximate 100 MEUR PROBA-3 mission (Llorente et al., 2013), launched by ESA in December 2024. Two SmallSats fly in formation with high-precision (millimeter-level) relative positions over separations of 25–250 meters. This mission explicitly demonstrates missing technologies needed for a large-scale mid-IR interferometer such as LIFE, including use of medium-scale (250 kg) spacecraft instead of CubeSats. The program also carries out science observations of the Sun in a coronographic configuration.

Lastly, there are two funded formation-flying missions meant to do actual radio space interferometry using formation-flying satellites. *Auroral Emissions Radio Observer - Vector Interferometry Space Technology using AERO* (Aero-Vista; Erickson & Hecht, 2018) is a NASA-funded mission nearing completion that aims to carry out actual stellar interferometry of the Sun at 15 MHz, using 2 x 6U CubeSats on a 450 km SSO. Originally expected to launch in 2023, the mission is currently delayed until end of 2026. The Sun Radio Interferometer Space Experiment (SunRISE*;* Lazio, J. C. Kasper, et al., 2022) is also a radio interferometer consisting of six CubeSats in super-geostationary orbit (GEO) to image coronal mass ejections at decametric wavebands. The spacings will be ~10 km with separations measured by space global position system (GPS) with correlation occurring on the ground. Originally scheduled for April 2024, the new launch date is later in summer 2026.

It is striking that nearly all funded formation-flying missions are targeting the Sun for study. After initial support for ST-3/Starlight, SIM, and TPF-I mission concepts, we note that neither the NASA Astrophysics Division nor NSF Astronomical Sciences has funded a formation-flying mission beyond Phase A in the past 20 years, including even the promising starshade coronagraph. The recent selections of the CubeSat STARI and SmallSat SEIRIOS missions are important steps in the right direction, and we hope that as technology is matured through the large number of pending missions, competitive missions will be supported by NASA Astrophysics, NSF Astronomical Sciences, and other national and international space agencies to study celestial objects other than the Sun, most especially exoplanets.

## 3.4 Technology items

LIFE would critically rely on many new technologies summarized in Table 3.1. Some required technologies are specific to interferometry and will not be readily demonstrated by the missions described in the following Section 3.5 below. However, we find there are several interferometry-specific technologies that need to be demonstrated in space (the relevant environment for force disturbances that cannot be reproduced on the ground) to achieve a medium TRL.

The most challenging technologies (listed at the top of Table 3.1):

1) transferring starlight between spacecraft

2) achieving and maintaining (relative) optical path difference between the interferometer arms with nanometer-level precision, and

3) achieving and maintaining the absolute orientation of the arms fixed on the target star.

Note that (2) will likely be achieved by optomechanical means onboard the spacecraft (e.g., space-borne delay lines), rather than formation flying, while (3) will require sub-arcsec guiding on two guide stars.

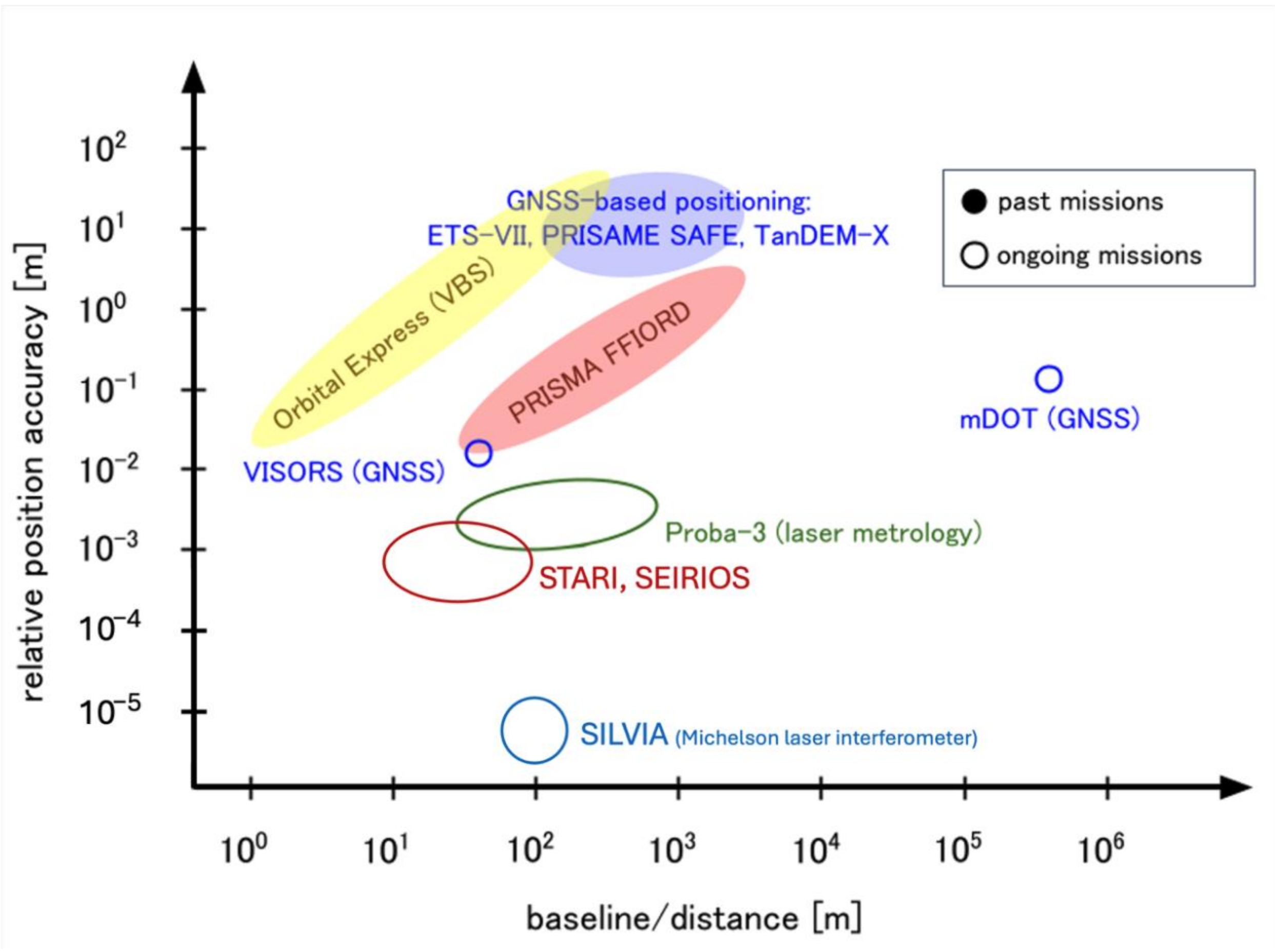

Figure 3.2: Accuracy of closed-loop positioning of past and planned missions. Reproduced with permission by T. Matsuo.

Going further down the list in Table 3.1, the millimeter-level formation flying (item 4) would be necessary to initially align the interferometer to begin fringe scanning. Precise formation flying and well as fiber coupling technologies (item 5) would presumably mature independently of space interferometers, as these are useful in other science and communication applications. Figure 3.2 summarizes the closed performance for maintaining a flying formation for recent and upcoming missions.

Some technologies (items 6 through 9) are primarily relevant to space interferometry but also to some ground-based applications (see Section 7); they can be tested in the lab environment. However, demonstrating a null depth at $10^{-5}$ contrast (item 6) in space will greatly increase the confidence in the LIFE mission.

**Table 3.1: Major missing technologies for LIFE.**

| # | Description | Specific to Interferometry | Medium TRL Requires Space |
|---|---|---|---|
| 1 | Starlight transfer between spacecraft | Yes | Yes |
| 2 | Fringe tracking at the 1-nm level in the dynamical environment of space | Yes | Yes |
| 3 | Pointing error of whole array to within 0.03 λ/D (delay lines) | Yes | Yes |
| 4 | Three-dimensional formation flying at the 1-mm precision level | No | Yes |
| 5 | High-throughput, single-mode fiber injection while formation flying | No | Yes |
| 6 | Null depth at $10^{-5}$ contrast | Yes | No |
| 7 | High-throughput broadband nulling over the mid-IR bandpass | Yes | No |
| 8 | Cryogenic-capable nulling beam combiner | Yes | No |
| 9 | Space-compatible, mid-IR integrated-optics combiner | Yes | No |
| 10 | Cryogenic-capable, deformable mirrors | No | No |

## 3.5 Science and technology pathfinders

We recommend demonstrating some of the critical technologies identified in Section 3.4 in space on smaller missions in parallel with designing a large LIFE-like mission. This approach would reduce the risk associated with combining multiple technologies that were not previously deployed in space, as was attempted during the development of TPF-I. In Table 3.2, we summarize a few possible small-mission architectures that would demonstrate a subset of the technologies in Table 3.1.

The trade space of the precursor mission is roughly covered by the architectures in Table 3.2. Low-Earth orbit (LEO) is cheaper to access than high-Earth orbit (HEO) and provides better GPS accuracy for navigation. However, LEO is also more challenging for station keeping and cryogenics. Smaller apertures (< 10 cm) can fit on CubeSats but limit science to very bright objects. Longer baselines enable higher resolution but also make light transfer between spacecraft more challenging. Longer observing wavelengths enable exciting science but increase mission complexity and development costs of new or miniaturized technologies, especially when it comes to cryogenics.

A free-flying CubeSats architecture could be flown in LEO and would demonstrate starlight transfer between spacecraft, fringe tracking, and pointing. Some orbital configurations (Hansen & Ireland, 2020; Pogorelyuk et al., 2022) maintain the optical path distance between each collector and the combiner nominally equal throughout the orbit without applying thrust, while disturbance might be handled by onboard delay lines. As a result, free-flying CubeSat missions might not require continuous formation-flying control and use alternatives to propulsion such as differential drag. One should also look into recent advances in test-mass accelerometers that might have powerful applications for formation-flying interferometry (Siu et al., 2023). One benefit of such space configurations is the relatively good *uv*-plane coverage even with two just elements, i.e., each orbit covers a line or an ellipse in the *uv* plane. The *uv* coverage might offset the relatively short baseline (tens of meters) of this architecture and might enable some shape measuring capabilities beyond radii and oblateness of stars. The small collection area of each CubeSat (< 10 cm mirrors diameter) would limit the targets to just bright stars in visible or near-IR light. The recently funded STARI and SEIRIOS missions will directly test these operational concepts in the 2029–2030 time frame.

**Table 3.2: Potential small-satellite demonstrator missions to address LIFE's technology gap list.**

| Architecture | Science Application | Technologies Demonstrated (from Table 3.1) |
|---|---|---|
| Free-flying CubeSats in LEO | Non-nulling interferometry in VIS or near-IR: stars radii and oblateness (high *uv*-plane coverage might enable more advanced shape measurements) | 1, 2, 3, potentially 4, 5 |
| Structurally connected SmallSat interferometer | Nulling interferometry to detect faint companions of bright stars | 3, 6, 7, potentially 8, 9 |
| SmallSat mini-LIFE in HEO | Visible nulling of exoplanets and microlensing events | 1 through 6 |

A single spacecraft with a launch volume of roughly a cubic meter could expand to tens of meters of baseline via booms that are additively manufactured in space (van Belle et al., 2022). Such a structurally connected interferometer would demonstrate complementary technologies to a CubeSat free-flying mission such as nulling. Optical alignment of this setup, fiber coupling, and nulling would be significantly less complex to test on the ground compared to a free-flying mission due to the reduced number of degrees of freedom and a much better initial constraints on mirror position. Nulling with several tens of meters baseline would resolve terrestrial planet-forming regions of nearby young stellar objects, and probe cores of active galactic nuclei for jet-launching structures and binarity.

The most ambitious precursor mission we propose is a free-flying interferometer that would demonstrate the technology readiness for LIFE and carry an exciting science mission. This will require larger apertures than the CubeSat demonstrator, as well as a more stable high-Earth orbit, but would avoid expensive cryogenics and mid-IR detector technologies. One conceivable demonstrator mission would complement *NASA’s Nancy Grace Roman Space Telescope* exoplanet theme, utilizing a visible-light nuller to augment the Roman Coronagraph observations and/or a near-IR phase-referencing combiner to follow-up bright Roman-detected microlensing events to directly measure Einstein Ring radii. Such a mission would also reinforce how interferometry and coronagraphy work together in powerful ways, as we expect for HWO and LIFE.

# Section 3 Recommendations

| Objectives | Recommendations | Near-term Actions (< 3 years) | Long-term Actions (>3 years) |
|---|---|---|---|
| 1. Advance the TRLs of the key technologies through lab and space missions. | **Generate more funding opportunities for potential breakthrough technologies like nulling and space interferometry** | Curate list of relevant TRLS, from which to define a technology gap list for an exoplanet nulling space interferometer | Encourage NASA/NSF/ESA to increase funding designed explicitly for technology maturation of transformative technologies, including lab and CubeSat flight opportunities. |
| 2. NASA involvement in formation-flying mission. | **Fly small-scale, likely CubeSat, technology demonstrator.** | Crowdsource essential components of Phase A study. | Develop international partnerships and MOUs. |
| 3. Seek funded mission to fly during Roman's operational lifetime, or soon thereafter, i.e., > $20M level, start < 2030. | **Develop a SmallSat science interferometer mission, taking advantage of *NASA’s Nancy Grace Roman Space Telescope*.** | Organize an international working group. | Develop science mission Phase A+; suitable topics could be exoplanet nulling + Roman coronagraph or microlensing follow-up. |

# 4 LIFE technology: Nulling architecture and backend approach

## 4.1 Introduction

In this section, we explore and evaluate the optimal nulling architecture for a LIFE-like mission. While the field has matured with many concepts having been proposed, few lab realizations have been demonstrated, and more study is needed before a final decision is made. To guide lab efforts, we emphasize the importance of conducting comprehensive simulations to trade off different system configurations, such as bulk versus single-mode waveguide beam combiners, and co-axial versus multi-axial approaches. These simulations should incorporate not only photon noise, but also the effects of beam combiner amplitude and phase errors, optical path delay errors, imperfect spatial filtering, and thermal emission diffraction during beam transport. Beyond the traditional double Bracewell (Bracewell & MacPhie, 1979; Angel & Woolf, 1997) and designs as discussed in Guyon et al. (2013), we also explore the concept of kernel nulling (Martinache & Ireland, 2018), which offers resistance to phase errors and could provide a robust solution for multi-telescope systems. Also, the potential of 2- and 3-telescope systems should not be entirely dismissed, as the LIFE project point design strives for the best balance between performance, complexity, and feasibility. There is an urgent need to curate a TRL-level maturation effort for the critical nulling subsystems and to generate a Technology Gap List to guide near-term research and development efforts.

## 4.2 System configuration/architecture definition

A very important open question regarding space-based nulling is the number of telescopes needed in an optimal mission design. The original spinning nulling interferometer approach (Bracewell & MacPhie, 1979) used only two telescopes, which resulted in three well-known issues:

1) A limited (~$10^{-5}$) achievable null depth due to residual “stellar leakage” (e.g., Angel & Woolf, 1997; Mennesson et al., 2005; Serabyn, 2000) when considering the finite size of nearby Sun-like stars.

2) Thermal background and detector gain fluctuations must remain constant—or be known—down to a small fraction of the exoplanet signal over the long time periods (~1 hour) required to modulate the planetary signal by baseline rotation.

3) Confusion between the planet signal and the signal produced by a spatially extended inclined exozodiacal dust cloud, as both emissions get modulated the same way when spinning the baseline of any architecture producing a centro-symmetric transmission pattern on the sky, such as a two-telescope nuller (Mennesson & Mariotti, 1997).

The final TPF-I and Darwin studies eventually favored a four-telescope X-array configuration, consisting of a crossed double Bracewell (see Figure 1.2), which—at the moment—is also considered the baseline choice for LIFE (Glauser et al., 2024). These studies led to a careful cataloging of both the necessary observational modes, such as, e.g., phase chopping (Mennesson, Léger & Ollivier, 2005) to separate planet

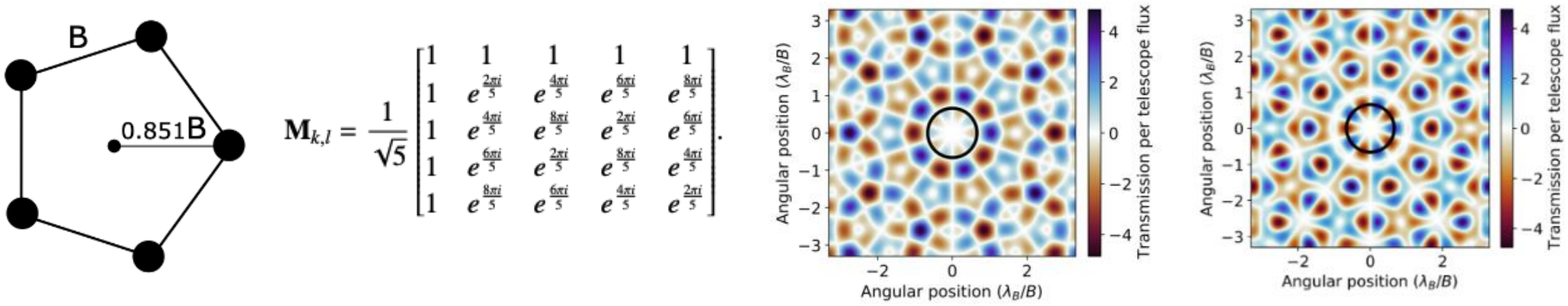

Figure 4.1: The five-telescope kernel-nuller design (from Hansen et al., 2022) and its transfer matrix are shown here in the left two panels. This configuration has two dark outputs, shown in the two right panels. Credit: ESO

signals from exozodiacal and background signals, and of the numerous potential instrumental performance limitations.

More recently, there has been progress investigating “kernel nulling” (Martinache & Ireland, 2018), which extends the idea of kernel phase (itself a generalization of the interferometric closure phase concept) to nulling interferometry. Fundamentally, this idea refers to designing a system that produces two outputs that, when subtracted, produce a differential observable that is resistant to second-order excursions in piston phase delay. The notion of phase chopping in the X-array design produces an identical effect, though kernel nulling is formulated around the difference between two static outputs, rather than swapping outputs on the detector over time. The idea of phase chopping can also be included in kernel-nulling formulations; indeed, it may be required for the same reasons as in the TPF-I and Darwin missions (detector inhomogeneities, gain and bias effects), as well as removing the effect of differing amounts of background zodiacal light emission in each output. Kernel nulling has allowed new designs of multi-telescope nullers to be conceptualized with the same resistance to phase excursions as the four-telescope Emma X-array design. Preliminary studies as presented in Hansen et al. (2022) (see also Figure 4.1) seem to indicate that five telescopes in a squashed pentagonal formation (similar to Mennesson & Mariotti (1997), but now with four different kernel-nulled outputs instead of one, may provide a larger number of Earth-like planet detections than the four telescope X-array, even for the same total collecting area. However, questions remain whether the added complexity and cost of an additional telescope, a non-rectangular array, and a more complex beam combiner design (of the Guyon et al. (2013) variety discussed below) are worthwhile trade-offs.

With the maturing yield calculations becoming available within the LIFE consortium for an expanding range of architectures, we should consider a thorough review of simple- and higher-throughput two-aperture nuller options (i.e., Fiber Nuller; Mennesson et al., 2010) before these simpler approaches are formally rejected. For instance, the finite stellar diameter leakage issue is common to the four-collector Emma X-array and LIFE designs—which are also nulling telescopes pairwise—and may still be tolerable, especially at the longer wavelengths where solar or exozodiacal emission dominates over stellar leakage. Also, accumulated experience with the Keck (Serabyn et al., 2012) and LBT (Defrère et al., 2016) interferometers have demonstrated the ability to estimate and subtract thermal backgrounds at the few - parts per million level (at LBTI). This is already better than the relative precision required in space given the much-reduced thermal background. A more challenging aspect is to stabilize (or calibrate) the response of mid-IR science detectors down to a fraction of the planet signal over periods of hours. But this is not any different from the developments required to do accurate mid-IR transit spectro-photometry of exo-Earths from space in the future. At this point, the most fundamental issue for a two-telescope system remains the inability to separate the planets and exozodiacal signals via baseline rotation. One solution to explore is spectral disambiguation, where the two signals might instead be separated using high enough spectral resolution (R ~ 1,000-10,000), leveraging the fact that only the planet signal may show deep spectral lines and absorption features from key molecules (Mennesson & Serabyn, priv. comm.). That high spectral resolution, two-telescope solution remains to be simulated in detail: first by assessing the maximum spectral resolution achievable while remaining (exo)zodi background limited (rather than detector noise dominated), and then by assessing its power to identify individual planet molecular content via nulling observations of realistic astrophysical scenes.

In any event, thorough and systematic studies are required to properly trade-off these designs, incorporating not just fundamental photon noise (as in the preliminary studies above), but also instrumental noise arising from beam combiner amplitude and phase errors, optical path delay (OPD) errors and thermal emission and diffraction during beam transport between spacecraft. A comprehensive theoretical instrumental noise analysis akin to Lay (2004), taking into account varying numbers of telescopes and configurations, will be a necessary complement to basic optical simulations of the various designs. The sensitivity and information content of the transmission maps of the various multi-combiner designs should also be considered in these trade-off studies, as the complexity (and hence information) of the transmission map of an odd number (e.g., five) of combiners may provide a stronger means of disentangling a planetary signal from other photon sources such as multiple planets and exozodiacal emission. Finally, the impact of data reduction algorithms on architecture is covered in Section 6 in this report.

## 4.3 Nulling beam combiner considerations and implementation details

In free-space optical systems, beams are usually combined pairwise, using individual beam splitters, unless a multi-axial (Fizeau) combiner is used for multiple beams. An ideal 50/50 intensity splitting ratio equalizes the combining field amplitudes to ensure high visibilities and deep nulls, while also providing a phase shift between the reflected (r) and transmitted (t) beams of exactly $\pi/2$ radians, due to symmetry and energy conservation considerations (Phillips & Hickey, 1995; Traub, 2000; Serabyn, 2003). However, as this implies fringe minima at non-zero, wavelength-dependent, optical path differences, broadband cancellation of stellar electric fields necessitates an additional $\pi/2$-radian phase shift to enable achromatic field subtraction, i.e., nulling. The extra phase shift can be supplied by various means, including passage through unbalanced dielectric media (Hinz et al., 1998, 2000) (i.e., a longitudinal phase shift), or by adding a second beam-splitter pass (Serabyn & Colavita, 2001).

These two options bring differing implementation issues: whereas beam combination at a single beam splitter must include an extra unbalanced reflection in one of the two beam trains to allow both polarization states to be co-phased simultaneously, a beam-splitter double-pass configuration (the modified Mach Zehnder configuration) actually yields symmetric constructive outputs at zero OPD, requiring the inclusion of a relative $\pi$ phase shift between the two beams to turn the constructive output destructive. Numerous methods to introduce an achromatic π-radian phase shift also exist (Serabyn, 2003), including, e.g., unbalanced dielectrics, anti-symmetric periscopes, the Gouy through-focus phase, a half-period lateral grating translation, orthogonally oriented half wave plates, and geometric phase. When working behind imperfect or unstable optical beam trains, the relative phase between two incoming beams will likely possess both chromatic and time-dependent phase errors, likely making fixed phase shifters inadequate. Moreover, beam amplitudes are also likely to differ after propagating down different beam trains. Active control of amplitude and phase (the latter likely at higher speeds than the former), as well as dispersion correction is then likely a necessity for deep, broadband nulling. Multiple dielectric plates or wedges with net thicknesses adjustable by means of rotation or translation, respectively, can supply a variable phase shift and dispersion (Koresko et al., 2003) and ultimately quasi-achromatic nulls over broad bandpasses (Mennesson et al., 2003). Residual dispersion can be addressed by spectrally dispersed nulling, i.e., nulling a set of narrow spectral channels individually, with each channel's null phase offset solved for by the "null self-calibration" technique (Hanot et al., 2011), but this is likely to have signal-to-noise ratio limitations. On the other hand, spectrally dispersed nulling can be made active, by means of an adaptive nuller (Peters, Lay & Jeganathan, 2008; Peters, Lay & Lawson, 2010) that disperses the nulled light along a deformable mirror that corrects each channel's mean phase by surface element pistons, and the amplitudes by surface tip-tilt deviations prior to beam combination.

Nulling implementations for more than two collecting apertures generally consider a beam-combining architecture based on a modified Mach-Zehnder (MMZ), a Fizeau combiner for a four-telescope double-Bracewell, or a Guyon et al. (2013) type of design for three, four, or five collecting apertures. Considering the implementation of a two-telescope nuller as falling within the same category as the double Bracewell, one can classify nulling beam combiner implementations into the following classes, none of which can immediately be ruled out, as all should possess similar first-order signal-to-noise ratios:

1) Double Bracewell, with a waveguide spatial filter and periscope for a baseline 180-degree phase shift, but likely augmented by an adaptive nuller for residual dispersion. One can envision four subtypes:

    a. Fiber spatial filter, MMZ nulling beam combiner

    b. Integrated optics version of an MMZ nulling beam combiner

    c. Fiber spatial filter, with 1$^{st}$ stage nuller being a Fizeau combiner

    d. Integrated optics spatial filter and 2$^{nd}$ stage combiner, with 1$^{st}$ stage nulling being a Fizeau combiner

2) Double Bracewell, with an MMZ nulling-beam combiner, periscope phase shifter, a pinhole or imperfect hollow conductive waveguide spatial filter, and a deformable mirror to ensure less than ~$10^{-5}$ coupling into low-order spatial modes (< 5 nm RMS in the sum of defocus and astigmatism).

3) Guyon et al. (2013) type of design, with an adaptive nuller. This has three subtypes:

    a. Fiber/waveguide spatial filter, but all beamsplitters in bulk optics.

b. Nulling beam splitters in bulk optics, followed by an integrated optics beam combiner for the final stage(s).

c. Completely integrated optics beam combiner.

Note that for option 1b, one can assume the cross combiner to also be an integrated optics combiner, as the more difficult nuller is already in integrated optics. For 1c, we did not consider a four-aperture Fizeau combiner, because the purely temporal phase chop loses a factor of $\sqrt{2}$ in signal to noise. In general, we did not consider beam splitter-based design without an adaptive nuller because of the ~0.1% tolerance on the beam splitter intensity ratio. There are of course further possibilities, but these capture the vast majority of realistic designs. Our recommendation is to compare these beam combiner designs in terms of both performance and work needed to reach high TRLs.

## 4.4 Curation of Technology Readiness and Gap List

The last section laid out the critical nulling combiner functional blocks that will be required to function over a wide wavelength range (for example, 4–18 $\mu$m), including spatial filtering, polarization, field amplitude and phase control. Further, they will need to operate at cryogenic temperatures. Our final recommendation is to establish a curated list of subsystem TRL levels based on current ground, lab-based performances. This will lead to prioritized set of technology gaps to guide future research investments. We note that progress on the first two recommendations (architecture and implementation strategy) will prune the required technology tree near the root and will allow the technology gap list to be efficiently guided toward the most promising architecture design.

While there are countless coronagraphic testbeds around the world, we have now only a few nulling testbeds in active development (e.g., the Nulling Interferometry Cryogenic Experiment/NICE at ETH Zurich; Birbacher et al., 2024; Ranganathan et al., 2024) and even fewer serious sky instruments. Additional efforts and collaborative work to solve the remaining obstacles to deep broadband infrared nulls for exoplanet characterization need to be encouraged.

# Section 4 Recommendations

| Objectives | Recommendations | Near-term Actions (< 3 years) | Long-term Actions (> 3 years) |
|---|---|---|---|
| 1. Determine high-level array architecture. | **Trade different architectures through detailed yield simulations including imperfect phasing, including simplest 2 or 3T designs.** | Simulate ability to identify and retrieve abundance of key molecules (e.g., $H_2O$, $CO_2$, $O_3$) in the atmosphere of an Earth analog for different architectures. | Estimate science (characterization) yield. Include further physical effects such as thermal and vibrations from spacecraft. |
| 2. Determine beam combination implementation strategy. | **Trade beam combiner options: bulk vs single-mode waveguides, co-axial vs. multi-axial, temporal vs. instantaneous phase modulation.** | • Perform yield simulations using throughput assumptions from bulk vs. SM optics.<br>• Demonstrate a sufficiently deep null obtained with or without SM waveguides in the lab. | • Lab demos of null depth demonstrated with multi-axial vs. co-axial nullers.<br>• Lab nuller demo of off-axis point source spectrum extraction. |
| 3. Guide technical development and testbed demonstrations. | • **Generate Technology Gap list**<br>• **Develop nulling benches up to the requested LIFE TRL level.** | • Curate TRLs for relevant nulling subsystems. Identify areas in need of R&D, e.g., mid-IR spatial filters and low-noise, stable mid-IR detectors.<br>• Support the ETH NICE bench design. | • Update every 2-3 years based on dedicated and ancillary technical developments.<br>• Contribute to the expansion benches with building blocks (waveguides, adaptive nullers, metrology, detectors [e.g., KIDS]). |

# 5 Integrated optics and interferometric functions

## 5.1 Introduction and context

Here, we report on the integrated optics (IO) and interferometric functions relevant for the LIFE mission, emphasizing the need to federate efforts in exploring photonic technologies for nulling. We find that a research and technology program is essential to demonstrate the feasibility of ultra-broadband single-mode waveguides for spatial filtering in the 4–18 $\mu$m region. In the near term, we recommend organizing an astro-photonics workshop focused on nulling, with the aim of fostering collaborations, building cross-knowledge, and maintaining a dynamic research community. Further, a bibliographic survey of existing mid-IR waveguide technologies should be curated, leading to support for one or more community-accessible test benches for evaluating performance of novel components. In the long term, we recommend demonstrating high-throughput mid-IR broadband spatial filtering, with a particular focus on the region above 12 $\mu$m. We also emphasize the importance of supporting ground-based projects like Asgard-NOTT (Defrère et al., 2024) to gain operational experience and to push the development of nulling benches to the required TRLs. Currently, no sufficiently deep achromatic null has been demonstrated (in lab or on sky) at any wavelength range to meet the minimum TRL for applying for a space mission at even the NASA Small Explorers (SMEX) level. We note with caution the demands for high-throughput, deep nulls, and broad wavelength coverage for the LIFE mission that are likely challenging to achieve with integrated optics; that said, IO components and fibers can still play crucial roles in LIFE subsystems such as fringe tracking and spatial filtering or for ground-based nulling instruments such as Asgard-NOTT (cf. Figure 5.1).

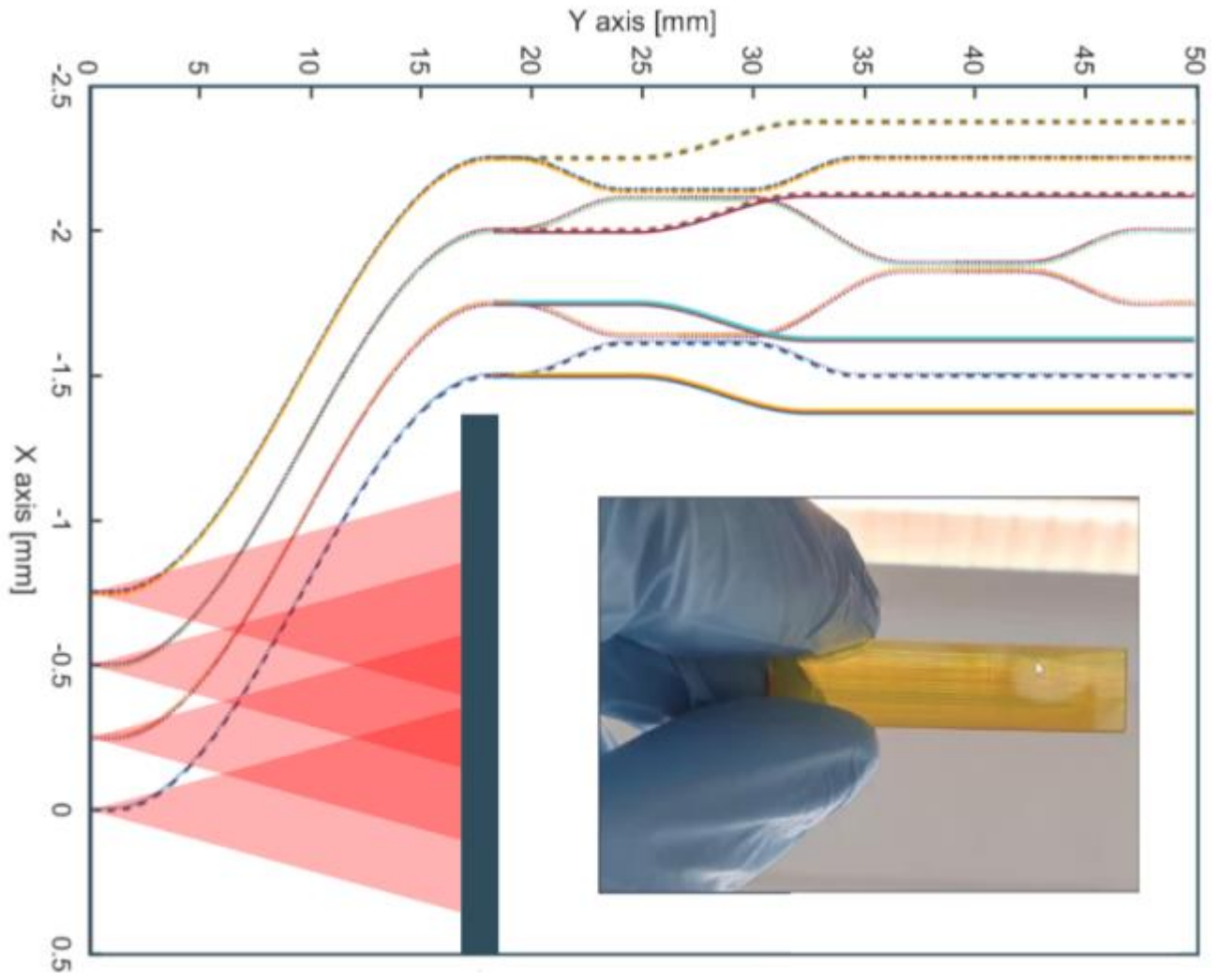

Figure 5.1: Prototype four-beam nuller photonic chip designed for the Asgard-NOTT instrument using chalcogenide-based technologies. Credit: Adapted from Sanny et al. (2022) with permission.

The interest in applying single-mode photonics to astronomical interferometry instrumentation dates back to more than four decades (Froehly, 1981). Fibers offer an efficient way to transport the light without the need for a complicated optical train. Single-mode waveguides (whether fibers or planar circuits) provide a natural spatial(modal) filtering that allows corrugated wavefronts to be flattened and lead to precisely calibrated observables (Coudé du Foresto & Ridgway, 1992; Perrin et al., 1998). Integrated circuits allow complex interferometric optical functions to be implemented on a single chip (Malbet et al., 1999).

Three decades of astronomical instrumental research have massively benefited from the huge R&D effort of the telecommunications industry and the generous availability of the Infrared Optical Telescope Array (IOTA) interferometer (SAO-CfA) for technological and scientific demonstrations. These efforts have led to the consequence that most of the operational interferometric instruments in the near-IR were or are single-mode including

- The Fiber Linked Unit for Optical Recombination (FLUOR) at the Infrared Optical Telescope Array (IOTA) and later at the Center for High Angular Resolution Astronomy (CHARA) array
- The Michigan Infrared Combiner (MIRC), Michigan Young Star Imager at CHARA (MYSTIC), Stellar Parameters and Images with a Cophased Array (SPICA) at CHARA array
- The Visible Imaging System for Interferometric Observations at NPOI (VISION) at the Navy Precision Optical Interferometer (NPOI)
- The VLT Interferometer Commissioning Instrument (VINCI), Astronomical Multi-Beam combiner (AMBER), Precision Integrated-Optics Near-infrared Imaging ExpeRiment (PIONIER), and GRAVITY at the Very Large Telescope Interferometer (VLTI)

PIONIER (Benisty et al., 2009) and GRAVITY (Perraut et al., 2018) have demonstrated a four-beam interferometric combination in a photonics-integrated chip, while MIRC was the first single-mode instrument to combine six telescopes (Monnier et al., 2006). However, none of the previous instruments were designed to reach contrast measurements much better than 1/1000.

The importance of wavefront single-mode spatial filtering for nulling interferometry was identified 20 years ago by Mennesson et al. (2002). In the early 2000s, the Darwin and TPF-I space mission projects (see Section 9.2) triggered a significant amount of research to develop

1) high-throughput, mid-IR photonics technologies (e.g., Labadie et al., 2011, Vigreux et al., 2015) and
2) visible/near-infrared, single-mode two/four-telescope nulling circuits (e.g., Buisset et al., 2007, Errmann et al., 2015).

The withdrawal of the missions nearly stopped the effort, but the advent of ground-based projects such as Asgard-NOTT (Defrère et al., 2024) in the short-wavelength mid-IR regime with new photonics developments (Sanny et al., 2022) or single-pupil near-IR nulling demonstration (Norris et al., 2020) is strongly re-energizing the field.

There is some doubt that mid-IR photonics technologies will reach the level of maturity in wavelength coverage and throughput requested for an interferometric space-based mission with four to five collector spacecraft. However, the potential complexity of the requested final system (see, e.g., Hansen et al., 2023), the time to mission launch, and the strong push of the industry to develop mid-IR photonics sensing technologies should be a strong motivation to encourage technological studies for on-chip nulling demonstrations as an alternative to high footprint bulk optics technologies.

Finally, a clear distinction should be made between photonics technologies needed for ground-based instrumentation and a space-nulling mission like LIFE. The first should focus on shorter wavelength photonics to minimize the sky background contribution and consider in the design the need to tackle atmospheric turbulence, while the latter should push high modal filtering mid-IR wide-band circuits.

## 5.2 Potential Photonic Technology Pathways

We propose the following three areas to federate efforts in exploring photonics technologies:

**1) Develop a research and development (R&D) program to demonstrate the feasibility of ultra-broadband single-mode waveguides for spatial/modal filtering purposes in the 4-18 $\mu$m region:**

There are currently several technological developments in that field that are worth pursuing: chalcogenide fibers, hollow-core fiber (Numkam Fokoua et al., 2023), endlessly single-mode photonics crystal waveguides (Ireland et al., 2024), Silicon/Germanium (Fédéli et al., 2018), Telluride (Vigreux et al., 2015), InGaAs/InP (Montesinos-Ballester, Glauser, et al., 2024; Montesinos-Ballester, Jöchl, et al., 2024) waveguides. One critical aspect is the region above 12 $\mu$m. The final applicability of such technologies should be based on wavefront filtering correction performance, overall throughput and ease of operation. Only after a significant R&D effort, that will include inventing efficient injection strategies, should the photonics solution be compared to a full high-order wavefront correction bulk-optics equivalent. Moreover, this might have implications on new mid-IR instruments based on the Palomar Fiber Nuller concept.

**2) Integrated optics building blocks:**

The current generation of mid-IR photonics are much less transmissive and broadband than bulk optics. However, they offer such considerable simplification of the instrument design that they should get the benefit of the doubt and be explored thoroughly. An effort to build and characterize high-throughput and broadband interferometric circuit building blocks (couplers, [active] achromatic phase shifters, achromatic Mach Zehnders) should be carried out. Moreover, even if a complete multi-channel nulling circuit is not available, it might be worth exploring hybrid solutions combining photonics and bulk-optics building blocks. The recent demonstration of an integrated optics chip capable of flattening the intensity response of a laser frequency comb should be a strong incentive to push the development of an integrated adaptive nuller, a key component in a nulling instrument.

It should be recalled that such technological efforts require significant development-characterization cycles. Therefore, they require long-term investment and technical partnerships that can guarantee the fabrication process is mastered over time. The fact that the mid-IR domain is increasingly appealing to the chemical/biological industry should be of significant help.

**3) Nulling optical benches:**

New laboratory setups, such as, e.g., the NICE experiment at ETH Zurich, and actual sky instruments capable of integrating photonics building blocks, such as Asgard-NOTT (led by KU Leuwen), should be supported to test the operational implications of using such technologies. Both types of benches/experiments are needed to test space concepts in the mid-IR and ground-based concepts that consider turbulence and fast fringe tracking.

# Section 5 Recommendations

| Objectives | Recommendations | Near-term Actions (< 3 years) | Long-term Actions (> 3 years) |
|---|---|---|---|
| 1. Build up experimental experience and trigger collaboration | **Federate the efforts to explore photonics technologies for nulling.** | Organize an astro-photonics "nulling" workshop. | Repeat at regular intervals. |
| 2. Demonstrate high throughput mid-IR (up to 20 microns) broadband spatial filters | **Develop a research and development (R&D) program to demonstrate the feasibility of ultra-broadband single-mode waveguides for spatial/modal filtering purposes in the 4–20 $\mu$m region.** | Carry out a bibliographic census of existing mid-IR waveguide technologies. | Develop a community-accessible bench capable of testing the spatial-filtering performance of mid-IR waveguides to the level request by life. |
| 3. Demonstrate the photonics building blocks necessary for a four to five beams nulling combiner including an adaptive nuller. | **Raise the maturity of mid-IR photonics by demonstrating Integrated optics building blocks.** | Identify promising technologies for different building blocks (couplers, MMI, phase shifters, AWG spectrometers). | Develop mid-IR nulling photonics circuits. |

# 6 Nulling data reduction and its impact on architecture

## 6.1 Introduction and context

As one of the main pillars of the direct detection of exoplanets, data reduction is essential in extracting reliable, high-contrast astronomical signals from noisy data or images. Data reduction techniques for nulling interferometric observables are intrinsically connected to the nulling beam combination strategy and the overall mission architecture. In addition to extracting the astronomical signal, they can also help relax instrumental constraints on some control subsystems as demonstrated with ground-based nulling experiments (see, e.g., Chingaipe et al., 2022; Cvetojevic et al., 2022; Defrère et al., 2016; Hanot et al., 2011; Martinod et al., 2021; Mennesson et al., 2011). In the context of the LIFE mission, it is therefore critical to consider data reduction approaches at the root of the error budget, linking the instrument design with its scientific return. Over the years, several data reduction methods have been developed to predict the scientific capabilities of different mission architectures. Using a frequency approach and signal cross-correlation, Lay (2004) showed that a combination of nulling architecture choices and data reduction could mitigate the impact of both instrumental noise and symmetric background sources (e.g., exozodiacal disk emission).

Data reduction efforts were also pushed by ground-based nulling experiments and the development of nulling self-calibration approaches (see, e.g., Mennesson et al., 2011; Defrère et al., 2016 for two-beam nullers) , including a new generic nulling data reduction package, Generic data Reduction for nulling Interferometry Package (GRIP, Martinod et al., 2025), or with advanced end-to-end simulations (e.g., Absil et al., 2006; Defrère et al., 2010; Laugier et al., 2023). With the advent of more sophisticated observing instruments and beam combination schemes (e.g., Guyon et al., 2013; Laugier et al., 2020; Martinache & Ireland, 2018) revisiting these data reduction approaches is necessary. The effectiveness of these methods should also be tested on more complex astrophysical scenes (see below). In the next section, we provide a few recommendations to make progress towards the definition of the best data reduction approach for the LIFE mission.

## 6.2 Main recommendations

### *6.2.1 Data challenges*

Our first recommendation is to set up data challenges to compare architectures and data reduction approaches on complex astrophysical scenes to identify data reduction methods that can prove optimal in practice. The new generation of data-reduction methods must be tested on more complicated astrophysical scenes, i.e., not only on single-planet simulations, and more complete instrumental error profiles. The most promising method should be the one that most effectively tackles these challenges, ensuring adequate relaxation of instrumental requirements (reducing costs and risks) and flexibility with adverse scenes (ensuring output quality, usability, and reliable interpretation). Similar to what has been done in the direct imaging exoplanet community for ground-based observations (Cantalloube et al., 2022), a data challenge engaging the wider community should be set up for nulling interferometry.

There are a few requirements for this data challenge:

1. **Common fiducial cases:**

   They are essential to compare performances between different instrument architectures and data reduction approaches. They should range from basic to more challenging cases, such as a single planet (e.g., Jupiter, Earth), multiple planets, and full planetary systems, e.g., including (perturbed or structured) asteroid belts and exozodiacal disks. A particularly relevant aspect in the context of studying the impact of multiple planets is a high-dynamic range between the objects. Also, orbital motion is to be considered for cases with slow signal-to-noise ratio buildup, i.e., the planets move significantly during exposures. Additional cases may also be prepared to test a potential imaging mode of LIFE.

2. **Data exchange standard for nulling data:**

   The current OIFITS standard is not adapted to nulling interferometry. This should be revised, and a set of requirements must be established. This would make the sharing and storing of nulling data more efficient. The challenge will be to accommodate numerous competing architectures. The new standard must be flexible to handle a variable number and layout of apertures, the number and nature of interferometric outputs, and different phasing modulation and chopping schemes, which must all be included in the new data file extensions. Telemetry data should also be included as metadata (i.e., phase or pointing errors, estimated covariance matrix, actual combiner scheme as measured by calibration, on the ground or in flight). The standard must grow and evolve continuously to the final flight standard and be compatible with ground-based nulling interferometry. As of 2024, a standard called NIFITS designed around these requirements has been drafted and is being tested for adoption in LIFE and in ground-based nullers[7].

3. **Common bank of instrumental errors, filters, etc., identified for different cases, including LIFE requirements:**

   The results of past and future ground-based or space-demonstration experiments must be compiled and made available for simulations. The same applies to the requirements of future experiments and missions.

### *6.2.2 Linking combiner precision, errors, and data reduction*

Our second recommendation is to investigate the three-way link between combiner precision, input errors and output errors, and including the effect of data reduction. In addition to a comprehensive report, one could also envision writing a peer-reviewed paper on the applicability and extension of nulling self-calibration (NSC) to the LIFE mission.

[7] github.com/rlaugier/nifits

Previous literature has mostly investigated the propagation of stochastic input errors through a perfect beam combiner. This is insufficient to build a realistic error budget. The situation has started to change recently with new studies (see, e.g., Hansen et al., 2023; Laugier et al., 2023), which include some of the consequences of an imperfect combiner on the output noise with Monte Carlo simulations. A more direct link, either analytical or numerical (non-stochastic), will be a cornerstone of the construction of an error budget that can be optimized with technical and financial budget constraints. The effectiveness of data reduction techniques to mitigate these instrumental errors should be investigated, including the correlated nature of these error and their deviation from Gaussian statistics (see, Dannert et al., 2025; Huber et al., 2025). For instance, NSC has been shown to relax instrumental constraints on control loops for ground-based Bracewell interferometers (Hanot et al., 2011). While it is expected that NSC could be ineffective in the noise/signal regimes relevant to the LIFE science case and challenging when using more than two apertures, this has yet to be established.

### *6.2.3 Linking architecture choice and data reduction with detector requirements*

Our third recommendation is to link the architecture choice of a LIFE-like mission with the data reduction approach and the requirements on detector stability. This will result in a review of state-of-the-art technologies for mid-IR, a report on the effect of detector instability on nulling interferometry, and a list of proposed solutions to mitigate the expected impact.

The stability of mid-IR detectors is a major concern in high-contrast nulling interferometry. It has been shown to limit the performance of ground-based nulling interferometers (see, e.g., Defrère et al., 2016). Several strategies can mitigate this effect, such as fast phase or background chopping. Earlier studies identified phase chopping as an effective countermeasure, especially the excess low-frequency noise present in mid-IR detectors. New generations of detectors (i.e., Teledyne's GEOSNAP) have shown improved performance on this front. An investigation into the evolution of this requirement and possible mitigation strategies is required.

# Section 6 Recommendations

| Objectives | Recommendations | Near-term Actions (< 3 years) | Long-term Actions (> 3 years) |
|---|---|---|---|
| 1. Facilitate nulling data exchange. | **Setting up a data challenge to compare architectures and data reduction/signal extraction approaches.** | • Define common fiducial cases.<br>• Define NIFITS data exchange standard for nulling data.<br>• Train community to use data standard. | Establish a common database of instrumental errors to standardize the input to instrument simulations |
| 2. Guide the design and requirements of nulling beam combiners. | **Investigating the three-way link between combiner precision, input errors and output errors and the effect of data reduction.** | Incorporate handling of correlated errors and deviations from Gaussian statistics. | • Check the relevance of NSC[8] for > 2 aperture space-based nullers.<br>• Derive instrumental requirements to alleviate the need for NSC. |
| 3. Assess the viability of current mid-IR detectors. | **Link architecture choice and data reduction with the requirements on detector stability.** | Review literature on detector stability | • Perform dedicated stability measurements on existing mid-IR detectors.<br>• Support or restart detector development (e.g., KIDs, Si:As) . |

[8] NSC: Nulling Self-Calibration

# 7 High-contrast interferometry from the ground

## 7.1 Introduction

Ground-based interferometric facilities can be used for the technical development of nulling interferometry, science demonstrations in support of the LIFE mission, community/momentum building via the direct interferometric characterization of (bright) exoplanets, and collection of ancillary data on targets under consideration for LIFE. In this section, we summarize the capabilities and sensitivities of existing instrumentation and provide an overview of future opportunities for high-contrast interferometry from the ground.

## 7.2 Available instrumentation and current capabilities

### *7.2.1 VLTI*

The VLTI located at Cerro Paranal in Chile combines the light from the four 8-meter Unit Telescopes (UT) or four 1.8-meter Auxiliary Telescopes (ATs) with baselines extending out to 130 meters for the UTs and 200 meters for the ATs.

The dual-beam mode of the GRAVITY beam combiner (GRAVITY Collaboration et al., 2017) at the VLTI can be used to phase reference on a bright fringe tracker star (K < 10 mag) and observe a nearby, faint science target (K < 17 mag) within 2' for the UTs and 4' for the ATs. The ExoGRAVITY project (Lacour et al., 2020) uses GRAVITY to survey known exoplanets within 100 to 200 mas from the host star. Using the host star as the fringe tracker, GRAVITY has demonstrated 45–100 $\mu$as astrometric precision on directly imaged planets to constrain orbital dynamics and collected medium-resolution K-band spectra (R ~ 500) to characterize the physical properties of exoplanets and detect molecular signatures within their atmospheres. The small diffraction limit of a single UT requires knowing the predicted location of the exoplanet based on prior radial velocity and astrometric measurements with high precision, ideally less than ~60 mas (or paving the focal plane). Exoplanets that have been detected and characterized by GRAVITY include HR 8799 e (ρ = 390 mas, ΔK = 10.7 mag; GRAVITY Collaboration et al., 2019), beta Pic b (ρ = 144-320 mas, ΔK = 10 mag; GRAVITY Collaboration et al., 2020; Lagrange et al., 2020), beta Pic c (ρ = 130 mas, ΔK = 10.8 mag; Lacour et al., 2021; Nowak et al., 2020), PDS 70 b (ρ = 170 mas; Wang et al., 2021), PDS 70 c (ρ = 216 mas; Wang et al., 2021), and HD 206893 c (ρ = 236 mas, ΔK = 10.2 mag; Hinkley et al., 2023; Kammerer et al., 2021). Figure 7.1 shows the orbits of beta Pic b and c.

The recent GRAVITY+ upgrade includes the implementation of wide-field off-axis fringe tracking, new adaptive optics on all Unit Telescopes, and laser guide stars (Gravity+ Collaboration et al., 2022). GRAVITY Wide increases the sky coverage up to tens of arcseconds for the UTs but limited by atmospheric turbulence (GRAVITY+ Collaboration et al., 2022), by using star separators located at the coudé focus of each telescope, bringing the subfields into the VLTI lab separately, and overlapping at the entrance of the GRAVITY beam combiner. The installation of new deformable mirrors, AO wavefront sensors, and laser guide stars enables GRAVITY+ to fringe track on objects as faint as K=13 mag and collect science data on objects as faint as K=22 mag.

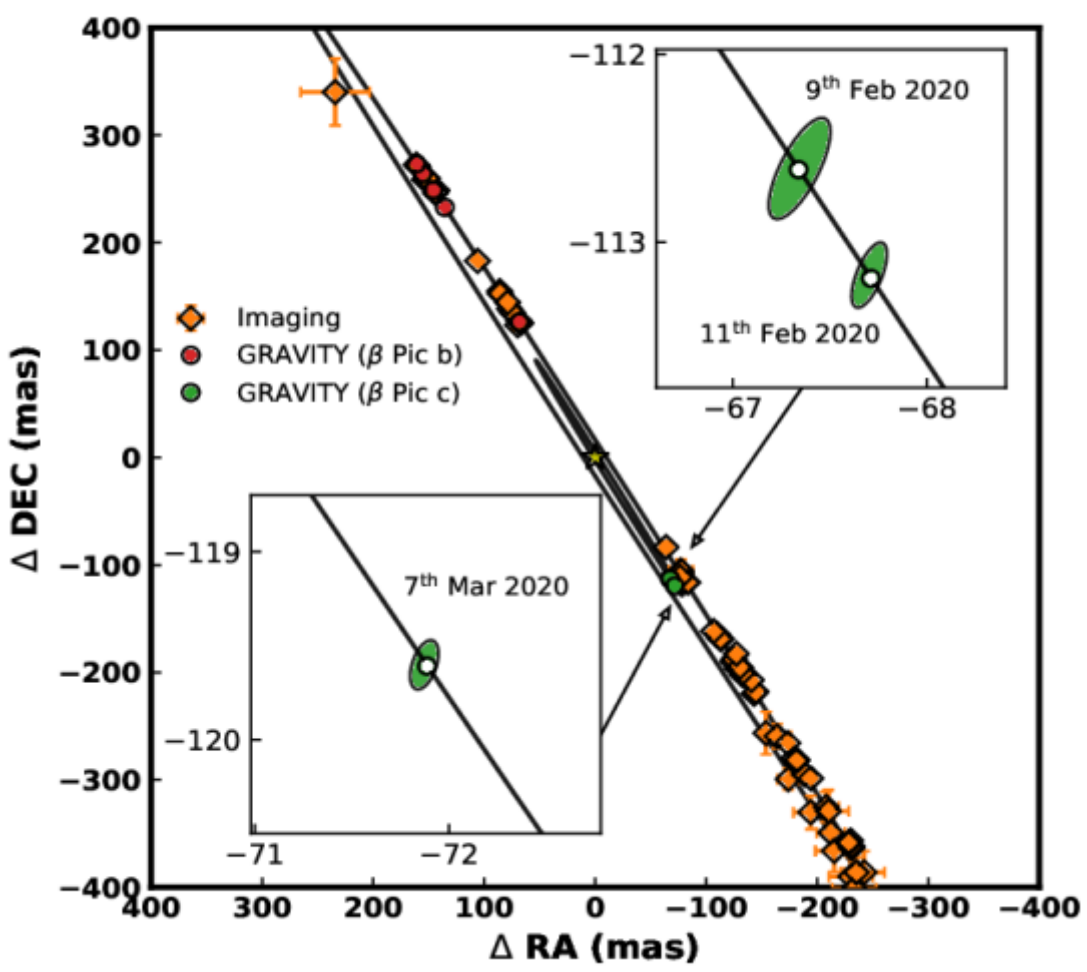

Figure 7.1: Orbital motion of planets b and c around beta Pictoris as measured with direct imaging and with VLTI/GRAVITY. Credit: Figure from Nowak et al. (2020); reproduced with permission © ESO

An important objective for VLTI—and other ground-based interferometers—is to break the current contrast limitation at the diffraction limit and access the bonanza of giant exoplanets located within 5 AU of nearby (< 100 pc) young stars, especially those showing proper motion anomaly in the Gaia vs. Hipparcos data. This could be done with GRAVITY, which already provides high-accuracy visibility and differential phase and phase closure measurements, but remains currently limited to the detection of exoplanets at ~2-3 λ/D because of instrument biases. A faint companion detection pilot program using GRAVITY, starting with known binaries on the ATs with high contrast (e.g., 100–1000:1), could serve as a powerful means to understand and reduce the impact of current systematics. The approach would be to benchmark and optimize dedicated data reduction codes for the detection of known targets, initially at modest contrast, before applying these codes to bona fide exoplanets at higher contrast, first on the ATs, and then on the UTs. The analysis of previous GRAVITY data obtained with the ATs on beta Pic b—when the planet was within the AT diffraction limit—indicates a detection limit of 2000:1 is already achievable as close as 0.1 λ/D using visibility amplitudes alone (Mennesson et al., priv. comm.). Leveraging the spectral dimension of the GRAVITY mid- (R = 500) and high- (R = 4000) resolution data may also further boost the detectability of exoplanets as high spectral resolution observations have the critical ability to distinguish intrinsic extrasolar planet spectral features from larger background or starlight residuals, even in high noise cases. This has already been demonstrated by single telescope instruments at the VLT and Keck, e.g., beta Pic b and PDS 70 b & c with VLT/SINFONI (Hoeijmakers et al., 2018; Haffert et al., 2019), beta Pic b with VLT/CRIRES (Snellen et al., 2014), HR 8799 c with Keck/OSIRIS (Konopacky et al., 2013) and for all the known HR 8799 planets with Keck/KPIC (J. J. Wang et al., 2021). This could be extended to interferometric observables, e.g., using visibility amplitude, spectral phase, or phase closure molecular mapping instead.

### 7.2.2 CHARA Array

The CHARA array located at the Mount Wilson Observatory in California combines the light of six 1-meter telescopes with baselines ranging from 30 to 330 meters (ten Brummelaar et al., 2005).

The six-beam MIRC-X H-band combiner (Anugu et al., 2020) and MYSTIC K-band combiner (Setterholm et al., 2023) at the CHARA array can reach contrasts of ΔH ~ 6 mag for detecting faint companions within 0.5 to 50 mas separations (Gallenne et al., 2015) in their low spectral resolution mode (R = 50). Project PRIME (PRecision Interferometry with Mircx for Exoplanets) seeks to push the detection limits even further using precision closure phases to detect non-transiting hot Jupiters. Initial results using the original MIRC combiner achieved an upper limit on the contrast ratio of $\upsilon$ And b of 2200: 1 (Zhao et al., 2011). CHARA, with the MIRC-X combiner, is reporting upper limits for the hot Jupiter $\upsilon$ And b at a contrast of ~5000:1 (see Figure 7.2), ruling out state-of-the-art atmospheric models and suggesting we are on the cusp of a definitive detection (Monnier et al., 2024b).

The ARrangement for Micro-Arcsecond Differential Astrometry (ARMADA) project at CHARA and VLTI has demonstrated astrometric precision at the 20–50 µas level in measuring the positions of binaries with separations < 200 mas. This level of precision is sufficient for detecting the astrometric wobble from substellar mass companions around intermediate-mass binaries (Gardner et al., 2018, 2021, 2022). This technique was originally pioneered with the PHASES project at the Palomar Testbed Interferometer (Muterspaugh et al., 2006; 2010a,b). However, the ARMADA program achieves significantly higher astrometric precision (see Figure 7.2 middle and right panels).

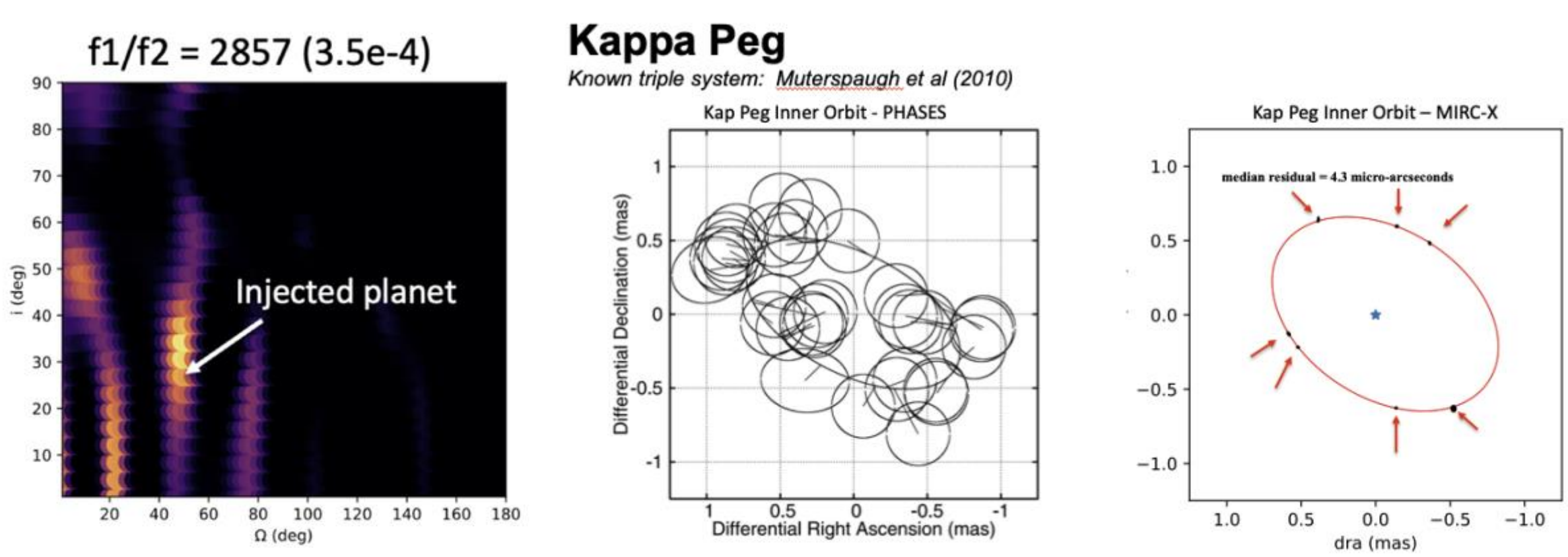

Figure 7.2: (Left) Using CHARA/MIRC-X, Monnier et al. (2024b) report upper limits of ~5000:1 for the hot Jupiter companion to ups And. (Middle and Right): CHARA/MIRC-X can detect < 10 micro-arcsecond wobbles, capable of detecting exoplanets with sufficient monitoring (Gardner et al., 2021). The left panel shows the initial astrometric precision for the inner orbit for kappa Peg from Muterspaugh et al. (2006) compared with the improved astrometric precision obtained through the ARMADA project. Reproduced with permission from American Astronomical Society (AAS).

### *7.2.3 LBTI*

The Large Binocular Telescope Interferometer (LBTI, Ertel, Hinz, et al., 2020; Hinz et al., 2016) combines the light from the two 8.4-meter primary apertures of the LBT. With the two apertures on a common mount, an edge-to-edge diameter of 22.7 meters, and a center-to-center separation of 14.4 meters, the LBTI makes the LBT a critical pathfinder for 30-meter class telescopes. The LBTI is also the only one of two thermal-infrared (3–13 $\mu$m) interferometers (the other being MATISSE/VLTI; Lopez et al., 2022) and the only one combining the light from eight-meter-class telescopes in the Northern Hemisphere.

The LBTI has completed the NASA-funded HOSTS survey for exozodiacal dust (Ertel et al., 2018; S. Ertel et al., 2020) using nulling interferometry in the N band and has demonstrated a post-processed null depth of $10^{-4}$ (Defrère et al., 2016). A sensitivity study has shown that background bias caused by imperfect thermal-infrared background subtraction and excess low-frequency noise (ELFN) of the detector are the main sensitivity limitations, together with telescope vibrations (Ertel et al., 2020). Post-processing of the data using a principal component analysis (PCA)-based background subtraction has been shown to improve the sensitivity by a factor of ~2 (Rousseau et al., 2024).

## 7.3 Science and future instrument opportunities

### *7.3.1 LBTI instrumentation*

In addition to nulling interferometry, the LBTI features a Fizeau (imaging) interferometry mode (Spalding et al., 2019, 2022; Isbell et al., 2024, 2025) and a non-redundant aperture-masking mode (Sallum et al., 2017, 2021). Both modes exploit the full 23-meter edge-to-edge diameter of the LBT's double aperture. In contrast to other optical long-baseline interferometers, the LBT has a high pupil-filling factor (in Fizeau mode) of approximately 25% for a circular 23-meter aperture and approximately 75% for an elongated aperture filling the space between the two primary mirrors and a high and dense instantaneous *uv*-plane coverage in aperture-masking mode. The Fizeau mode is available in the L to N bands and can, in principle, be combined with integral field and prism spectroscopy. Both modes are so far experimental, but the team is making a major push to make 23-meter LBTI observations more routine. Major limitations are currently a bright limiting magnitude of the fringe tracker and the excess low-frequency noise of the Aquarius detector used in the N band (Ertel et al., 2022). A fringe tracker upgrade is currently in progress (Ertel et al., 2020). Funding for a NOMIC detector upgrade replacing the current AQUARIUS detector with a GeoSnap detector is now funded by Breakthrough Initiatives with a target deployment on sky in fall 2026 (PI: Kevin Wagner). The detector upgrade will also pioneer new mid-IR detector technology, which will be deployed on ELT/METIS, and is crucial for LIFE to achieve its high stability for precision N-band interferometry. An option to extend the LBTI's wavelength range to the visible for maximum angular resolution is being explored (Hinz et al., 2014).

### *7.3.2 LBTI science*

LBTI has been developed for sensitive, high-contrast interferometry at thermal infrared wavelengths. The main science case has originally been the search and characterization of exozodiacal dust using nulling interferometry. With the planned GeoSnap upgrade, an improved sensitivity by a factor of a few is possible and a similar factor can be achieved from improved background subtraction (Ertel et al., 2020; Rousseau et al., 2024). This will allow for a combined factor of up to 10 in improved sensitivity to exozodiacal dust, which would allow for the detection and characterization of individual systems, with dust levels down to a few times the solar system zodiacal dust. The Fizeau and aperture masking interferometry modes, in combination with an improved limiting magnitude of the fringe tracker, can be exploited to image young exoplanets and protoplanetary disks in nearby star-forming regions with ELT-class angular resolution and at thermal-infrared angular resolutions comparable to current visible and near-IR instruments (Ertel et al., 2020). The Fizeau imaging mode can further be leveraged to detect giant planets around nearby stars down to the habitable zone and even potentially rocky, habitable-zone planets around the closest, suitable Sun–like stars in the N band (Wagner et al., 2021; Ertel et al., 2022). The Fizeau mode is also being further developed to allow more general observations, e.g., of evolved stars, active galactic nuclei, and solar-system bodies (moons, asteroids) at 30-meter-class resolution (Isbell et al., 2024, 2025).

### *7.3.3 VLTI/NOTT instrumentation*

NOTT (Defrère et al., 2024) is the nulling interferometer of Asgard, an instrument suite recently proposed to ESO for the visitor focus of the VLTI (Martinod et al., 2023). It will operate at L' band (3.5–4.0 $\mu$m), a key wavelength to image young planetary systems and exozodiacal disks. The system is optimized for high-contrast and high-sensitivity imaging within the diffraction limit of a single UT/AT telescope. It is designed as a double-Bracewell nulling instrument producing spectrally dispersed (R = 40, 400), complementary nulling outputs and simultaneous photometric outputs for self-calibration purposes. The instrument will use a cryogenically cooled GLS-based (gallium lanthanum sulphide) photonic chip (Sanny et al., 2022), optics (Dandumont, Mazzoli, et al., 2022; Garreau et al., 2022), and a HAWAII 2RG detector to reach the sensitivity and contrast required for exoplanet imaging (see end-to-end performance simulations in Laugier et al., 2023).

### *7.3.4 VLTI/NOTT exoplanet science*

The main scientific goal of NOTT is to characterize the chemical composition of known giant exoplanets discovered by radial velocity or expected to be found soon with the new release of the astrometric measurements made with Gaia (approximately 25% of planets more massive than ~0.3 $M_{Jup}$ present in nearby moving groups, Wallace et al., 2021). These exoplanets are located in uncharted territory, i.e., close to the water-ice line, which is inaccessible to current direct-imaging instruments and future ELTs. Although the Gaia mission provides valuable information about the masses of giant gas exoplanets and the ages of the systems, it only brings partial knowledge about the formation and evolution of exoplanetary systems. By measuring the effective temperatures of the exoplanets, NOTT will better constrain their planet-formation models (Wallace, Ireland and Federrath, 2021). The NOTT exoplanet target list currently includes nearly 1500 young (< 250 Myr) stars within 150 pc (Dandumont, Laugier, et al., 2022). In parallel, NOTT will also enable the direct detection and characterization of hot Jupiters (see target list in Defrère et al., 2018). Low-resolution spectroscopic observations of such planets in the thermal near-IR will provide their radius and effective temperature as well as critical information to study the non-equilibrium chemistry of their atmosphere via the $CH_4$ and CO spectral features.

### 7.3.5 VLTI/NOTT exozodi science

Hot exozodiacal dust has been commonly detected using constructive optical long-baseline interferometry in the near-IR (Absil, di Folco, et al., 2006; Absil et al., 2013, 2021; Ertel et al., 2014, 2016, Nunez et al., 2017). The origin of this dust is enigmatic, and it represents a risk to detecting exo-Earths by creating coronagraph leakage at visible wavelengths and null variations at LIFE wavelengths. Studying the dust's origin and abundance is thus a critical aspect of preparing the mission. NOTT's improved contrast and longer observing wavelength (where the dust-to-star contrast is more favorable than in the near-IR, Kirchschlager et al., 2020) combined will result in a factor of ~50 in improved sensitivity to hot exozodiacal dust. This will revolutionize our understanding of hot dust, allowing for determining a luminosity function, studying variability, and reconstructing images of the dust distribution.

### 7.3.6 CHARA and MROI nulling

Both the mature CHARA Array interferometer and the emerging Magdalena Ridge Observatory Interferometer (MROI; Creech-Eakman et al., 2022) can advance the technology of nulling, as the VLT/NOTT program, but focusing on shorter wavelengths, namely H and K bands. It was shown for the Planet Formation Imager (PFI) studies (Monnier et al., 2018; Wallace and Ireland, 2019) that K-band nulling could detect many young giant exoplanets, since the thermal background from the atmosphere is so much lower than in the L, M, and N bands. With advances in wavefront controls with deformable mirrors, high-speed sensitive detectors, high-transmission integrated optics, and new nulling architectures, CHARA and MROI should pursue a science-capable nuller in the near future. The scientific goal will be to detect and characterize nearby exo-Jupiters that are too close for 8-meter-class telescopes. Technically, this will pave the way to maturing technologies needed for precursor space interferometer nullers.

### 7.3.7 KPIC/VFN instrumentation

The Keck Planet Imaging and Characterizer instrument (KPIC, Mawet et al., 2016) was commissioned at the Keck II telescope in 2019. It consists of two modes:

- the off-axis "direct spectroscopy" (DS, non-coronagraphic) KPIC mode, where a single-mode fiber is placed at the location of a previously known planet orbiting several diffraction beamwidths away from the central star, and
- a "vortex fiber nulling" (VFN, Echeverri et al., 2019) mode, where a charge-1 or charge-2 vortex mask is placed in an intermediate pupil plane to cancel out starlight, and where the resulting starlight plus any exoplanet light located within or close to the diffraction limit is injected into an on-axis single-mode fiber.

In both cases, the single-mode fiber feeds the NIRSPEC high-resolution spectrograph (R ~ 37,500) operating at K-band. The regular DS mode has already successfully observed all previously known extrasolar giant planets accessible from Keck, identifying key molecules in their atmosphere and measuring their radial velocity and spin velocity (for known inclination). These current science capabilities are well illustrated by, e.g., KPIC high-resolution observations of the four HR 8799 planets (Figure 7.3;

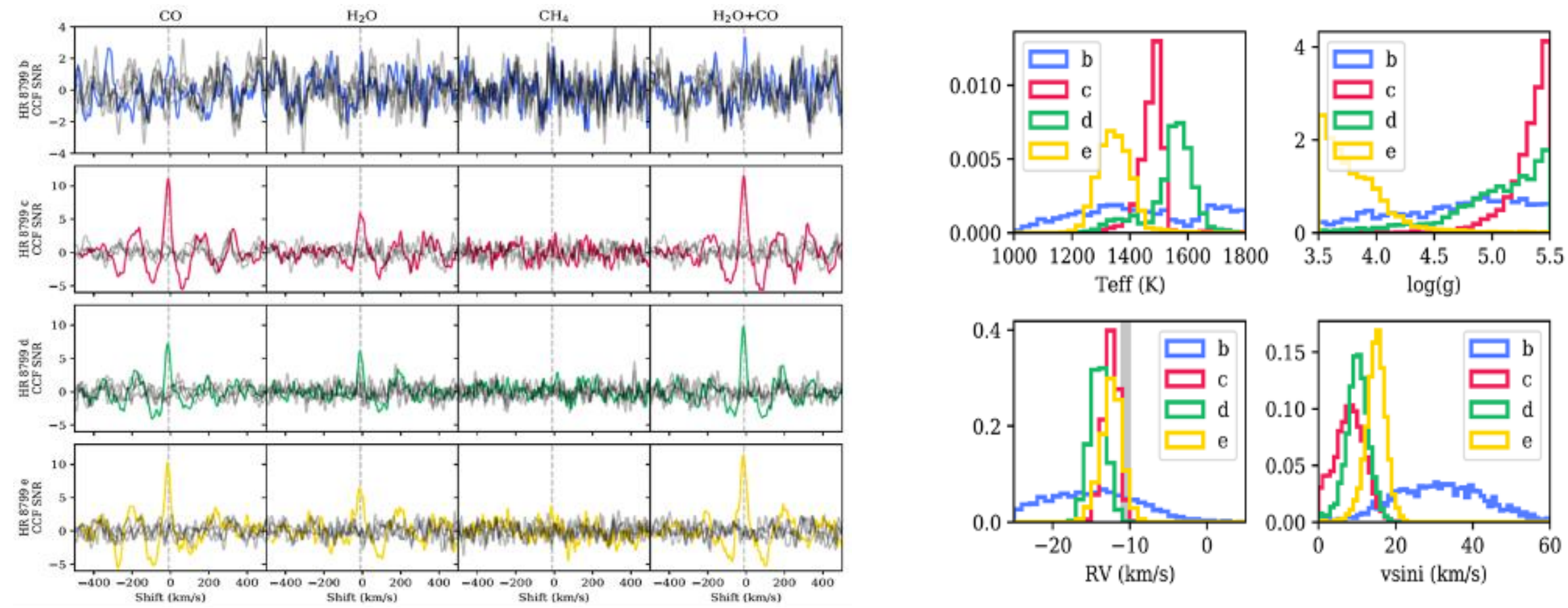

Figure 7.3: (Left) Cross-correlation functions with individual molecular species obtained by KPIC for all four HR 8799 planets. HR 8799 c, d, and e have strong detections of CO and $H_2O$ individually, whereas HR 8799 b is only weakly detected when combining $H_2O$ and CO molecular templates. (Right) Posterior distributions of the main planetary parameters for each HR 8799 planet, based on KPIC observations. (From Wang et al., 2021, reproduced with permission from AAS.)

Wang et al., 2021), where posterior probabilities were obtained for the planets' effective temperatures, surface gravities, radial velocities, and spins.

The VFN mode, which basically operates as a continuous interferometric nuller where opposite parts of the entrance pupil are nulled pairwise, was commissioned at Keck II in 2022. The faintest companion detected by the VFN to date (Echeverri et al., 2024) has a flux ratio of 400:1 and is located at a separation of 1.1 λ/D. With further improvements in pointing, wavefront control and detector fringing, the goal is to move closer to the exoplanet regime by targeting young stars with Gaia/Hipparcos proper motion anomalies compatible with the presence of planetary companions within the VFN 10–100 mas sensitivity region. Interestingly, we note here there is potential for synergistic observations of the same targets with the K-band Keck VFN, which provides high spectral resolution measurements, and the H-band CHARA phase closure instruments, which can pin down the planet location within the VFN field of view. We recommend that a joint Keck VFN/CHARA MIRCX/MYSTIC observing program be established for these targets, working our way from the modest contrast required for 100–1000:1 binaries to the one required for the detection of extrasolar giant planets. Such a pilot observing program to follow-up Keck VFN and Gaia DR3 low-mass companions was recently selected for observations from CHARA with the MIRCX/MYSTIC beam combiners (PI: Aniket Sanghi, 2024).

### *7.3.8 Ancillary science cases*

In addition to being testbeds for technology development and data reduction strategies, ground-based interferometers can provide important contributions of ancillary data for selecting the optimal targets for LIFE. This includes measuring stellar radii, characterizing stellar surfaces (star spot activity), searching for companions, and studying exozodiacal light. The community should seek to take advantage of synergies with ongoing and future projects. For example, SPICA is a new six-telescope visible light combiner being commissioned at the CHARA Array (Mourard et al., 2022). The goal of the SPICA project is to conduct a large and homogeneous survey of stellar parameters across the HR diagram. This provides a potential avenue for interferometrically surveying all stars under consideration by LIFE.

# Section 7 Recommendations

| Objectives | Recommendations | Near-term Actions (< 3 years) | Long-term Actions (> 3 years) |
|---|---|---|---|
| 1. Test concepts needed for the technical development of LIFE.<br>2. Optimize data reduction codes to mitigate systematics. | **Use ground-based infrastructure for technical development of nulling interferometry and science demonstrations in support of the LIFE mission.** | Seek funding to:<br>• build nullers at VLTI/CHARA/MROI, and<br>• improve sensitivity and wavelength coverage at LBTI. | Science demonstrations:<br>• Push sensitivity limits for ground-based instrumentation.<br>• Detection and characterization of brown dwarfs, hot Jupiters, and exozodiacal light. |
| 3. Provide ancillary data for selecting the optimal targets for LIFE. | **Use ground-based interferometers to obtain ancillary data on targets under consideration for LIFE.** | Seek opportunities and synergies with CHARA/SPICA survey and EPRV support. | Characterize stellar surfaces, measure radii, search for companions, and study exozodiacal light. |
| 4. Pool resources for ambitious ground-based work, technology development, and observing programs. | **Create a Nulling Interferometry Center.** | Hold a virtual workshop to plan such a center. | • USA: ATI/MRI/MSIP proposal for major nulling instrument.<br>• White papers for next Decadal Survey and other opportunities. |
| 5. Push the development of Asgard-NOTT to a performance level where it can detect and characterize a massive young planet from the ground. | **Make Asgard-NOTT a success and learn from the operational experience.** | Support the instrument design process. | Helping the consortium with the exploitation of technical/commissioning data. |

# 8 General (non-exoplanet) astrophysics with a LIFE-like mission

## 8.1 Introduction

In future decades, astronomers will take for granted the high-infrared sensitivities made possible by JWST and ELTs, but the angular resolution will nevertheless be lacking compared to long baseline interferometers such as VLTI, CHARA, and other future arrays, as well as ELTs. The prospect of a space interferometer like LIFE would have the potential to combine the best of both worlds—with infrared sensitivity on par with JWST (from four or more 3-meter class space apertures) and baselines > 200 meters that would roughly match the angular resolution of current ground-based interferometers. This section explores some of the potential areas of scientific inquiry beyond exoplanet research that could benefit from LIFE's capabilities. It also examines the impacts on the fundamental architecture of the LIFE mission needed to accommodate these exciting research avenues.

## 8.2 Science opportunities from an infrared interferometer in space

The science potential of LIFE for non-exoplanet science remains largely unexplored, although early studies of TPF-I and Darwin and recent work on the *Planet Formation Imager* (PFI, J. D. Monnier et al., 2018) form a starting basis. LIFE will excel at the traditional interferometry strengths of stellar diameter measurements, binary orbits, and imaging of stellar surfaces, although much of this can be done on the ground for bright objects. Measuring diameters of exciting new classes of objects, such as young stars, cool stars, brown dwarfs, and interacting binaries, will require < 0.1 milliarcsecond resolution, translating to baseline requirements of more than a kilometer.

The PFI Project has explored the potential for near and mid-IR interferometers in studying protoplanetary disks and directly detecting the forming exoplanets along with their circumplanetary disks. LIFE could detect disk gaps made by forming protoplanets on a scale beyond the Atacama Large Millimeter/submillimeter Array (ALMA) or ELTs while directly detecting the accreting planets themselves that are making the gaps and spatially resolving circumplanetary disks at a fraction of the Hill radius. Even circumplanetary disk accretion lines could be measured with the LIFE combiners' medium spectral resolution. The PFI science case is very compatible with LIFE in terms of angular resolution, sensitivity, and field of view. This science is still "exoplanet"-related, but would add a rich new dimension that would also be synergistic with ALMA, ELTs, JWST, and the Square Kilometre Array (SKA).

Active galactic nuclei (AGN) are excellent targets for infrared interferometry due to their relatively compact emission that doesn't require a huge field of view. LIFE would be able to resolve the parsec-scale dusty environments from near- to mid-IR. This would build on the work of VLTI/GRAVITY by allowing lower surface brightness emission to be mapped and observing many more targets at smaller angular scales. In combination with reverberation mapping, a program to measure black hole masses and distances on a large sample will be a powerful new probe of galaxy formation. With the proper spectral resolution, LIFE could study how jets are launched and collimated in a powerful new way. Lastly, the population of binary

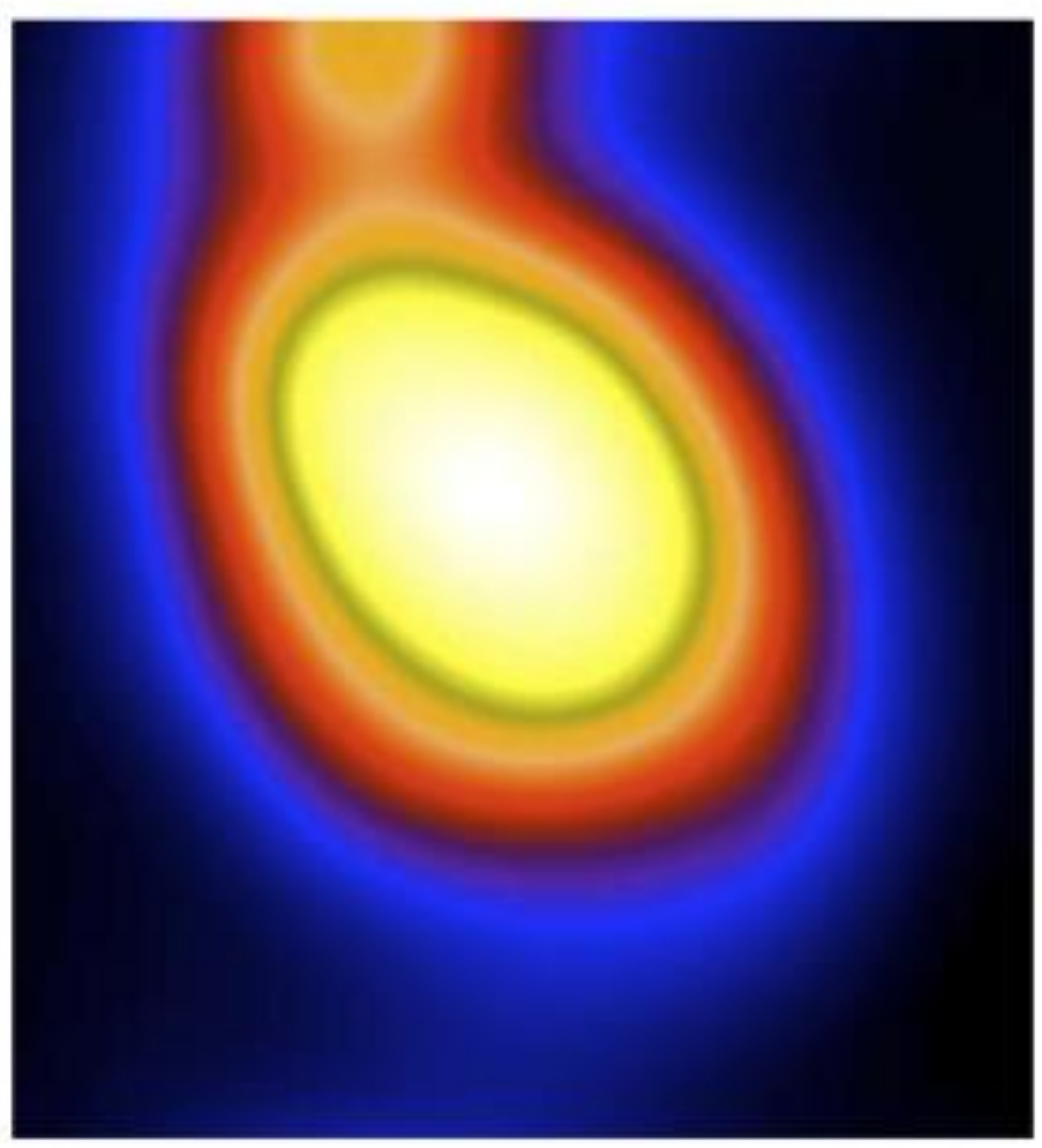

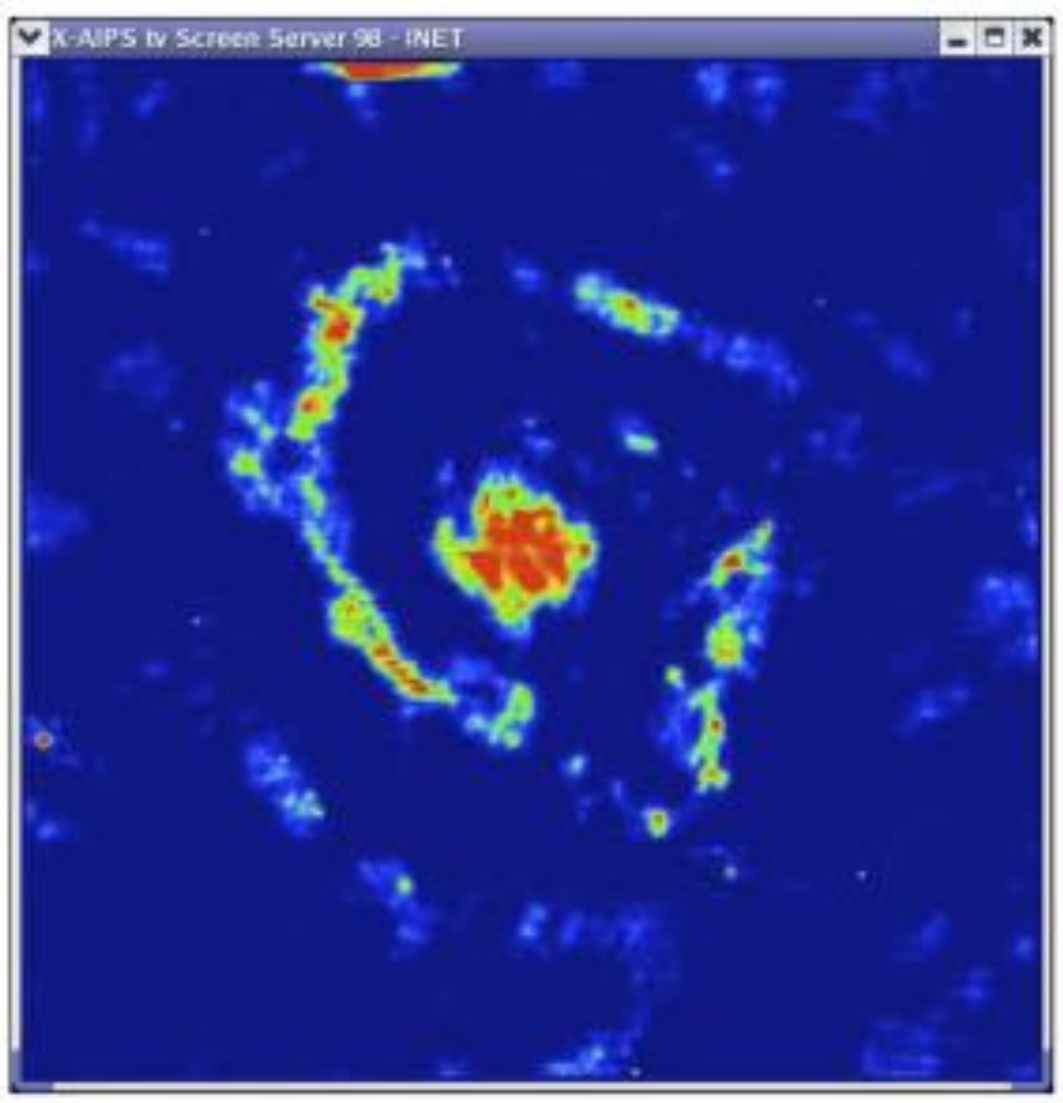

Figure 8.1: The original TPF-I science working group report (Lawson et al., 2007) explored how a mid-IR space interferometer could image the distant extragalactic universe. Even with a small field of view, a LIFE mission with imaging could prove potent. Here, we see simulated images of an M51-type galaxy at z = 3 as observed with JWST (left) and a space interferometer like LIFE (right). However, enabling this imaging power for LIFE incurs additional cost and complexity that might risk the primary exoplanet mission, requiring a careful trade study as requirements are traded against cost, schedule, and risk. Credit: NASA

supermassive black holes (SMBH) could be spatially resolved for the first time as they harden their orbits from the parsec scale, yet before they merge as sources of low-frequency gravitational waves recently detected by the NanoGrav collaboration (Agazie et al., 2023).

LIFE will have close to the sensitivity of JWST in some bands, with orders of magnitude finer angular resolution. JWST has now resolved complex substructures in distant galaxies, but the sizes and details of the individual star-forming regions remain too small to resolve with a single aperture. Operating in a kind of closely packed Fizeau mode, the 4+ LIFE apertures could make complex, high-dynamic range images with many times better resolution than JWST (< 0.1" in mid-IR) with excellent imaging fidelity (see, for example, Figure 8.1). Alternatively, the LIFE array could spread apart and could also move to longer baselines and image even small scales (< 0.01") as required, but with smaller FOV and limited *uv* coverage. For reference, star-cluster-forming molecular clumps of order 0.5 pc would subtend 0.01” at a distance of 10 Mpc.

Other less developed, though powerful, new science cases for an infrared space interferometer include measuring Einstein-ring radii of microlensing events, observing the evolution of dust in supernovae ejecta, dissecting individual protostars in the unique environments of the Large and Small Magellanic Clouds (low metallicity, irregular galaxies), and measuring the diameter/binarity of asteroids and comets in our solar system.

## 8.3 Technical impacts on LIFE design for non-exoplanet science

The impact on the design of the LIFE mission might be substantial to accomplish all the non-exoplanet science goals mentioned above. We find four main difficulties:

1) LIFE plans to use the bright exoplanet host stars as a fringe tracker, leaving faint targets unavailable to observe without a different way to do fringe tracking required for longer coherent exposures.

2) Higher angular resolutions will need longer baselines, up to many kilometers.

3) Spreading only four apertures over hundreds of meters limits the *uv* coverage, allowing only images of simple objects or very long observations to collect the necessary *uv* coverage.

4) Most interferometric combiners have a small field of view—only the size of the primary beam or slightly larger.

Here, we briefly outline initial thoughts on how to address these challenges.

a) One strong selling point for LIFE is that it does not need an elaborate method for stabilizing the interferometric array for long, coherent exposures because the prime science targets are bright enough for fringe tracking. Adding the ability for long exposures on faint targets will require either phase referencing on a nearby bright star or a highly precise way to monitor the relative spacecraft positions (at the micrometer level!). For phase referencing, the system could either chop between a nearby reference star and the faint target to monitor static drifting between the spacecraft for phase referencing, or might need simultaneous dual-star mode, as is done on the ground. The latter option would require a much larger change in architecture, including a star separator at the collectors, separate delay lines, and beam combiners, etc. We note the fringe tracking on the guide star could be done by the near-IR delay line/combiner, allowing long exposures on the mid-IR combiner, for instance. This could limit the wavelength ranges but could be done with a less dramatic impact on infrastructure.

   Another option is to develop extremely precise inertial sensors/accelerometers that could keep the system phased up over long periods without the need for active fringe tracking. Current accelerometers cannot achieve this, though perhaps a modification of the LISA pathfinder test masses concept could be used as an inertial sensor, allowing the spacecraft positions to be known with << 1 micrometers relative to an ideal co-moving inertial reference frame (Davila Alvarez et al., 2021).

b) The *uv* coverage of only four (or five) apertures is poor when covering 100+ meters. This fundamentally limits the complexity of a snapshot image that can be attained. However, for slowly changing science, the *uv* coverage can be accumulated by moving the telescopes around in baseline lengths and angles, though at a very high cost in observing time. Thus, imaging complex objects will be very expensive in time, negating the possibilities of surveys.

   One exception to this would be a Fizeau imager, where the telescope light is combined in an image plane. This would enable interesting imaging capabilities if the spacings were only a few times the telescope diameter. Note, however, that this could pose some safety risks for formation flying, as well as require a totally different beam combiner to be added along with associated subsystems.

c) Most of the beam combiners that are envisioned for nulling interferometry involve a single-mode fiber. This fiber is generally coupled to the diffraction-limited field of view of a single aperture, approximately to the arcsec level in mid-IR and 2–4 times smaller in the near-IR. Some compelling science cases, such as imaging complex star-forming regions in distant galaxies from JWST follow-up, would benefit from a wider field of view. One solution contemplated is to "mosaic," where many pointings are stitched together over time. This is not practical for many objects, though, since even one individual field of view is itself expensive, requiring time to cover the *uv* plane through reconfiguring the telescope separations. An alternative solution is to give up some of the angular resolution and to combine in a Fizeau mode, as mentioned above. Fizeau combination directly creates an image plane with a synthetic aperture and can inherently have a wide field of view.

It is likely some science cases can be accommodated without major changes to the LIFE design, while others require extensive changes to the fundamental design. Trading these off will be an important job in the coming years.

## 8.4 Conclusions

An infrared space interferometer has the potential for revolutionary discoveries, but tapping into its full capabilities may necessitate significant alterations to the LIFE mission's design and technical specifications. While some scientific objectives, such as imaging protoplanetary disks and active galactic nuclei, can fit within the existing architecture, others like measuring brown dwarf diameters, imaging young galaxies, or large-scale reverberation mapping programs will require substantial changes, including longer baselines, dual-star phase referencing or Fizeau mode combination. To ensure an optimal mission design, it is crucial for the LIFE mission to conduct a trade study early to determine the feasibility and scope of non-exoplanet science.

Expanding LIFE’s capabilities for non-exoplanet science could significantly reduce the mission time available for exoplanet research, as many of these topics, such as protoplanetary disk studies, AGN imaging, and galactic structure surveys, also require time-intensive observing campaigns. Addressing such objectives would likely demand a mission specifically tailored to their unique requirements, including a wider field of view and wide-angle phase referencing—design elements fundamentally different from those optimized for exoplanet nulling in LIFE’s current concept. Furthermore, since the days of TPF-I and Darwin, astronomy has evolved significantly, with exoplanet science emerging as a dominant field. Given its importance, we advocate that LIFE prioritizes the core exoplanet mission and ancillary science achievable with its baseline design, avoiding the added complexity and risk of incorporating non-core objectives.

# Section 8 Recommendations

| Objectives | Recommendations | Near-term Actions (< 3 years) | Long-term Actions (> 3 years) |
|---|---|---|---|
| 1. Verify LIFE's ability to do non-nulling science. | **Develop a space-based interferometric imaging simulator.** | Leverage expertise at the Jean-Marie Mariotti Center (JMMC) to help extend the capabilities of current ground-based imaging simulators to a space-based variety. | Utilize simulator to trade-off science cases (see below). |
| 2. Evaluate science impact to decide on LIFE non-nulling observation modes. | **Determine whether non-nulling "faint" modes should be supported and incorporate this into the LIFE architecture design.** | • Develop best small field-of-view (FOV) science scenarios if phase referencing is feasible.<br>• Invite expert input.<br>• Provide a prominent platform to non-exoplanet science discussions during LIFE meetings.<br>• Publish a paper in the LIFE series on these capabilities . | • List possible modes with a cost-benefit analysis, including:<br>- NIR phase referencing<br>- Chopping<br>- Dual beam<br>- Fizeau<br>• Identify compelling non-exoplanet science that aligns with the baseline mission design but avoids major added complexity. |

# 9 Building a community for LIFE

## 9.1 A bit of history

The genesis of a nulling mid-IR interferometer to detect and characterize terrestrial planets dates back to a 1978 *Nature* article by Ron Bracewell (Bracewell, 1978). The European Space Agency held a set of conferences between 1985 and 1991 that concerned interferometry in space, and most especially from the Moon (ESA SP-1150[9]). In these studies, the search for planets that could be abodes of life figured prominently and set the scene for the discovery of the first exoplanets. As the reality of planets beyond the solar system became apparent with the discovery of 51 Pegasi b (Mayor & Queloz, 1995) and dozens of other gas giants, the Darwin project was born in Europe (Léger et al., 1996a,b) and in the US, a version of what was known as the Terrestrial Planet Finder (Angel & Woolf, 1997). In 1998, a first success of mid-IR nulling on the ground was achieved using sub-apertures of the Multiple Mirror Telescope (MMT; Hinz et al., 1998).

Promising studies were initiated on both sides of the Atlantic. For nearly a decade into the mid-2000s, NASA supported the Terrestrial Planet Finder-Interferometer (TPF-I), and the European Space Agency (ESA) supported the Darwin program. Darwin received repeated endorsements from ESA's advisory bodies resulting in an extensive technological program that was initiated in 2001, aiming towards the reality of destructive ("nulling") interferometry being the technology implemented for Darwin. This program was to last until about 2010 and several of the technologies developed have been used on experiments launched into space—most spectacularly through the Swedish technical demonstrator of formation flying in space. PRISMA consists of the two satellites, Mango and Tango, developed under Swedish leadership with participation mainly by France and Germany. TPF-I received an endorsement for further study and technology development in the 2000 US National Academy of Sciences Decadal Report *Astronomy and Astrophysics in the New Millennium*.

NASA accepted a recommendation to spend $200 million on precursor studies and technology development and established the Origins program. ESA quickly opened a study office for the Darwin mission that lead to a collaboration between the two space agencies, which culminated in the design (Lay, Martin and Hunyadi, 2007) that forms the basis of the LIFE mission today. During this period, significant technology development on both continents led to a laboratory demonstration of the required ~$10^{-6}$ nulling performance (Martin et al., 2006) and formation flying (later published in Delpech et al., 2014).

However, despite this excellent technical progress in the lab, the successful demonstration of a four-beam (dual Bracewell) astronomical nuller at the Keck interferometer, and the continuing advance of exoplanet discovery, enthusiasm for both TPF-I and Darwin began to fade within the astronomical community and the space agencies. The 2010 National Academies' Decadal Survey *New Worlds, New Horizons in Astronomy and Astrophysics* supported the continuing development of JWST and proposed the new *Wide*

[9] The report is available here: https://esamultimedia.esa.int/multimedia/publications/SP-1150/SP-1150.pdf

*Field InfraRed Survey Telescope* (WFIRST) mission (now *NASA's Nancy Grace Roman Space Telescope*). But the report rescinded earlier Decadal Survey recommendations for the Space Interferometer Mission (1990 and 2000). The report did recognize the excitement of detecting and characterizing other Earths, recommending continuing technology funding for a future mission, but was agnostic about the design of the facility:

> *"The optimum strategy depends strongly on the fraction of stars with Earth-like planets orbiting them. If the fraction is close to 100 percent, then astronomers will not need to look far to find an Earth-like planet, but if Earth-like planets are rare, then a much larger search extending to more distant stars will be necessary. With this information in hand, ambitious planning can begin to find, image, and study the atmospheres of those Earth-like planets that are closest to our own…The culmination of the quest for nearby, habitable planets is a dedicated space mission…It is too early to determine what the design of that space mission should be…It is not even clear whether searches are best carried out at infrared, optical, or even ultraviolet wavelengths."*

Shortly after the 2010 report, work on the interferometric version of an exoplanet detection and characterization mission came to a halt with the technology funding in the US going almost exclusively to the visible light options, either a coronagraph or an external starshade.

## 9.2 Lessons from Darwin and TPF-I

As explained and motivated in the previous section, today we are in a completely different situation compared to ~15 years ago, when TPF-I and Darwin were put on hold. First and foremost, the exoplanet community came a long way in understanding the exoplanet population, which allows us to quantify the expected science return of a LIFE-like (or HWO-type) mission. We *know* there is a significant population of temperate terrestrial exoplanets out there; we *know* some of the nearest stars do harbor planetary systems including rocky worlds. Furthermore, key technologies for (nulling) interferometry—on ground and in space—did mature and new approaches became available that did not exist in the 2000s (see, e.g., Defrère, Léger, et al., 2018). Finally, and what has not addressed in previous sections, we have learned how to build large, complex, cryogenic systems as the JWST. Significant experience was especially gained at mid-infrared wavelengths with the MIRI instrument (Wright et al., 2023) and also with other cryogenic missions such as Herschel, featuring a 3.5-meter primary mirror in space (Pilbratt et al., 2010).

Yet, it is important to be aware of the challenges that TPF-I/Darwin faced, so that going forward, a LIFE-like mission can continue to proactively address them knowing they do not represent showstoppers.

Key lessons learned include:

- **Instrument complexity.** Although the underlying physics experiment of nulling starlight had been demonstrated in the laboratory at the $10^{-6}$ level, on both sides of the Atlantic, the implementation was complex and the transfer to cryogenic temperatures daunting. (We note the NICE experiment at ETH Zurich (Ranganathan et al., 2024) is addressing this challenge head-on.)

- **System complexity.** The nulling interferometer specified for TPF-I/Darwin required at least five spacecraft, four collectors as well as a beam-combining spacecraft carrying the complex nulling beam combiner. This appeared to be more expensive and complex than either a single spacecraft carrying an internal coronagraph or a two-spacecraft telescope plus starshade concept.

- **Beam Transport.** The nulling interferometer design requires maintenance of low background as multiple diffraction-limited beams traverse tens to hundreds of meters between the spacecraft. Furthermore, ideally, the baselines between the spacecraft can be freely adapted for each target system and the whole array rotates around the central axis to modulate the signal. (We note that a fixed set of baseline lengths as well as alternative signal modulation schemes are being investigated.)

- **Limited information content.** The *uv*-plane coverage of the interferometer is limited by the small number of telescopes, as well as by limits to their separations and angular distribution. It remains to be demonstrated that complex planetary systems with multiple planets of different brightness and potentially clumpy exozodiacal clouds can be successfully analyzed to yield the desired information. (We note the LBTI/HOSTS survey provided promising results concerning the statistically expected level of exozodiacal emission at the very wavelength of LIFE observations (S. Ertel et al., 2020) which suggest clumpy disks may not pose a severe challenge according to Defrère et al., (2010); also first analyses concerning the extraction of multiple planet signals have been carried out (Matsuo et al., 2023).

- **Lack of a broad program of astrophysics.** Darwin/TPF-I examines the universe through a single diffraction-limited beam, perhaps an arcsecond in diameter. While achieving excellent angular resolution within that single patch of sky, the information within that patch is limited by the UV coverage while the rest of the field of view of each aperture is unused. This is to be contrasted by the broad range of astrophysical possibilities a standard flagship observatory equipped with multiple instruments typically offers. (We note that one could consider LIFE in a manner similar to a single-purpose mission, such as cosmic microwave background facilities that are designed, optimized, and operated to address a singular, critical scientific question.)

- **Programmatic imperatives.** Leading up to and following the 2010 Decadal Survey, there was an overriding imperative to complete JWST. This led to clearing away missions like SIM and TPF-I while demanding flagship-sized budgets appealed to only niche communities. (We note the exoplanet community can no longer be considered a niche community in modern astrophysics; also, both the 2020 Decadal Survey as well as ESA's Voyage 2050 survey put the search for habitable and inhabited worlds at the top of their priorities.)

## 9.3 Towards the future

Going forward, it will be important that LIFE builds up and then leverages a large support base that helps turn the vision into reality. Considering classic implementation schemes, a preferred approach would be to keep pushing on ESA to seriously consider a LIFE-like mission. As mentioned in Section 1, the scientific theme of LIFE—the detection and atmospheric characterization of temperate terrestrial exoplanets at mid-IR wavelength—was given top priority in the Voyage 2050 report[10] that was given to the ESA Director for Science. However, not much has happened since and it must be understood that it is still a long way from being a top priority in a report to becoming an adopted ESA mission. Still, there is an opportunity, and one could imagine a scenario where ESA takes the lead and teams up with other agencies (e.g., NASA and JAXA) to develop the LIFE mission, similar to having European contributions to JWST and HWO. What is appealing here is the idea that LIFE and HWO could be seen as a joint and coordinated international effort to address one of humankind's oldest questions in a truly collaborative manner driven by a unified scientific vision. Making progress in this direction would require a concerted effort on addressing and educating ESA member states, industry partners, ESA advisory bodies, and also potential partner agencies.

In parallel, given the somewhat unclear situation with ESA and the financial situation of its Science Program that only recently saw a budget increase for the first time in many years, alternative development and implementation scenarios should be considered. New players have entered the international space arena and also the boundaries between agency-funded and privately-funded (space science) missions have started to blur (French et al., 2022) as prominently showcased by the Lazuli Space Observatory, a planned deep-space telescope featuring a primary mirror larger than that of the Hubble Space Telescope, fully funded through a charitable gift from Eric and Wendy Schmidt's philanthropic organization, Schmidt Sciences[11]. Hence, at this point in time, various options to implement LIFE should be actively explored.

Irrespective of possible implementation strategies, LIFE will require a broad support base, and the following sub-sections discuss different aspects in more detail.

### *9.3.1 Coordinating technology and instrument opportunities*

Implementing a LIFE-like mission requires a great many technological innovations plus a cadre of engineers and scientists capable of bringing them to fruition. Given the expense and long-term scales of space missions, many of these innovations will have to happen on the ground, either in laboratory testbeds or at telescopes being used to demonstrate the power of interferometry in its many forms. Section 7 provided many details and examples. Likewise, Section 3 presented several opportunities for upcoming and future space missions that will help pave the way towards a LIFE-like mission. What is missing at the moment are more coordinated and concerted efforts that link the various activities systematically and synergistically. Many important and impactful puzzle pieces exist, but there is a lack of trying to put them together in a grander picture that tells a greater story. A possible way to address this shortcoming is the creation of an international (virtual) Center for Nulling Interferometry that would serve as a platform to exchange knowledge and expertise and enable coordination of international collaborations.

[10] https://www.cosmos.esa.int/web/voyage-2050

[11] https://www.schmidtsciences.org/schmidt-observatory-system/

### *9.3.2 Coordinating and preparing science opportunities*

As discussed in Section 1, both LIFE and HWO need to have their own unique discovery space, but since they are driven by similar scientific objectives, which they address from two distinct directions (i.e., different wavelength ranges), their combined diagnostic power will outperform each individual mission. Given the similar scientific motivation, many preparatory activities on the scientific side are very similar. This can be leveraged to the benefit of both missions as a larger number of international scholars will actively engage in preparing both missions in parallel. This will make the preparation more robust, efficient, and comprehensive. To this end, one could imagine setting up joint working groups that exchange knowledge and tools and coordinating efforts, such as preparatory observations, at an international scale.

### *9.3.3 Building an inclusive and long-lasting international community for LIFE*

Finally, a LIFE-like mission that seeks to serve all humankind needs to be based on a strong, vibrant and inclusive community. This requires:

- **Professional outreach, recruiting, and retention:**

  The development and execution of the LIFE mission will engage more than a single generation of scholars and engineers. Considering that some people will work for only parts of the project, it is important to make the project appealing and rewarding, particularly for people who join the project at an early stage. Therefore, we need to work on methods to attract, train and retain the necessary cadre of scientists and engineers, including early-career researchers. It is also essential to develop a plan to transfer knowledge from the current generation to those who will carry the vision through the programmatic and technical challenges to come. This will involve offering near-term opportunities in scientific research, e.g., observing with ground/space-based facilities, working on models of exoplanet atmospheres to identify key requirements on the LIFE mission and its instrumentation (such as sensitivity, targets, spectral resolution, wavelength range), and developing laboratory testbeds and new technologies (e.g., mid-IR photonics) to challenge engineers and physicists.

- **Building broad-based and interdisciplinary community:**

  As we develop the community for LIFE, it is important to engage as many scholars as possible from a diverse set of research fields, for example, planetary science, (astro)biology, chemistry, and geosciences. Leveraging the inherent interdisciplinarity of a mission like LIFE will help design the best possible mission and increase its scientific impact. Therefore, when considering the main science objectives, and a potential ancillary science program, one should try to broaden the scientific scope towards other disciplines. Support in this direction can come from the large number of interdisciplinary science centers addressing questions related to origin and occurrence of life in the that have been founded at various academic institutions over the last years, such as the Leverhulme

Centre for Life in the Universe (University of Cambridge, UK)[12], the Origins of Life Initiative (Harvard University, USA)[13], the Center for the Origins of Life (University of Chicago, USA)[14], the Centre for Origin and Prevalence of Life (ETH Zurich, CH)[15], or the Earth–Life Science Institute (ELSI) in Tokyo[16].

- **Outreach to the astronomical community:**

  An endeavor of this magnitude will need the support of the broad astronomical community. It is important to bring the vision of LIFE to all fields in astronomy, whether they directly benefit from it or not. Constant presence at dedicated conferences is a must, but also participation in broad-based conferences where the relevance of the LIFE mission is highlighted. We need the full support of the exoplanet community, and therefore we should take into consideration their scientific interests—for example, exoplanet demographics and how LIFE can contribute to their understanding. At the same time, we should be confident in pursuing our vision knowing that exoplanet science, understanding other planetary systems, and the quest for life outside the solar system are topics that continue to significantly increase in importance and attractiveness among all fields of modern astrophysics (Mérand et al., 2021).

- **Outreach to the public:**

  A strategy to maintain the public interest in the search for habitable and inhabited worlds is to build on successes of Kepler, TESS, CHaracterising ExOPlanet Satellite (CHEOPS), and JWST, and soon Roman, PLAnetary Transits and Oscillations of stars (PLATO), and Ariel and other missions through articles, social media, and other communication channels. We must highlight the importance of multi-mission, multi-technique approaches to refine our understanding of our place in the universe. We need to bring the general public on board by creating outreach programs that captivate them already during the mission preparation, both from the scientific and engineering perspective. Also, the LIFE mission is inspiring to the public and a great vehicle to engage diverse groups of young people in science, technology, engineering and mathematics (STEM) disciplines.

- **Promote diversity, equity and inclusion:**

  Cultivating an intersectional team—spanning different nationalities, identities, and generations—strengthens mission integrity by ensuring a wider array of lived experiences are applied to problem-solving (Mandt, 2023). The LIFE mission should work towards a fully diverse and inclusive team not only across the consortium but also within each team. To get the most out of the money and time invested in such an undertaking, the LIFE team needs to be diverse at all levels. It should also be remarked that it is not enough to guarantee diversity to "just" bring people from different cultures together. There is a diversity of skills, expertise, experience, and perspective that is needed to create a diverse team. We also need to favor collaborative work and a friendly and tolerant environment. We will only be able to overcome challenges in the best, and most creative ways by enabling all who can contribute to fully participate, at all stages of project development.

---

[12] https://www.lclu.cam.ac.uk

[13] https://origins.harvard.edu

[14] https://originsoflife.psd.uchicago.edu

[15] https://copl.ethz.ch

[16] https://www.elsi.jp/en/

# Section 9 Recommendations

| Objectives | Recommendations | Near-term Actions (< 3 years) | Long-term Actions (> 3 years) |
|---|---|---|---|
| 1. Establish a Center for Nulling Interferometry. | **Create a virtual platform for the international community to coordinate knowledge exchange and collaborations.** | • Send out a survey to understand the needs and desires of the community.<br>• Establish a website and communication platform to initiate the exchange. | • Provide routes to joint funding opportunities.<br>• Host meeting series to encourage knowledge transfer. |
| 2. Advocate to national space agencies to support developing a large-scale space-based interferometer mission. | • **Engage with space agencies and continue to push for a space interferometric exoplanet mission.**<br>• **Encourage the agencies to pursue these science cases with urgency and ambition.** | • Continue engaging with ESA to encourage the selection of a LIFE-type mission for L5.<br>• Strongly support efforts by JAXA to implement space interferometry missions (SEIRIOS, SILVIA).<br>• Support the small interferometric mission study led by TU Delft. | Answer any relevant calls for mission studies and technology development focused on interferometry. |
| 3. Build a community surrounding interferometry for exoplanets. | • **Engage the academic community via workshops and conferences, including SPIE telescopes.**<br>• **Promote outreach to the wider public about the benefits of interferometry.** | • Continue building on the growing success of the interferometry conference at SPIE.<br>• Offer new conferences targeting the use of interferometry for the study of exoplanets.<br>• Engage with science communicators and the public press to "demystify" the use of interferometry as a measurement technique | • Encourage the study of astronomical optical/IR interferometry as a topic for university courses and even high school curricula.<br>• Offer fellowships for early career researchers focused on interferometry. |
| 4. Investigate alternative funding routes and implementation schemes. | **Engage with foundations and other non-traditional funding sources.** | Discuss with foundations and private donors to garner interest. | Establish a stable and reoccurring funding environment. |

# 10 Epilogue

At the time of writing, considerable progress has been made in the field of interferometry and in the conception of search-for-life missions since the KISS workshop in late 2022.

The LIFE project and its community have continued to advance the fidelity of a mission concept for a full-scale mid-infrared (mid-IR) space-based nulling interferometer. The LIFE science objectives are being refined in detail, while ongoing work on terrestrial planet retrieval methods has provided important insights into the correct treatment of water vapor, and novel population-based approaches are under exploration. The feasibility and scientific performance of the LIFE mission have been elevated to a new level through the development of two new instrument simulators capable of predicting the impact of instability noise on measurements. Based on these advances, top-level mission requirements are continually re-evaluated and converging toward a significant upscaling of mission sensitivity.

On the hardware side, these achievements are mirrored by the first mission-specification deep null achieved with the warm precursor of the NICE bench, which will soon transition to cryogenic operation. In parallel, a second nulling test bench with a four-beam design is under development at TU Delft. The same team is also leading, in collaboration with ESA, a study into a single-spacecraft interferometer concept aimed at mitigating technological risks associated with formation flying.

In addition, numerous small space missions outside Europe have been funded or are under active consideration since the time of the workshop. In the United States, missions such as STARI (University of Michigan) and the compact delay line project (RPI) have been funded as CubeSat-scale demonstrators. These will test key technologies including formation-flying beam transport and fiber injection, while also aiming to relax stability requirements and reduce complexity in future missions. In Japan, two formation-flying missions—SEIRIOS and SILVIA—have been proposed for launch in the early 2030s. SEIRIOS has been funded, pending a final review prior to launch, and aims to perform one of the first demonstrations of fringe tracking in space. The *Space Interferometer Laboratory Voyaging towards Innovative Applications* (SILVIA), one of the final candidates for a JAXA M-class mission, will push the boundaries of achievable stability in multi-element formation-flying constellations. Other mission concepts worldwide also aim to advance the technological readiness of space interferometry.

On the ground, the NOTT instrument for the VLTI is nearing completion and has received funding to continue the work of the HOSTS survey, studying exozodiacal dust that represents a potential noise source for mid-IR space interferometers. The LBTI nuller is also being upgraded and will continue contributing to this science case.

Since the recommendation of the HWO by the Decadal Survey on Astronomy and Astrophysics 2020, the HWO community has rapidly grown. Under the GOMAP framework, the START team has concluded its work and delivered its final recommendations to NASA, focusing on the scientific exploitation of HWO. With the establishment of the HWO Project Office at GSFC, science and technical assessment groups have begun iterating on initial mission design concepts.

The primary purpose of this report, however, is to present the recommendations of the participants as formulated during the 2022 workshop and to document their underlying rationale. Where relevant, subsequent progress has been incorporated to better reflect the current state of the field. Consequently, some of the recommendations in this report have already been implemented or are currently underway. References to relevant literature have been added throughout the text.

The coming years will be crucial for the development of a mid-IR, space-based interferometer. Recent discussions—for instance, those surrounding K2-18 b—have shown that the era in which biosignature detections can be seriously debated within the scientific community is rapidly approaching. At the same time, these developments underscore the importance of comprehensive, multi-wavelength, and multi-instrument datasets in reliably identifying and interpreting biosignatures in exoplanetary spectra. The need for a LIFE-like mission is increasingly evident.

To enable a coordinated LIFE-HWO partnership in the 2040s, the maturation of key technologies must begin now. In parallel, the design of the most suitable nulling interferometer architecture should converge by the end of the decade, ensuring sufficient time for detailed mission definition. This will require sustained investment in nulling interferometry and related scientific and technological fields. With the remarkable progress of the past year, the strong heritage from the Darwin and TPF-I mission studies, and broad community support, the LIFE project is well positioned to take on this once-per-generation challenge.

# 11 Acronyms and Abbreviations

| | |
|---|---|
| AAS | American Astronomical Society |
| AERO-VISTA | Auroral Emissions Radio Observer - Vector Interferometry Space Technology using AERO |
| AFKG | A-type star, F-type star, a G-type star, K-type |
| AGN | Active galactic nuclei |
| ALMA | Atacama Large Millimeter/submillimeter Array |
| AMBER | Astronomical Multi-Beam combiner |
| AO | adaptive optics |
| ARMADA | ARrangement for Micro-Arcsecond Differential Astrometry |
| ASTERIA | Arcsecond Space Telescope Enabling Research in Astrophysics |
| AT | Auxiliary Telescope |
| ATI | National Science Foundation’s Advanced Technologies and Instrumentation program |
| AU | astronomical units |
| AWG | arrayed waveguide grating (spectrometer) |
| CGI | Coronagraphic Instrument on NASA’s Nancy Grace Roman |
| CHARA | Center for High Angular Resolution Astronomy |
| CHEOPS | CHaracterising ExOPlanet Satellite |
| CLICK | CubeSat Laser Infrared CrosslinK |
| CRIRES | Cryogenic Infrared Echelle Spectrometer |
| DRM | Design Reference Mission |
| DS | direct spectroscopy |
| ELFN | excess low-frequency noise |
| EPRV | extreme precision radial velocity |
| ESA | European Science Agency |
| ESO | European Southern Observatory |
| ESPA (class) | EELV Secondary Payload Adapter |
| ETS-VII | Engineering Test Satellite No. 7 |
| ExEP | NASA's Exoplanet Exploration Program |
| FF | formation flying |
| FFRF | formation-flying radio frequency |
| FLUOR-IOTA | Fiber Linked Unit for Optical Recombination (at the) Infrared Optical Telescope Array |
| GEO | geostationary orbit |
| GLS | gallium lanthanum sulphide (photonic chip) |
| GMT | Giant Magellan Telescope |
| GOMAP | Great Observatories Maturation Program |
| GPS | global positioning system |
| GRACE | Gravity Recovery and Climate Experiment |
| GRIP | Generic data Reduction for nulling Interferometry Package |
| GSFC | NASA's Goddard Space Flight Center |
| HabEx | NASA's Habitable Exoplanets Observatory |
| HEO | high-Earth orbit |
| HITRAN | High Resolution Transmission (molecular spectroscopic database) |
| HOSTS | Hunt for Observable Signatures of Terrestrial Systems survey |
| HWO | NASA's Habitable Worlds Observatory |
| HZ | habitable zone |
| IO | integrated optics |
| JAXA | Japan Aerospace Exploration Agency |
| JMMC | Jean-Marie Mariotti Center |
| JWST | James Webb Space Telescope |
| KID | kinetic inductance detector |
| KISS | Keck Institute for Space Studies |
| KPIC | Keck Planet Imager and Characterizer |
| LBTI | Large Binocular Telescope Interferometer |
| LEO | low-Earth orbit |
| LIFE | Large Interferometer For Exoplanets |
| LOE | low-Earth orbit |
| LRI | Laser Ranging Interferometer |
| LUCA | Last Universal Common Ancestor |
| LUVOIR | Large Ultraviolet Optical Infrared Surveyor |
| MATISSE | Multi AperTure mid-Infrared SpectroScopic Experiment |
| MC | Monte Carlo |
| mDot | Miniaturized Distributed Occulter Telescope |
| METIS | Mid-infrared ELT Imager and Spectrograph |
| MEUR PROBA-3 | Million Euro Project for On-Board Autonomy-3 |
| Mid-IR | mid-infrared |
| MIRC | Michigan Infrared Combiner |
| MIRI | Mid-Infrared Instrument (for NASA's James Webb Space Telescope) |
| MiXI | Miniature Xenon Ion |
| MMI | multimode interference |
| MMT | Multiple Mirror Telescope |
| MMZ | modified Mach-Zehnder |
| MOU | memorandum of understanding |
| MRI | National Science Foundation's Major Research Instrumentation program |
| MROI | Magdalena Ridge Observatory Interferometer |
| MSIP | National Science Foundation's Mid-Scale Innovations program |
| MYSTIC | Michigan Young Star Imager at CHARA |
| NASA | National Aeronautics and Space Administration |
| NCCR | National Centre of Competence in Research |
| NExSS | Standards of Evidence for Life Detection report |
| NfoLD | NASA Network for Life Detection |
| NICE | Nulling Interferometry Cryogenic Experiment |
| NIFITS | Nulling Interferometry Flexible Image Transport System |
| NIR | near-infrared |
| NIRSPEC | Near-Infrared Echelle Spectrograph (Keck Observatory instrument) |
| NOMIC | Nulling Optimized Mid-Infrared Camera |
| NOTT | Nulling Observations for exoplaneTs and dusT |
| NSC | nulling self-calibration |
| NSF | National Science Foundation |
| OIFITS | Optical Interferometry Flexible Image Transport System |
| OPD | optical path difference |
| pc | parsec |
| PCA | principal component analysis |
| PFI | Planet Formation Imager |
| PHASES | Palomar High-precision Astrometric Search for Exoplanet Systems |
| PIONIER | Precision Integrated-Optics Near-infrared Imaging ExpeRiment |
| PLATO | ESO’s PLAnetary Transit and Oscillations of stars |
| PRIME | PRecision Interferometry with Mircx for Exoplanets |
| PRISMA | PRecursore IperSpettrale della Missione Applicativa |
| R&D | research and development |
| RF | radio frequency |
| RMS | root mean square |
| SEIRIOS | Space Experiment of InfraRed Interferometric Observation Satellites |
| SILVIA | Space Interferometer Laboratory Voyaging towards Innovative Applications |
| SIM | Space Interferometry Mission |
| SKA | Square Kilometre Array |
| SMEX | NASA Small Explorers class of missions |
| SNR | signal-to-noise ratio |
| SPICA | Stellar Parameters and Images with a Cophased Array |
| STARI | STarlight Acquisition and Reflection toward Interferometry |
| START | Science Technology Architecture Review Team |
| SunRISE | Sun Radio Interferometer Space Experiment |
| SWARM-EX | Space Weather Atmospheric Reconfigurable Multiscale Experiment |
| TanDEM-X | Combination of TerraSAR-X and Digital Elevation Management |
| TESS | NASA’s Transiting Exoplanet Survey Satellite |
| TMT | Thirty Meter Telescope |
| TPF-I | Terrestrial Planet Finder Interferometer |
| TRL | Technology Readiness Level |
| TU Delft | Dutch University of Technology |
| UV | ultraviolet |
| VFN | vortex fiber nulling |
| VINCI | VLTI Interferometer Commissioning Instrument |
| VIS | visibility |
| VISION-NPOI | The Visible Imaging System for Interferometric Observations at NPOI (at the) Navy Precision Optical Interferometer |
| VISORS | Virtual Super-resolution Optics with Reconfigurable Swarms |
| VLTI | Very Large Telescope Interferometer |
| VTXO | Virtual Telescope for X-ray Observations |